\documentclass[SIG,biber]{nowfnt} 

\usepackage[utf8]{inputenc} 
\usepackage{acronym,subcaption}
\usepackage{amsmath,dsfont}
\usepackage{amssymb,amsthm}
\usepackage{tcolorbox}
\usepackage{graphicx}
\usepackage[figuresright]{rotating}

\definecolor{NewColor}{rgb}{0.2,0,0.5}

\newcommand{\E}{\mathds{E}}		 			
\newcommand{\myVec}[1]{{\boldsymbol{#1}}}
\newcommand{\myMat}[1]{{\boldsymbol{#1}}}
\newcommand{\mySet}[1]{\mathcal{#1}}

\newcommand{\Input}{\myVec{x}}
\newcommand{\InputSpace}{\mySet{X}}

\newcommand{\Label}{\myVec{s}}
\newcommand{\LabelSpace}{\mySet{S}}

\newcommand{\myS}{\Label}			 		
\newcommand{\Nusers}{K}

\newcommand{\Niter}{Q}

\newcommand{\Ntraining}{n_t}

\newcommand{\PdfNew}[1]{p}

\newcommand{\SigW}{\sigma_w^2}

\newcommand{\Blklen}{T}			 			

\newcommand{\ObjParam}{\myVec{\theta}^{\rm o}}
\newcommand{\HypParam}{\myVec{\theta}^{\rm h}}

\newcommand{\dnnLabel}{\myVec{s}}
\newcommand{\dnnParam}{\myVec{\theta}}
\newcommand{\dnnFunc}{f_{\dnnParam}}

\newcommand{\csSignal}{\myVec{s}}

\newcommand{\csLatent}{\myVec{z}}
\newcommand{\csMatrix}{\myMat{H}}

\newcommand{\Distribution}{\mySet{P}}
\newcommand{\Data}{\mySet{D}}
\acrodef{cnn}[CNN]{convolutional neural network} 
\acrodef{relu}[ReLU]{rectified linear unit}
\acrodef{dnn}[DNN]{deep neural network}

\usepackage[ruled,linesnumbered]{algorithm2e}  
\SetKwInput{KwData}{\textbf{Initialization}} 
\usepackage{algpseudocode}

\acrodef{ai}[AI]{artificial intelligence}
\acrodef{ekf}[EKF]{extended Kalman filter}
\acrodef{adc}[ADC]{analog-to-digital convertor}
\acrodef{ista}[ISTA]{iterative soft thresholding algorithm}
\acrodef{cs}[CS]{compressed sensing}
\acrodef{bp}[BP]{belief propagation}
\acrodef{doa}[DoA]{direction-of-arrival}
\acrodef{bpsk}[BPSK]{binary phase shift keying}
\acrodef{dtft}[DTFT]{discrete-time Fourier transform}
\acrodef{dnn}[DNN]{deep neural network} 
\acrodef{gan}[GAN]{generative adversarial network} 
\acrodef{gnn}[GNN]{graph neural network} 
\acrodef{gru}[GRU]{gated recurrent unit} 
\acrodef{lstm}[LSTM]{long short-term memory} 
\acrodef{csi}[CSI]{channel state information}
\acrodef{map}[MAP]{maximum a-posteriori probability}
\acrodef{snr}[SNR]{signal-to-noise ratio}
\acrodef{bs}[BS]{base station} 
\acrodef{em}[EM]{expectation maximization} 
\acrodef{iot}[IOT]{Interent of Things}
\acrodef{mimo}[MIMO]{multiple-input multiple-output}
\acrodef{mse}[MSE]{mean-squared error}
\acrodef{mmse}[MMSE]{minimal \ac{mse}}
\acrodef{pdf}[PDF]{probability density function}
\acrodef{rv}[RV]{random variable}
\acrodef{fec}[FEC]{forward error correction} 
\acrodef{lti}[LTI]{linear time-invariant}
\acrodef{wss}[WSS]{wide-sense stationary}
\acrodef{psd}[PSD]{power spectral density}
\acrodef{ser}[SER]{symbol error rate} 
\acrodef{ber}[BER]{bit error rate} 
\acrodef{sgd}[SGD]{stochastic gradient descent}  
\acrodef{awgn}[AWGN]{additive white Gaussian noise} 
\acrodef{ut}[UT]{user terminal}  
\acrodef{ml}[ML]{machine learning}  
\acrodef{rnn}[RNN]{recurrent neural network} 
\acrodef{fc}[FC]{fully-connected}
\acrodef{sic}[SIC]{soft interference cancellation}
\acrodef{pmf}[PMF]{probability mass function}
\acrodef{sp}[SP]{sum-product} 
\acrodef{ista}[ISTA]{iterative soft thresholding algorithm}
\acrodef{pca}[PCA]{principal component analysis}
\acrodef{evd}[EVD]{eigenvalue decomposition}
\acrodef{svd}[SVD]{singular value decomposition}
\acrodef{gan}[GAN]{generative adversarial network}
\acrodef{gcn}[GCN]{graph convolutional network}
\acrodef{admm}[ADMM]{alternating direction method of multipliers}
\acrodef{music}[MUSIC]{multiple signal classification}

\title{Model-Based Deep Learning}

\maintitleauthorlist{
Nir Shlezinger \\
School of Electrical and Computer Engineering,\\ Ben-Gurion University of the Negev, Be'er Sheva, Israel \\
nirshl@bgu.ac.il
\and
Yonina C. Eldar \\
Faculty of Mathematics and Computer Science,\\ Weizmann Institute of Science, Rehovot, Israel \\
yonina.eldar@weizmann.ac.il
}

\issuesetup
{%
 copyrightowner={N.~Shlezinger and Y.~Eldar},
 volume        = xx,
 issue         = xx,
 pubyear       = 2023,
 isbn          = xxx-x-xxxxx-xxx-x,
 eisbn         = xxx-x-xxxxx-xxx-x,
 doi           = 10.1561/XXXXXXXXX,
 firstpage     = 1, 
 lastpage      = 18
 }

\usepackage{mwe}

\author[1]{Shlezinger,Nir}
\author[2]{Eldar,Yonina C.}

\affil[1]{School of Electrical and Computer Engineering, Ben-Gurion University of the Negev; nirshl@bgu.ac.il}
\affil[2]{Faculty of Mathematics and Computer Science, Weizmann Institute of Science; yonina.eldar@weizmann.ac.il}

\articledatabox{\nowfntstandardcitation}

\begin{document}

\makeabstracttitle

\begin{abstract}
Signal processing traditionally relies on classical statistical modeling techniques. Such model-based methods utilize mathematical formulations that represent the underlying physics, prior information and additional domain knowledge. Simple classical models are useful but sensitive to inaccuracies and may lead to poor performance when real systems display complex or dynamic behavior. 
 More recently,  deep learning approaches that use highly  parametric \acp{dnn} are becoming increasingly popular. Deep learning systems do not rely on mathematical modeling, and learn their mapping from data, which allows them to operate in complex environments. However, they lack the interpretability and reliability of model-based methods, typically require large training sets to obtain good performance,  and tend to be computationally complex.
 
  Model-based signal processing methods and data-centric deep learning each have their pros and cons. These  paradigms can be characterized  as edges of a continuous spectrum varying in specificity and parameterization. The methodologies that lie in the middle ground of this spectrum, thus integrating model-based signal processing with deep learning, are referred to as {\em model-based deep learning}, and are the focus  here. 
  
  This monograph provides a tutorial style presentation of model-based deep learning methodologies. These are families of algorithms  that combine principled mathematical models with data-driven systems to benefit from the advantages of both approaches. Such model-based deep learning methods exploit both partial domain knowledge, via mathematical structures designed for specific problems, as well as learning from limited data.  We accompany our presentation with  running signal processing examples, in super-resolution, tracking of dynamic systems, and array processing. We  show how they are expressed using the provided characterization and specialized in each of the detailed methodologies.  Our aim is to facilitate the design and study of future systems at the intersection of signal processing and machine learning that incorporate the advantages of both domains. The source code of our numerical examples are available and reproducible as Python notebooks.

\end{abstract}

\acresetall

\chapter{Introduction}
The philosophical idea of \ac{ai}, dating back to the works of McCarthy from the 1950's \cite{mccarthy2006proposal}, is nowadays evolving into reality. The growth of \ac{ai} is attributed to the emergence of \ac{ml} systems, which learn their operation from data, and particularly to deep learning, which is a family of \ac{ml} algorithms that utilizes neural networks as a form of brain-inspired computing \cite{lecun2015deep}.
 Deep learning is demonstrating unprecedented success in a broad range of applications: \acp{dnn} surpass human ability in classifying images \cite{he2015delving}; reinforcement learning allows computer programs to  defeat human experts in challenging games such as Go \cite{silver2017mastering} and Starcraft \cite{vinyals2019alphastar}; generative models translate text into images \cite{ramesh2022hierarchical} and create images of fake people which appear indistinguishable from true ones \cite{karras2019style}; and large language models generate sophisticated documents and textual interactions~\cite{openai2023gpt4}.

    While deep learning systems rely on data to learn their operation, traditional signal processing is dominated by algorithms that are based on simple mathematical models which are hand-designed from domain knowledge. Such knowledge can come from statistical models based on measurements and understanding of the underlying physics, or from fixed deterministic representations of the particular problem at hand. These knowledge-based processing algorithms, which we refer to henceforth as {\em model-based methods},  carry out inference based on domain knowledge of the underlying  model relating the observations at hand and the desired information. Model-based methods, which form the basis for many classical and fundamental signal processing techniques, do not rely on data to learn their mapping, though data is often used to estimate a small number of parameters.  
Classical statistical models rely on simplifying assumptions (e.g., linear systems, Gaussian and independent noise, etc.) that make inference tractable, understandable, and computationally efficient. Simple models frequently fail to represent nuances of high-dimensional complex data, and dynamic variations, settling with the famous observation made by statistician George E. P. Box that {\em ``Essentially, all models are wrong, but some are useful"}. The usage of mismatched modeling tends to notably affect the performance and reliability of classical methods.

The success of deep learning in areas such as computer vision and natural language processing made it increasingly popular to adopt methodologies geared towards data for tasks traditionally tackled with model-based techniques. It is becoming common practice to replace principled task-specific decision mappings with abstract purely data-driven pipelines, trained with massive data sets.  In particular, \acp{dnn} can be trained in a supervised way end-to-end to map inputs to predictions.
The benefits of data-driven methods over  model-based approaches are threefold: 
\begin{enumerate}
	\item Purely data-driven techniques do not rely on analytical
	approximations and thus can operate in scenarios where
	analytical models are not known. This property is key to the success of deep learning systems in computer vision and natural language processing, where accurate statistical models are typically scarce.
	\item For complex systems, data-driven algorithms are able to recover features from observed data which are needed to carry out inference \cite{Bengio09learning}. This is sometimes difficult to achieve analytically, even when complex models are perfectly known, e.g., when the enviornement is characterized by a fully-known complex simulator or a partial differential equation. 
	\item  The main complexity in utilizing \ac{ml} methods is in the training stage. In most signal processing domains, this procedure is carried out offline, i.e., prior to deployment of the device which utilizes the system. Once trained, they often implement inference at a  lower  delay compared with their analytical model-based counterparts~\cite{gregor2010learning}. 
\end{enumerate}

Despite the aforementioned advantages of deep learning methods, they are subject to several drawbacks. These drawbacks may be limiting factors particularly for various signal processing, communications, and control applications, which are traditionally tackled via principled methods based on statistical modeling.
For one, the fact that massive data sets, i.e., large number of training samples, and high computational resources are typically required to train such \acp{dnn} to learn a desirable mapping, may constitute major drawbacks. Furthermore, even using pre-trained \acp{dnn} often gives rise to notable   computational burden due to their immense parameterization. This is highly relevant for hardware-limited  devices, such as mobile phones, unmanned aerial vehicles, and \acl{iot} systems, which are often limited in their ability to utilize highly-parametrized \acp{dnn} \cite{chen2019deep}, and require adapting to dynamic conditions.   Furthermore, the abstractness and extreme parameterization of \acp{dnn} results in them  often being  treated as black-boxes;   {understanding how their predictions are obtained and characterizing confidence intervals tends to be quite challenging. As a result, deep learning does} not  offer the interpretability, flexibility, versatility, reliability, and generalization capabilities of model-based methods~\cite{monga2021algorithm}.

The limitations associated with model-based methods and conventional deep learning systems gave rise to a multitude of techniques for combining model-based processing and \ac{ml}, aiming to benefit from the best of both approaches. These methods are typically application-driven, and are thus designed and studied in light of a specific task. For example, the combination of \acp{dnn} and model-based sparse recovery algorithms was shown to facilitate sparse recovery \cite{gregor2010learning, ongie2020deep} as well as enable compressed sensing beyond the domain of sparse signals \cite{bora2017compressed,yang2018admm}; Deep learning was used to empower regularized optimization methods \cite{gilton2019neumann,ahmad2020plug,dong2018denoising}, while model-based optimization contributed to the design and training of \acp{dnn} for such tasks \cite{aggarwal2018modl,mataev2019deepred,zhang2020deep}; Digital communication receivers used \acp{dnn} to learn to carry out and enhance symbol detection and decoding algorithms in a data-driven manner \cite{shlezinger2019viterbinet, shlezinger2019deepSIC, nachmani2018deep}, while symbol recovery methods enabled the design of model-aware deep receivers \cite{samuel2019learning, he2018model,khani2020adaptive}. The proliferation of  hybrid model-based/data-driven systems, each designed for a unique task, motivates establishing a concrete systematic framework for combining domain knowledge in the form of model-based methods and deep learning, which is the focus here.

In this monograph we present strategies for designing algorithms that combine model-based methods with data-driven deep learning techniques. While classic model-based inference and deep learning are typically considered to be distinct disciplines, we  view them as edges of a continuum varying in specificity and parameterization. We build upon this characterization to provide a tutorial-style presentation of the main methodologies which lie in the middle ground of this spectrum, and combine model-based optimization with \ac{ml}.
 This hybrid paradigm, which we coin {\em model-based deep learning}, is relevant to a multitude of research domains where one has access to some level of reliable mathematical modelling. While the presentation here is application-invariant, it is geared towards families of problems typically studied in the signal processing literature. This is reflected in our running examples,  which correspond to three common signal processing tasks of compressed signal recovery, tracking of dynamic systems, and \ac{doa} estimation in array processing. These running examples are repeatedly specialized throughout the monograph for each surveyed methodology, facilitating the comparison between the considered approaches. 
 
 We begin by providing a unified characterization for inference and decision making algorithms in Chapter~\ref{ch:Inferece}. There, we discuss different types of inference rules, present the running examples, and discuss the main pillars of designing inference rules, which we identify as selecting their type, setting the objective, and their evaluation procedure. Then, we show how classical model-based optimization as well as data-centric deep learning are obtained as special instances of this  unified characterization in Chapters~\ref{ch:MB} and \ref{ch:Deep}, respectively. We there also review relevant basics that are core to the design of many model-based deep learning systems, including fundamentals in convex optimization (for model-based methods) and in neural networks (for deep learning). We identify the components dictating the distinction between model-based and data-driven methodologies in the formulated objectives, the corresponding decision rule types, and their associated parameters.
 
 The main bulk of this monograph, which builds upon the fundamental aspects presented in Chapters~\ref{ch:Inferece}-\ref{ch:Deep}, is the review of hybrid model-based deep learning methodologies in Chapter~\ref{ch:MBDL}. A core principle of model-based deep learning is to leverage data by converting classical algorithms into trainable models with varying levels of abstractness and specificity, as opposed to the more classical model-based approach where data is used to characterize the underlying model. These  two rationales are highly related to the \ac{ml} paradigms of generative and discriminative learning \cite{ng2001discriminative, jebara2012machine}. Consequently, we commence this part by presenting a spectrum of decision making approaches which vary in specificity and parameterization, with model-based methods and deep learning constituting its edges, followed by a review of generative and discriminative learning. Based on these concepts, we provide a systematic categorization of model-based deep learning techniques as concrete strategies positioned along the continuous spectrum.  
 
 We categorize model-based deep learning methods into three main strategies:
 \begin{enumerate}
 	\item {\em Learned optimization}: This approach is highly geared towards classical optimization and aims at leveraging data to fit model-based solvers. In particular, learned optimizers use automated deep learning techniques to tune parameters conventionally configured by hand.
 	\item {\em Deep unfolding}: This family of techniques converts iterative optimizers into trainable parametrized architectures. Its instances notably vary in their parameterization and abstractness based on the interplay imposed in the system design between the trainable architecture and the model-based algorithm from which it originates.
 	\item {\em \ac{dnn}-aided inference}: These schemes augment model-based algorithms with trainable neural networks, encompassing a broad family of different techniques which vary in the module being replaced with a \ac{dnn}.
 \end{enumerate} 
  
  We exemplify the considered methodologies for the aforementioned running examples via both analytical derivations as well as simulations. By doing so, we provide a systematic qualitative and quantitative comparison between representative instances of the detailed approaches for signal processing oriented scenarios. The source code used for the results presented in this monograph is available as Python Notebook scripts\footnote{The source code and Python notebooks can be found online at \url{https://github.com/ShlezingerLab/MBDL_Book}.}, detailed in a pedagogic fashion such that they can be presented alongside lectures, either as a dedicated graduate level course, or as part of a course on topics related to \ac{ml} for signal processing.

\chapter{Inference Rule Design}
\label{ch:Inferece}
The first part of this monograph is dedicated to reviewing both classical inference that is based on knowledge and modeling, as well as deep learning based inference. We particularly adopt a unified perspective which allows these distinct approaches to be viewed as edges of a continuous spectrum, as elaborated further in Chapter~\ref{ch:MBDL}. To that aim, we begin with the basic concept of {\em inference} or {\em decision making}.

\section{Decision Mapping}
\label{sec:Mapping}
\begin{figure}
	\centering
	\includegraphics[width=0.8\columnwidth]{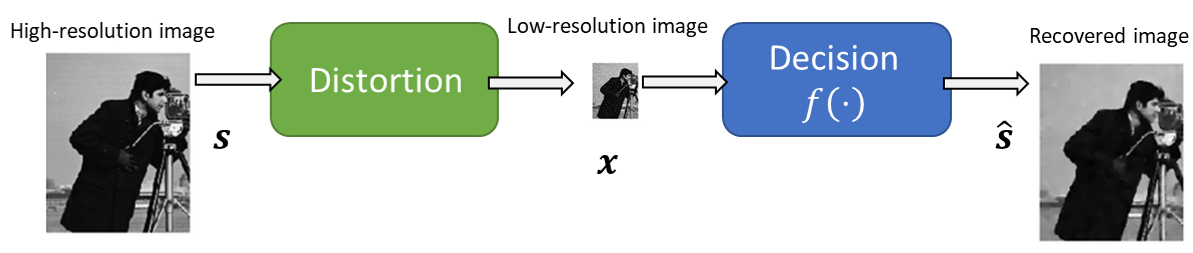}
	\caption{Super-resolution recovery illustration.}
	\label{fig:ImageCompress1}
\end{figure}

	We consider a generic setup where the goal is to design a decision mapping. A decision rule $f$ maps the input, denoted $\Input\in\InputSpace$, which is the available observations, into a decision denoted $\hat{\Label}\in \LabelSpace$, namely,
\begin{equation}
\label{eqn:DecisionRule}
f:\InputSpace\mapsto\LabelSpace.   
\end{equation}
	This generic formulation encompasses a multitude of settings involving estimation, classification, prediction, control, and many more. Consequently, it corresponds to a broad range of different applications. The specific task dictates the input space $\InputSpace$ and the possible decisions $\LabelSpace$.  
To keep the presentation focused, we repeatedly use henceforth three concrete running examples:
\begin{example}[Super-Resolution]
	\label{exm:Inverse}
	Here, $\hat{\Label}$ is some high-resolution image,  while $\Input$ is a distorted low-resolution version of $\hat{\Label}$. Thus, $\InputSpace$ and $\LabelSpace$ are the spaces of low-resolution and high-resolution images, respectively. The goal of the decision rule is to reconstruct $\Label$ from its noisy compressed version $\Input$, as illustrated in Fig.~\ref{fig:ImageCompress1}.
\end{example}

\begin{figure}
	\centering
	\includegraphics[width=0.8\columnwidth]{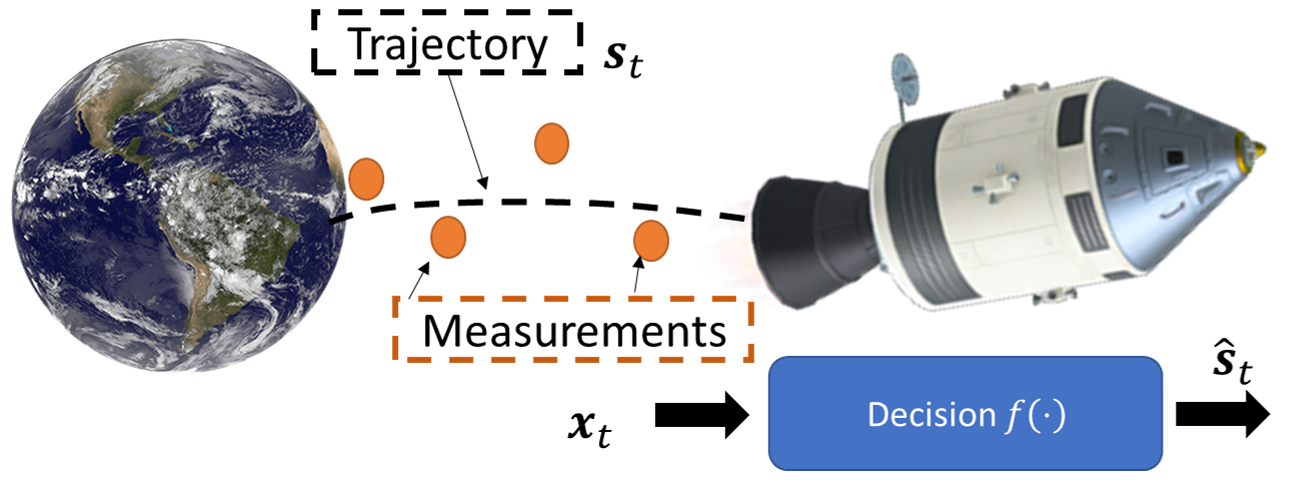}
	\caption{Tracking of dynamic system illustration.}
	\label{fig:Control1}
\end{figure}

\begin{example}[Tracking of Dynamic Systems]
	\label{exm:Control}
	In our second running example, we consider a dynamic system. At each time instance $t$, the goal is to map the noisy state observations $\Input_t$, with $\InputSpace$ being the space of possible sensory measurements, into an estimate if the underlying state $\hat{\Label}_t$ within some possible  space $\LabelSpace$. In particular, the latent state vector $\Label_t$  evolves in a random fashion that is related to  the previous state $\Label_{t-1}$, while being partially observable via the noisy $\Input_t$. The setup is illustrated in Fig.~\ref{fig:Control1}. 
\end{example}

\begin{figure}
	\centering
	\includegraphics[width=0.6\columnwidth]{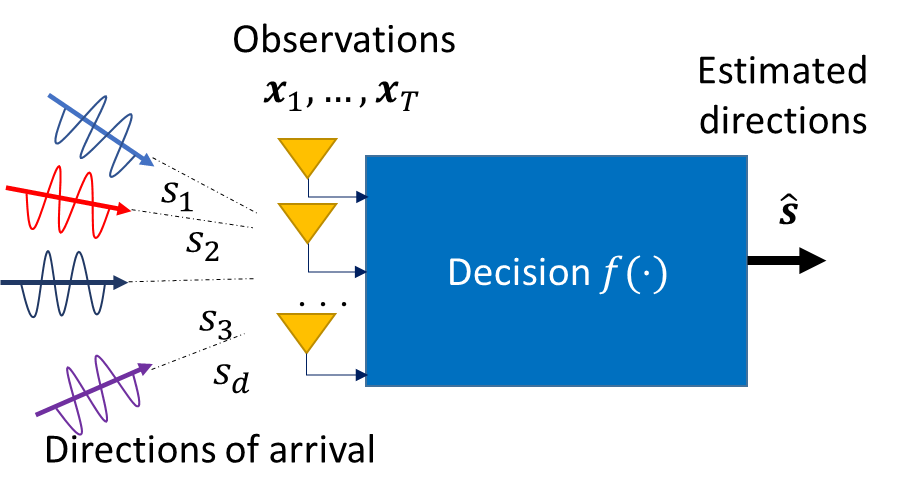}
	\caption{\ac{doa} estimation setup illustration.}
	\label{fig:DoA1}
\end{figure}

\begin{example}[Direction-of-Arrival Estimation]
	\label{exm:DoA}
	Our third running example considers the array signal processing problem of  \ac{doa} estimation. Here, the observations are a sequence of $T$ multivariate array measurements, i.e., $\Input_1,\ldots, \Input_T$, which correspond to a set of $d$ impinging signals arriving from different angles, as illustrated in Fig.~\ref{fig:DoA1}. The goal is to recover the angles, and thus $\Label$ is the \acp{doa}. Accordingly, $\InputSpace$ is the set of vectors whose cardinality is determined by the number of array elements, while $\LabelSpace$ is the set of valid angle values, e.g., $\LabelSpace = \big[\frac{-\pi}{2},\frac{\pi}{2}\big]^d$. 
\end{example}

The design of an inference rule is typically a two-step procedure. The first step involves selecting the type of inference rule, while the second step tunes its parameters based on some design objective. We next elaborate on each of these steps.

\section{Inference Rule Types}
\label{sec:InfType}

\begin{figure}
	\centering
	\includegraphics[width=0.8\columnwidth]{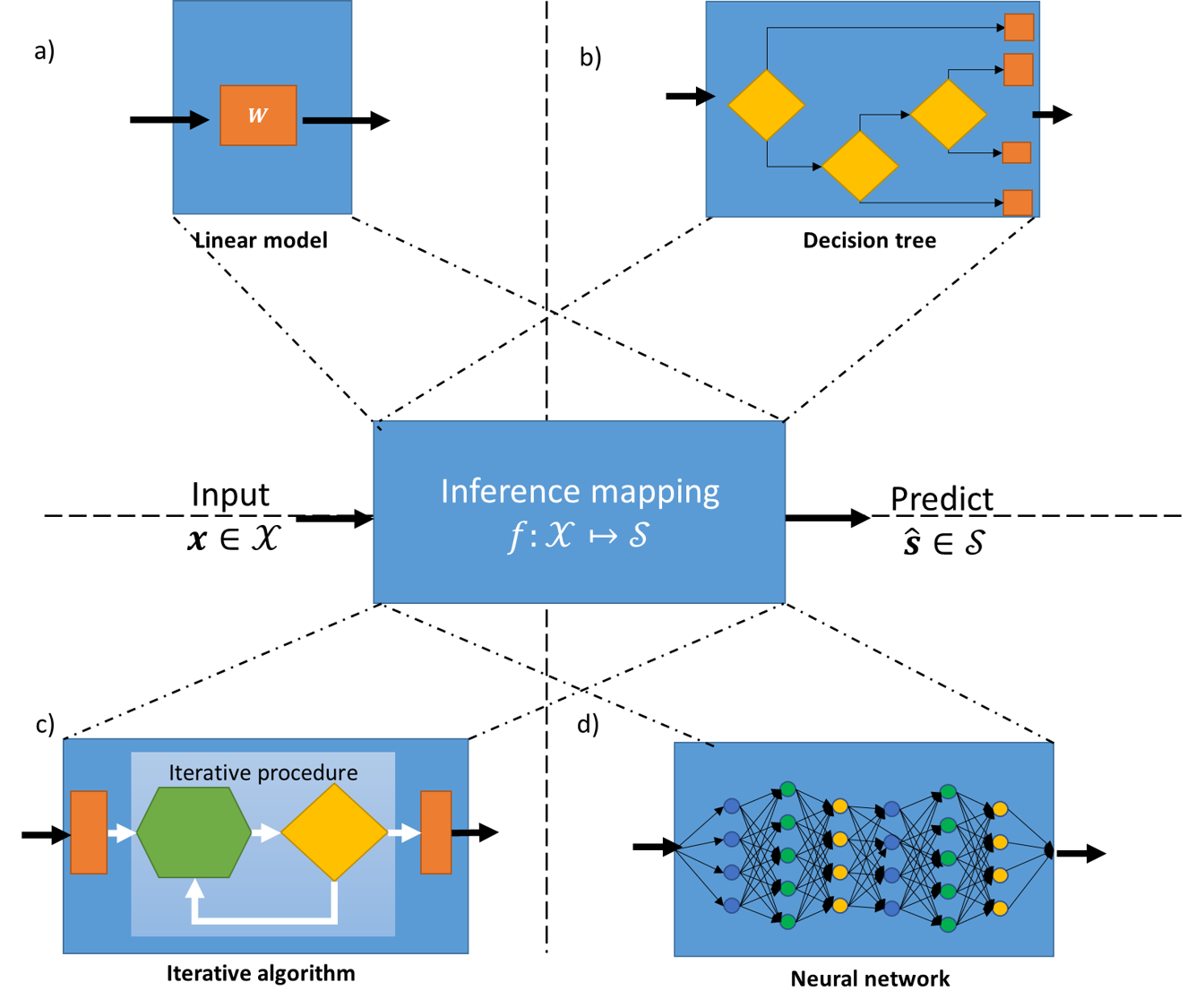}
	\caption{Examples of different inference rule types: $a)$ linear model; $b)$ decision tree; $c)$ iterative algorithm; $d)$ neural network.}
	\label{fig:InfRuleType}
\end{figure}

The above generic formulation allows the inference rule $f$ to be any mapping from $\InputSpace$ into $\LabelSpace$. In practice, inference rules are often carried out using a structured form, i.e., there exists some set of mappings  $\mySet{F}$ from which $f$ is selected. Some common types of inference rules are:
\begin{itemize}
	\item A {\em linear model} given by $\hat{\Label} = \myMat{W} \Input$ for some matrix $\myMat{W}$. 
	\item A {\em decision tree} chooses $\hat{\Label}$ from some possible decisions $\{\Label_k\}_k$ by examining a set of nested conditions $\{{\rm cond}_k\}_k$, e.g., if ${\rm cond}_1(\Input)$ then $\hat{\Label} = \Label_1$; else inspect ${\rm cond}_2(\Input)$, and so on.   
	\item An {\em iterative algorithm} refines its decision using a mapping $g: \LabelSpace\times\InputSpace \mapsto \LabelSpace$, repeating 
	\begin{equation}
	\hat{\Label}^{(k+1)} = g\big(\hat{\Label}^{(k)}; \Input\big), \quad k=0,1,2,\ldots
	\end{equation}	
  from some initial guess $\hat{\Label}^{(0)}$ until convergence. These types of inference rules are discussed in detail in Chapter~\ref{ch:MB}. 
	\item A {\em neural network} is typically given by a concatenation of $K$ affine layers and non-linear activations, such that 
	\begin{equation}
	\hat{\Label} = h_{K}\big(h_{K-1}\big(\cdots h_1(\Input)\big)\big),
	\end{equation}
	where each $h_k(\cdot)$ is given by $h_k(\myVec{z}) = \sigma\big(\myMat{W}_k\myVec{z}+\myVec{b}_k\big)$ with $\sigma(\cdot)$ being an activation and $(\myMat{W}_k,\myVec{b}_k)$ are parameters of the affine transformation. These types of inference rules are discussed extensively in Chapter~\ref{ch:Deep}. 
\end{itemize}
The above inference rules are illustrated in Fig.~\ref{fig:InfRuleType}.

	The boundaries between inference rule types are not always strict, so there may be  overlap between the  categories. For instance, an iterative algorithm with $K$ iterations can often be viewed as a neural network, as we further elaborate on in Chapter~\ref{ch:MBDL}.

\section{Tuning an Inference Rule}
Inference rules are almost always {\em parameterized}, and determine how the context is processed and mapped into $\hat{\Label}$ based on some predefined structure. In some cases the number of parameters is small, such as decision trees with a small number of conditions involving comparison to some parametric threshold. In other cases, such as when $f$ is a \ac{dnn}, decision mappings involve a massive number of parameters.   These parameters capture the different mappings one can represent as decision rules. 

In general, the more parameters there are, the broader the family of mappings captured by $\mySet{F}$, which in turn results in the inference rule capable of accommodating more diverse and generic functions.
Decision rules with fewer parameters are typically more specific,  capturing a limited family of mappings. 
Let ${\Theta}$  denote the parameter space for a decision rule family $\mySet{F}$, such that for each $\myVec{\theta}\in\Theta$, $f(\cdot; \myVec{\theta})$ is a mapping in $\mySet{F}$. 
We refer to the setting of the inference rule parameters as {\em tuning}. The preference of one parameterization over another involves the formulation of an objective function.


A decision rule is evaluated using a loss function. The objective can be given by a cost or a loss function which one aims at minimizing, or it can be specified by an application utility or reward, which we wish to maximize. We henceforth formulate the objective as a loss function, which evaluates the inference rule for a given input compared with some desired decision; such a loss function is formulated as a mapping \cite{shalev2014understanding}
\begin{equation}
\label{eqn:lossfunction}
l:\mySet{F}\times\InputSpace\times\LabelSpace \mapsto \mySet{R}^+.
\end{equation}
Broadly speaking, \eqref{eqn:lossfunction} dictates the success criteria of a decision mapping for a given context-decision pair. 
For instance, in classification tasks, candidate losses include  the error rate (zero-one) loss
\begin{equation}
\label{eqn:ErrRateLoss}
l_{\rm Err}(f,\Input,\Label) =  {\boldsymbol 1}_{f(\Input) \neq \Label},
\end{equation}
 while the $\ell_2$ loss given by 
 \begin{equation}
 \label{eqn:L2Loss}
 l_{\rm Est}(f,\Input,\Label) = \|\Label - f(\Input)\|_2^2,
 \end{equation}
 is suitable for estimation tasks. In addition, prior knowledge on the domain of $\Input$ is often incorporated into loss measures in the form of regularization. For instance,  $\ell_1$ regularization is often used to encourage sparsity, while regularizing by the total variation norm is often used in image denoising to reduce noise while preserving details such as edges. 

It is emphasized that the objective function that is used for tunning the inference rule is often surrogate to the true system objective. In particular, loss functions as in \eqref{eqn:lossfunction}  often include simplifications, approximations, and regularizations, introduced for tractability and to facilitate tuning.   Furthermore, the true goal of the system is often challenging to capture in mathematical form. For instance, in medical imaging evaluation often involves inspecting the outcome by human experts, and can thus not be expressed as a closed-form mathematical expression. 

\begin{figure}
	\centering
	\includegraphics[width=0.8\columnwidth]{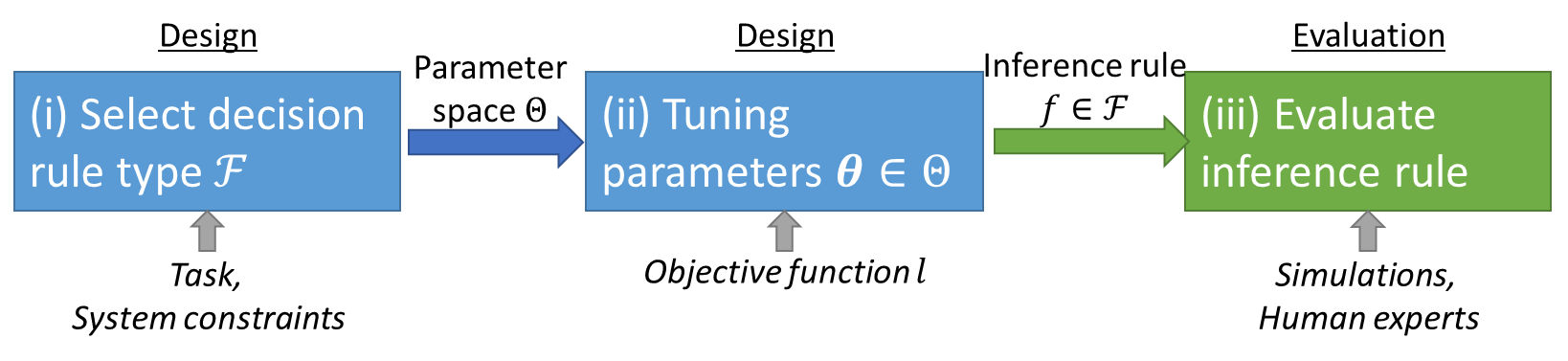}
	\caption{Inference rule selection procedure illustration.}
	\label{fig:DesignProc1}
	\vspace{-0.2cm}
\end{figure}

\smallskip
The overall design and evaluation procedure is illustrated in Fig.~\ref{fig:DesignProc1}. In the next chapters we discuss strategies to carry out the design procedure. We begin in Chapter~\ref{ch:MB} with traditional  approaches, referred to as {\em model-based} or {\em classic} methods, that are based on modeling and knowledge. Then, in Chapter~\ref{ch:Deep} we discuss the data-centric approach which uses \ac{ml}, particularly focusing on deep learning being a leading family of \ac{ml} techniques.

\chapter{Model-Based Methods}
\label{ch:MB}

In this section we review the model-based approach for designing decision mappings based on domain knowledge. Such methods can be generally studied based on the objective used to set the decision rule, and the solver designed based on the objective function. 

\section{Objective Function}
In the classic model-based approach, knowledge of an underlying model relating the input $\Input$ with the desired decision $\Label$ is used along with the loss measure $l(\cdot)$ in \eqref{eqn:lossfunction} to formulate an analytical objective, denoted $\mySet{L}:\mySet{F}\mapsto \mySet{R}$. The objective can then be applied to select the decision rule from $\mySet{F}$, which can either be a pre-defined inference rule type or even from the entire space of mappings from $\InputSpace$ to $\LabelSpace$.

 A common model-based approach is to model the inputs as being related to the targets  via some statistical distribution measure $\Distribution$ defined over $\InputSpace \times \LabelSpace$. The stochastic nature implies that the targets are not fully determined by the input we are observing.     Formally, $\Distribution$ is a joint distribution over the domain of inputs and targets. One can view such a distribution as being composed of two parts: a distribution over the  input $\Distribution_x$ (referred to as the {\em marginal distribution}), and the conditional probability of the targets given the inputs $\Distribution_{s|x}$ (referred to as the {\em inverse distribution}). {Similarly, $\Distribution$ can be decomposed into the distribution over the target $\Distribution_s$ (coined the {\em prior distribution}), and the  probability of the inputs conditioned on the targets $\Distribution_{x|s}$ (referred to as the {\em likelihood}).}

 Model-based methods are typically divided into two main frameworks: the first is the {\em Bayesian} setting, where it is assumed there exists a joint distribution $\Distribution$; the second paradigm is  the {\em non-Bayesian} one, where the target is viewed as a deterministic unknown variable. In the non-Bayesian case there is no prior distribution $\Distribution_s$, and the statistical model underlying the problem is that  of $\Distribution_{x|s}$.

In a Bayesian setting, given knowledge of a statistical distribution measure $\Distribution$, one can formulate the risk  
 \begin{equation}
 \label{eqn:Risk}
\mySet{L}_{ \Distribution}(f) =\E_{(\Input,\Label)\sim \Distribution} \{l(f,\Input,\Label)\},
 \end{equation}
 and aim at setting the mapping $f$ to the one which minimizes the risk $\mySet{L}_{ \Distribution}(f)$. For some commonly-used loss measures, one can analytically characterize the risk-minimizing inference rule, as discussed below. 
 
 \paragraph{Error-Rate Loss}
When using the error rate loss \eqref{eqn:ErrRateLoss}, the inference rule which minimizes the risk among all mappings from $\InputSpace$ to $\LabelSpace$ is the {\em \ac{map}} rule, given by:
\begin{align}
f_{\rm MAP}(\Input) &=   \mathop{\arg \min}\limits_{f(\cdot)}  \E_{(\Input,\Label)\sim \Distribution} \{l_{\rm Err}(f,\Input,\Label)\} \notag \\ 
&=  \mathop{\arg \max}\limits_{\Label \in \LabelSpace } \Pr(\Label | \Input) \notag \\
&= \mathop{\arg\min}\limits_{\Label \in \LabelSpace} -\log \Pr(\myVec{s}|\myVec{x}) \notag \\
&= \mathop{\arg\min}\limits_{\Label \in \LabelSpace} -\log \Pr(\myVec{x}|\myVec{s}) -\log \Pr(\myVec{s}).
\label{eqn:Map}
\end{align}

 \paragraph{Cross-Entropy Loss}
    An alternative widely-used loss function for classification problems is the cross-entropy loss. Here, the inference rule does not produce a label ({\em hard decision}), but rather a probability mass function over the label space ({\em soft decision}). Namely, $f(\Input)$  is a $K\times 1$ vector $f_1(\Input),\ldots, f_K(\Input)$ with non-negative entries such that $\sum_k f_k(\Input) = 1$. For this setting, the cross entropy loss is given by:
\begin{equation}
l_{\rm CE}(f,\Input,\Label) = -\sum_{k=1}^{K} \mathds{1}_{\Label = \Label_k} \log f_k(\Input),
\end{equation}
where $\mathds{1}_{(\cdot)}$ is the indicator function. 

While the main motivation for using the cross-entropy loss stems from its ability to provide a measure of confidence in the decision, as well as  computational reasons, it turns out that the optimal inference rule in the sense of minimal cross-entropy risk is the true conditional distribution. To see this, 
we write $p(\Label_k|\Input) = f_k(\Input)$, and note that for any distribution measure $p(\Label|\Input)$ over $\LabelSpace$ it holds that
\begin{align}
\mySet{L}_{\Distribution}(f) &= \E_{\Distribution}\left\{ l_{\rm CE}(f,\Input,\Label) \right\} \notag \\
&= \E_{\Distribution}\left\{  -\log p(\Label|\Input)  \right\} \notag \\
&= \E_{\Distribution}\left\{  -\log \frac{p(\Label|\Input)}{\Pr(\Label | \Input)}  \right\} + \E_{\Distribution}\left\{  -\log {\Pr(\Label | \Input)}  \right\} \notag \\
&= D_{\rm KL}\left(\Pr(\Label | \Input) || p(\Label|\Input)\right) + H(\Label | \Input),
\label{eqn:ProofCE1}
\end{align}
where $H(\Label | \Input)$ is the true cross-entropy of $\Label$ conditioned on $\Input$, and $D_{\rm KL}(\cdot || \cdot)$ is the Kullback-Leibler divergence \cite{cover2012elements}. As $D_{\rm KL}( \Pr(\Label | \Input) || p(\Label|\Input) )$ is non-negative and equals zero only when $p(\Label|\Input) \equiv \Pr(\Label | \Input)$, it holds that \eqref{eqn:ProofCE1} is minimized when the inference output is the true conditional distribution, i.e.,
\begin{equation}
\label{eqn:CEminimize}
f_{\rm CE}(\Input) = [\Pr(\Label_1 | \Input), \ldots, \Pr(\Label_K | \Input)].
\end{equation}
We note that \eqref{eqn:CEminimize} implies that the \ac{map} rule can be obtained by taking the $\arg\max$ over the entries of $f_{\rm CE}(\cdot)$.

 \paragraph{$\ell_2$ Loss}
Similarly, under the $\ell_2$ loss \eqref{eqn:L2Loss}, the risk objective becomes the \ac{mse}, which is minimized by the conditional expectation
\begin{align}
f_{\rm MMSE}(\Input) &=   \mathop{\arg \min}\limits_{f(\cdot)}  \E_{(\Input,\Label)\sim \Distribution} \{l_{\rm Est}(f,\Input,\Label)\} \notag \\ 
&= \E_{\Label\sim \Distribution_{\Label | \Input}} \{\Label | \Input\}.
\label{eqn:MMSE}
\end{align}

In the non-Bayesian setting, the risk function as given in \eqref{eqn:Risk} cannot be used, since there is no prior distribution $\Distribution_s$ which is essential for formulating the stochastic expectation. Several alternative objectives that are often employed in such setups are discussed below.

\paragraph{Negative Likelihood Loss} 
A common inference rule used in the non-Bayesian setting is the one which minimizes the negative likelihood, i.e., the maximum likelihood rule, given by
\begin{align}
f_{\rm ML}(\Input) &=  \mathop{\arg \min}\limits_{\Label  \in \LabelSpace } -\Pr(\Input | \Label ) \notag \\
&=  \mathop{\arg \max}\limits_{\Label \in \LabelSpace}  \log \Pr(\Input | \Label ). 
\label{eqn:MaxLikelihood}
\end{align}
Note that the maximum likelihood rule in \eqref{eqn:MaxLikelihood} coincides with the \ac{map} rule in \eqref{eqn:Map} when the target obeys a uniform distribution over a discrete set $\LabelSpace$.

\paragraph{Regularized Least Squares}
Another popular inference rule employed in non-Bayesian setups aims at minimizing the $\ell_2$ loss while introducing  regularization to account for some prior knowledge on the deterministic target variable $\Label$. This inference rule is given by
\begin{align}
f_{\rm RegLS}(\Input) = \mathop{\arg \min}\limits_{\Label  \in \LabelSpace } \left\|\Label - f(\Input)  \right\|_2^2 + \rho \phi(\Label), 
\label{eqn:RegLS}
\end{align}
where $\phi(\Label)$ is a  predefined regularization function describing a property that the target is expected to exhibit, and $\rho >0$ is a regularization coefficient.

\color{black}

\subsection{Running Examples}
\label{ssec:MBExamples}
The formulation of the objective function is dictated by the model imposed on the underlying relationship between $\Input$ and the desired decision $\Label$. This objective typically contains parameters of the model, which we denote by $\ObjParam$, and henceforth write $\mySet{L}_{\ObjParam}(f)$. We next show the objective parameters which arise in the context of the running examples introduced in Chapter~\ref{ch:Inferece}.

   \begin{example}
	\label{exm:InverseModel0}
	A common approach to treat the super-resolution problem in Example~\ref{exm:Inverse} is to assume the compression obeys a linear Gaussian model, i.e., 
	\begin{equation}
	\label{eqn:linearGuassian}
	\myVec{x} = \myMat{H}\myVec{s} + \myVec{w}, \qquad \myVec{w} \sim \mathcal{N}(\myVec{0}, \sigma^2 \myMat{I}).
	\end{equation}
	The matrix $\myMat{H}$ in \eqref{eqn:linearGuassian} represents, e.g., the point-spread function of the system. 
	In this case, the   log likelihood in \eqref{eqn:Map} is given by
	\begin{align}
	-\log \Pr(\myVec{x}|\myVec{s}) &=    -\log \left( (2\pi)^{-k/2} {\rm det} (\sigma^2 \myMat{I})^{-1/2} e^{- \frac{1}{2\sigma^2}\|\myVec{x}-\myMat{H}\myVec{s}\|^2 }\right) \notag \\
	&= \frac{1}{2\sigma^2}\|\myVec{x}-\myVec{s}\|^2  +{\rm const}.
	\end{align}
	Consequently, the \ac{map} rule in \eqref{eqn:Map} becomes 
	\begin{equation}
	\label{eqn:Recovery1}
	f_{\rm MAP}(\Input) = \mathop{\arg\min}\limits_{\myVec{s}} \frac{1}{2}\|\myVec{x}-\myMat{H}\myVec{s}\|^2 +\sigma^2\phi(\myVec{s}),
	\end{equation}
	where $\phi(\myVec{s}) := -\log p(\myVec{s})$. The resulting objective requires imposing a prior on $\mySet{S}$ encapsulated in  $\phi(\myVec{s})$.   The parameters of the objective function in \eqref{eqn:Recovery1} are thus $\myMat{H}$ and any parameters used for representing the prior $\phi(\cdot)$. 
\end{example}

The formulation of the \ac{map} mapping in \eqref{eqn:Recovery1} coincides with the regularized least-squares rule in \eqref{eqn:RegLS}. However, there is a subtle distinction between the two approaches: In Example~\ref{exm:InverseModel0}, the rule \eqref{eqn:Recovery1} is derived from the underlying model $\Distribution$ and the regularization is dictated by the known prior distribution $\Distribution_s$. In the non-Bayesian setting, there is no prior distribution, and $\phi(\cdot)$ is selected to match some expected property of the target, as exemplified next.

   \begin{example}
	\label{exm:InverseModel1}	
	For the super-resolution setup, a popular approach is to impose sparsity in some known domain $\myMat{\Psi}$ (e.g., wavelet), such that $\myVec{s} = \myMat{\Psi}\myVec{r}$, where $\myVec{r}$ is sparse. This boils down to an objective  defined on $\myVec{r}$, which for the regularized least-squares formulation is given by
	\begin{equation}
	\label{eqn:InvesObj1}
	\mySet{L}_{ \ObjParam}(\myVec{r}) =  
	\frac{1}{2}\|\myVec{x}-\myMat{H}\myMat{\Psi}\myVec{r}\|^2_2 +\rho\|\myVec{r}\|_0.
	\end{equation}
 The parameters of the   objective  in \eqref{eqn:InvesObj1} are 
	\begin{equation}
	\label{eqn:vPsi}
	\ObjParam = \{\myMat{H},\myMat{\Psi},\rho\}.
	\end{equation}
	The parameters $\myMat{H}$ and $\myMat{\Psi}$ in \eqref{eqn:vPsi} follow from the underlying statistical model and the sparsity assumption imposed on the target. However, the parameter $\rho$ does not stem directly from the system model, and has to be selected so that a solver based on \eqref{eqn:InvesObj1} yields satisfactory estimates of the target. 
\end{example}

\color{black}
    The above examples shows how one can leverage domain knowledge to formulate an objective, which is dictated by the parameter vector   $\ObjParam$. They also demonstrate  two key properties of model-based approaches: $(i)$ that surrogate models can be quite unfaithful to the true data, since, e.g., the Gaussianity of $\myVec{w}$ implies that $\myVec{x}$ in \eqref{eqn:linearGuassian} can take negative values, which is not the case for, e.g., image data; and $(ii)$ that simplified models allow translating the task into a relatively simple closed-form objective, as in \eqref{eqn:InvesObj1}.  A similar Bayesian approaches can be used to tackle the dynamic system tracking setting of Example~\ref{exm:Control}.

\begin{example}
	\label{exm:ControlModel1}
	Traditional state estimation considers \ac{mse} recovery in dynamics that take the form of a state-space model, where
	\begin{subequations}
		\label{eqn:ssmodel1}
		\begin{align}
		\myVec{s}_{t+1} &= g(\myVec{s}_t,\myVec{v}_t), \\
		\myVec{x}_t &=  h(\myVec{s}_{t},\myVec{w}_t).
		\end{align}
	\end{subequations}
	The noise sequences $\myVec{v}_t, \myVec{w}_t$ are assumed to be i.i.d. in time.  The objective at each time instance $t$ is given by
	\begin{equation}
	\label{eqn:LQGObj1}
	\mySet{L}_{ \ObjParam}(f) = \E\{\|\myVec{s}_t - \hat{\myVec{s}}_t\|^2\}, \qquad \hat{\myVec{s}}_t = f\left(\{\myVec{x}_\tau\}_{\tau\leq t}\right).
	\end{equation}
	An important special case of the state-space model in \eqref{eqn:ssmodel1} is the linear-Gaussian model,  where
	\begin{subequations}
		\label{eqn:ssmodel}
		\begin{align}
		\myVec{s}_{t+1} &= \myMat{F}\myVec{s}_t + \myVec{v}_t, \\
		\myVec{x}_t &=  \myMat{H}\myVec{s}_{t} +\myVec{w}_t.
		\end{align}
	\end{subequations}
	Here, the noise sequences $\myVec{v}_t, \myVec{w}_t$ are zero-mean Gaussian signals, i.i.d. in time, with covariance matrices $\myMat{V}, \myMat{W}$, respectively.
	The parameters of the objective function \eqref{eqn:LQGObj1} under the state-space model in \eqref{eqn:ssmodel} are thus
	\begin{equation}
	\label{eqn:LQGObj2}
	\ObjParam = \{\myMat{F}, \myMat{H}, \myMat{V}, \myMat{W}\}.
	\end{equation}
\end{example}

An additional form of domain knowledge often necessary for formulating faithful objective functions is reliable modeling of hardware systems. Such accurate modeling is core in the \ac{doa} estimation setting of Example~\ref{exm:DoA}.

\begin{example}
	\label{exm:DoaModel1}
	A common treatment of  \ac{doa} estimation considers {a non-Bayesian setup where the input is obtained using} a uniform linear array with $N$   half-wavelength spaced elements. In this case, assuming that the $d$ transmitted signals are narrowband and in the far-field of the array, the received signal at time instance $t$ is  modeled as obeying the following relationship
	\begin{equation}
	\label{eqn:DoAModel}
	\Input_t = \myMat{A}(\Label)\myVec{y}_t + \myVec{w}_t.
	\end{equation}
	In \eqref{eqn:DoAModel},   $\myVec{y}_t$ is a $d\times 1$ vector whose entries are the source signals, $\myVec{w}_t$ is i.i.d. noise with covariance matrix $\myMat{W}$, and $\myMat{A}(\theta) = [\myVec{a}(s_1),\ldots,\myVec{a}(s_d)]$ is the steering matrix, where
	\begin{equation}
	\label{eqn:SteeringVecs}
	\myVec{a}(\psi) \triangleq [1, e^{-j\pi \sin(\psi)}, ... , e^{-j\pi(N-1)\sin(\psi)}].
	\end{equation}
	As in Example~\ref{exm:DoA}, we use $\myVec{s}$ to represent the $d$ \acp{doa} of the impinging signals. 
	
	Since the array is known to be uniform with half-wavelength spacing, the steering vectors in \eqref{eqn:SteeringVecs} can be computed for any $\psi$. Thus, as 
	the \acp{doa} are assumed to be deterministic and unknown, the model parameters
	$ \ObjParam$ include the distributions of the sources $\myVec{y}_t$ and the noises $\myVec{w}_t$. For instance, when further imposing a zero-mean Gaussian distribution on $\myVec{y}_t$ and $\myVec{w}_t$,  
	\begin{equation}
	\label{eqn:DoAObj2}
	\ObjParam = \{\myMat{Y}, \myMat{W}\},
	\end{equation}
	with $\myMat{Y}$ being the covariance matrix of $\myVec{y}_t$. 
		When the covariance matrix $\myMat{Y}$ is assumed to be diagonal, the sources are said to be {\em non-coherent}.

In order to formulate the maximum likelihood inference rule of $\Label$ given $\{\Input_t\}_{t=1}^T$, one has to model the temporal dependence of $\myVec{y}_t$ and $\myVec{w}_t$. For instance, in the simplistic case where both are assumed to be i.i.d. in time, by writing the covariance of $\Input_t$ as $\myMat{\Sigma}(\Label) = \myMat{A}(\Label)\myMat{Y}\myMat{A}^H(\Label) + \myMat{W}$, the maximum likelihood rule \eqref{eqn:MaxLikelihood} becomes \cite{jaffer1988maximum}
	\begin{align}
&f_{\rm ML}(\Input) 
=  \mathop{\arg \min}\limits_{\Label \in \big[\frac{-\pi}{2},\frac{\pi}{2}\big]^d} \sum_{t=1}^T \Input_t^H \myMat{\Sigma}^{-1}(\Label)  \Input_t + T \log \det\left(\myMat{\Sigma}(\Label)  \right) \notag \\
&\quad=  \mathop{\arg \min}\limits_{\Label \in \big[\frac{-\pi}{2},\frac{\pi}{2}\big]^d} 
{\rm tr}\left(\myMat{\Sigma}^{-1}(\Label)\left(\frac{1}{T}\sum_{t=1}^T\Input_t\Input_t^H \right)  \right) + \log \det\left(\myMat{\Sigma}(\Label)  \right).
\label{eqn:DoAObj1}
	\end{align}	
\end{example}

\color{black}


The formulation of the objectives in Examples~\ref{exm:InverseModel0}-\ref{exm:DoaModel1} rely on full domain knowledge, e.g., one has to know the    prior $\phi(\cdot)$, or the covariance matrices and the location of the array elements in order to express the objectives in \eqref{eqn:Recovery1},  \eqref{eqn:LQGObj1}, and \eqref{eqn:DoAObj1}, respectively.

\section{Explicit Solvers}
\label{subsec:MBExplicit}
 Model-based methods determine the parametric objective based on domain knowledge, obtained from measurements and from understanding of the underlying physics. Data and simulations are often used to estimate the parameters of the model, e.g., covariances of the noise signals. Once the objective is fixed, setting the decision rule boils down to an optimization problem, where one often adopts {\em highly-specific} types of decision mappings whose structure follows from the optimization formulation. In some cases, one can even obtain a closed form {\em explicit solution}. These solvers arise in setups where the objective takes a relatively simplified form, such that one can characterize the optimal mapping. 

\begin{example}
	\label{exm:LQGPoliciy}
	The minimal \ac{mse} estimate for the objective \eqref{eqn:LQGObj1} is known to be given by 
	\begin{equation}
	\hat{\Label}_t=f_{\rm MMSE}(\{\myVec{x}_\tau\}_{\tau\leq t}) = \E\{\Label_t|\{\myVec{x}_\tau\}_{\tau\leq t}\}.
	\end{equation}
	This implies that the decision rule should recover the mean of the conditional probability of $\Label_t|\{\myVec{x}_\tau\}_{\tau\leq t}$, which   can be written using Bayes rule as
	\begin{align}
	\Pr(\Label_t | \Input_1,\ldots,\Input_t) &=\frac{\Pr(\Input_t|\Label_t,\Input_1,\ldots,\Input_{t-1})\Pr(\Label_t|\Input_1,\ldots,\Input_{t-1})}{\Pr(\Input_t|\Input_1,\ldots,\Input_{t-1})} \notag \\
	&\stackrel{(a)}{=} \frac{\Pr(\Input_t|\Label_t)\Pr(\Label_t|\Input_1,\ldots,\Input_{t-1})}{\Pr(\Input_t|\Input_1,\ldots,\Input_{t-1})}.
	\label{eqn:CK1}
	\end{align}
	Here, $(a)$ holds under \eqref{eqn:ssmodel1}, and 
	\begin{align}
	&\Pr(\Input_t|\Input_1,\ldots,\Input_{t-1}) 
	= \int \Pr(\Input_t, \Label_t|\Input_1,\ldots,\Input_{t-1})d\Label_t \notag \\
	&\qquad = \int \Pr(\Input_t| \Label_t,\Input_1,\ldots,\Input_{t-1}) \Pr(\Label_t|\Input_1,\ldots,\Input_{t-1}) d\Label_t \notag \\
	&\qquad = \int \Pr(\Input_t| \Label_t) \Pr(\Label_t|\Input_1,\ldots,\Input_{t-1}) d\Label_t.
	\label{eqn:CK1a}
	\end{align}
	Similarly, we can write $\Pr(\Label_t|\Input_1,\ldots,\Input_{t-1})$  as
	\begin{align}
	&\Pr(\Label_t|\Input_1,\ldots,\Input_{t-1}) 
	= \int \Pr(\Label_t,\Label_{t-1}|\Input_1,\ldots,\Input_{t-1})d\Label_{t-1} \notag \\
	&\qquad = \int \Pr(\Label_t|\Input_1,\ldots,\Input_{t-1},\Label_{t-1}) \Pr(\Label_{t-1}|\Input_1,\ldots,\Input_{t-1}) d\Label_{t-1} \notag \\
	&\qquad =  \int \Pr(\Label_t|\Label_{t-1}) \Pr(\Label_{t-1}|\Input_1,\ldots,\Input_{t-1}) d\Label_{t-1}.
	\label{eqn:CK2}
	\end{align}
	Equations \eqref{eqn:CK1}-\eqref{eqn:CK2}, known as Chapman-Kolmogorov equations, show how to obtain the optimal estimate by adaptively updating the posterior $\hat{p}_t \triangleq   \Pr(\Label_t | \Input_1,\ldots,\Input_t)$ as $t$ evolves. In particular, given  $\hat{p}_{t-1}$, the Chapman Kolmogorov equations indicate how  $\hat{p}_{t}$ is obtained via
	\begin{equation}
	\label{eqn:recursion}
	\hat{p}_{t-1} \stackrel{\eqref{eqn:CK2}}{\Rightarrow}  \Pr(\Label_t|\Input_1,\ldots,\Input_{t-1}) \stackrel{\eqref{eqn:CK1a}}{\Rightarrow}\Pr(\Input_t|\Input_1,\ldots,\Input_{t-1}) \stackrel{\eqref{eqn:CK1}}{\Rightarrow}  \hat{p}_{t}.
	\end{equation}
	While the above relationship indicates a principled solver, it is not necessarily analytically tractable (in fact, the family of particle filters \cite{arulampalam2002tutorial} is derived to tackle the challenges associated with computing \eqref{eqn:recursion}).
	
	For the special case of a linear Gaussian state-space model as in \eqref{eqn:ssmodel}, all the considered variables are jointly Gaussian, and thus we can write
	\begin{equation}
	\Label_t | \Input_1,\ldots,\Input_t \sim \mathcal{N}\left(\hat{\Label}_{t|t}, \myMat{\Sigma}_{t|t}\right).
	\end{equation}
	Since $\Label_t | \Label_{t-1} \sim \mathcal{N}(\myMat{F} \Label_{t-1}, \myMat{V})$ by \eqref{eqn:ssmodel}, then \eqref{eqn:CK2} implies that
	\begin{align}
	\Label_t | \Input_1,\ldots,\Input_{t-1} \sim \mathcal{N}\Big(&\hat{\Label}_{t|t-1} = \myMat{F}\hat{\Label}_{t-1|t-1}, \notag \\
	& \myMat{\Sigma}_{t|t-1} = \myMat{F}\myMat{\Sigma}_{t-1|t-1}\myMat{F}^T + \myMat{V}   \Big).
	\label{eqn:CProb1}
	\end{align}
	Similarly, since $\Input_t | \Label_{t} \sim \mathcal{N}(\myMat{H} \Label_{t}, \myMat{W})$ by \eqref{eqn:ssmodel},   combining \eqref{eqn:CProb1} and \eqref{eqn:CK1a} implies that 
	\begin{align}
	\Input_t|\Input_1,\ldots,\Input_{t-1} \sim      \mathcal{N}\Big(&\hat{\Input}_{t|t-1} = \myMat{H}\hat{\Label}_{t|t-1}, \notag \\
	& \myMat{Q}_{t} = \myMat{H}\myMat{\Sigma}_{t|t-1}\myMat{H}^T + \myMat{W}   \Big).
	\label{eqn:CProb2}
	\end{align}
	Finally, substituting \eqref{eqn:CProb1} and \eqref{eqn:CProb2} into \eqref{eqn:CK1} results in
	\begin{align}
	\label{eqn:Kalman1}
	\hat{\Label}_{t|t} &= \hat{\Label}_{t|t-1} +\myMat{K}_t(\Input_t - \hat{\Input}_{t|t-1}), \\
	\myMat{\Sigma}_{t|t} &= \myMat{\Sigma}_{t|t-1} - \myMat{K}_t\myMat{H}\myMat{\Sigma}_{t|t-1}, \label{eqn:SigmaKal}\\
	\myMat{K}_t &= \myMat{\Sigma}_{t|t-1} \myMat{H}^T\myMat{Q}_{t}^{-1}.
		\label{eqn:Kalman3}
	\end{align}
	The matrix  $\myMat{K}_t$ is referred to as the Kalman gain. Equations~\eqref{eqn:Kalman1}-\eqref{eqn:Kalman3} are known as the Kalman filter \cite{kalman1960new}, which is one of the most celebrated and widely-used algorithms in signal processing and control. 
\end{example}

    Example~\ref{exm:LQGPoliciy} demonstrates how the modeling of a complex task using a simplified linear Gaussian model, combined with the usage of a simple  quadratic objective, results in an explicit solution, which here takes a linear form. The resulting Kalman filter achieves the minimal \ac{mse} when the state-space formulation it uses faithfully represents the underlying setup, and its parameters, i.e., $\ObjParam$, are accurate. However, mismatches and approximation inaccuracies in the model and its parameters can notably degrade performance.
    
\begin{figure}
	\centering
		\begin{subfigure}{0.42\textwidth}
		\centering
		{\includegraphics[width=\columnwidth]{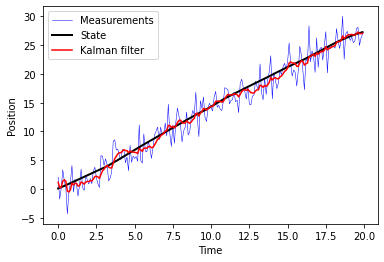}} 
		\caption{ Kalman filter with full domain knowledge.
		}
		\label{fig:KalmanFullCSI} 	
	\end{subfigure}
	$\quad$
	\begin{subfigure}{0.42\textwidth}
		\centering
		{\includegraphics[width=\columnwidth]{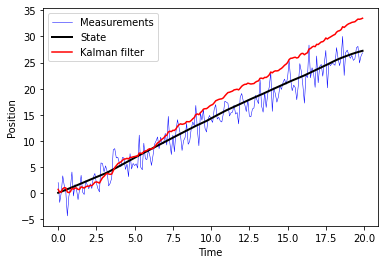}} 
		\caption{ Kalman filter with rotated observations.
		}
		\label{fig:KalmanMismatch} 
	\end{subfigure}
	\caption{Tracking a dynamic system from noisy observations using the model-based Kalman filter.}
	\label{fig:Kalman}
\end{figure}

    To see this, we simulate the tracking of a two dimensional state vector from noisy observations of its first entry obeying a linear Gaussian state-space model, where the state evolution matrix is $\myMat{F}  = \left[\begin{array}{c c} 1 & 0.1 \\ 0 & 1 \end{array}\right]$. When the state-space model is fully characterized, the Kalman filter successfully tracks the state, and we visualize the estimated first entry compared with its true value and noisy observations in Fig.~\ref{fig:KalmanFullCSI}. However, when instead of the data being generated from a state-space model with state evolution matrix $\myMat{F}$, it is generated from the same model with the matrix rotated by  $0.01$ [rad], i.e., there is a slight mismatch in the state-space model, then there is notable drift in the performance of the Kalman filter, as illustrated in Fig.~\ref{fig:KalmanMismatch}. These results showcase the dependency of explicit solvers on accurate modeling.

\section{Iterative Optimizers}
\label{subsec:MBIterative}

So far, we have discussed model-based decision making, and saw that objectives are often parameterized (we used $\ObjParam$ to denote these objective parameters). We also noted that in some cases, the formulation of the objective indicates  how it can be solved, and even saw scenarios for which a closed-form analytically tractable solution exists. However, closed-form solutions are often scarce, and typically arise only in simplified models. When closed-form solutions are not available, the model-based approach resorts to iterative optimization techniques. 

Optimization problems take the generic form of 
\begin{align*}
&{\rm minimize} \quad  \mySet{L}_0(\myVec{s}),\\
&{\rm subject ~to}\quad  \mySet{L}_i(\myVec{s}) \leq 0, \quad i=1,\ldots m,
\end{align*}
where $\myVec{s}\in\mySet{R}^n$ is the optimization variable, the function $\mySet{L}_0:\mySet{R}^n\mapsto\mySet{R}$ is the objective, and $\mySet{L}_i:\mySet{R}^n\mapsto\mySet{R}$ for $i=1,\ldots, m$ are constraint functions. 
By defining the constraint set $\mySet{S} = \{\myVec{s}: \mySet{L}_i(\myVec{s}) \leq 0, \forall i=1,\ldots m\}\subset \mySet{R}^n$, we can write the global solution to the optimization problem as 
\begin{equation}
\label{eqn:Opt1}
\myVec{s}^* = \mathop{\arg \min}\limits_{\myVec{s} \in \mySet{S}}\mySet{L}_0(\myVec{s}).
\end{equation}
While we use the same notation ($\Label$) for the optimization variable and the desired output of the inference rule, in the context of inference rule design the optimized variable is not necessarily the parameter being inferred. 

As the above formulation is extremely generic, it subdues many important families. One of them is that of {\em linear programming}, for which the functions $\{\mySet{L}_i(\cdot)\}_{i=0}^m$ are all linear, namely
\begin{equation*}
\mySet{L}_i(\alpha \myVec{s}_1 + \beta \myVec{s}_2) = \alpha \mySet{L}_i( \myVec{s}_1) + \beta \mySet{L}_i (\myVec{s}_2), \qquad \forall \alpha, \beta \in \mySet{R}, \myVec{s}_1, \myVec{s}_2 \in \mySet{R}^n.
\end{equation*}
A generalization of linear programming is that of {\em convex optimization}, where for each $i=0,1,\ldots,m$ it holds that
\begin{equation}
\label{eqn:convex}
\mySet{L}_i(\alpha \myVec{s}_1 + \beta \myVec{s}_2) \leq \alpha \mySet{L}_i( \myVec{s}_1) + \beta \mySet{L}_i (\myVec{s}_2), \qquad \forall \alpha, \beta \in \mySet{R}, \myVec{s}_1, \myVec{s}_2 \in \mySet{R}^n.
\end{equation}
In particular, if \eqref{eqn:convex} holds for $i=0$, we say that the optimization problem \eqref{eqn:Opt1} has a {\em convex objective}, while when it holds for all $i=1,\ldots m$ we say that $\mySet{S}$ is a {\em convex set}. 

Convex optimization theory provides useful iterative algorithms for tackling problems of the form \eqref{eqn:Opt1}.  Iterative optimizers typically give rise to additional parameters which affect the speed and convergence rate of the algorithm, but not the actual objective being minimized. We refer to these parameters of the solver as {\em hyperparameters}, and denote them by $\HypParam$. As opposed to  the objective parameters $\ObjParam$ (as in, e.g., \eqref{eqn:vPsi}), they often have no effect on the solution when the algorithm is allowed to run to convergence, and so are of secondary importance.   But when the iterative algorithm is stopped after a predefined number of iterations, they affect the decisions, and therefore also the decision rule objective. Due to the surrogate nature of the  objective, such stopping does not necessarily degrade the evaluation performance. We next show how this is exemplified for several common iterative optimization methods.

\subsection{First-Order Methods}
We begin with unconstrained optimization, for which $\mySet{S}=\mySet{R}^n$, i.e., 
\begin{equation}
\label{eqn:Opt1a}
\myVec{s}^* = \mathop{\arg \min}\limits_{\myVec{s}}\mySet{L}_0(\myVec{s}).
\end{equation}
When the objective is convex,  $\mySet{L}_0(\cdot)$ has a single minima, and thus if one can iteratively update over $\myVec{s}^{(k)}$ and guarantee that $\mySet{L}_0(\myVec{s}^{(k+1)}) < \mySet{L}_0(\myVec{s}^{(k)})$ then this procedure is expected to converge to $\myVec{s}^*$. The family of iterative optimizers that operate this way are referred to as {\em descent methods}.

Descent methods that operate based on first-order derivatives of the objective, i.e., using gradients, are referred to as {\em first-order methods}. Arguably the most common first-order method is gradient descent, which dates back to Cauchy (in the 19th century).

\paragraph{Gradient Descent}
Gradient descent stems from the first-order multivariate Taylor series expansion of $\mySet{L}_0(\cdot)$ around $\myVec{s}^{(k)}$, which implies that 
\begin{equation}
\label{eqn:Taylor}
\mySet{L}_0(\myVec{s}^{(k)} + \Delta \myVec{s}) \approx \mySet{L}_0(\myVec{s}^{(k)}) + \Delta \myVec{s}^T \nabla_{\myVec{s}} \mySet{L}_0(\myVec{s}^{(k)}),
\end{equation}
which holds when $\myVec{s}^{(k)} + \Delta \myVec{s}$ is in the proximity of $\myVec{s}^{(k)}$, i.e., the norm of $\Delta \myVec{s}$ is bounded by some $\epsilon > 0$. In this case, we seek the setting of $\Delta \myVec{s}$ for which the objective is minimized subject to $\|\Delta \myVec{s}\|_2 \leq \epsilon$. 
The right hand side of \eqref{eqn:Taylor} is minimized under the constraint when $\Delta \myVec{s} = -\mu \nabla_{\myVec{s}} \mySet{L}_0(\myVec{s}^{(k)})$ by the Cauchy-Schwartz inequality, where $\mu$ is set to $\mu = \frac{\epsilon}{\|\nabla_{\myVec{s}} \mySet{L}_0(\myVec{s}^{(k)})\|_2}$, guaranteeing that the norm of $\Delta \myVec{s}$ is bounded and thus the first-order Taylor series representation in \eqref{eqn:Taylor} approximately holds. The resulting update equation, which is repeated over the iteration index $k=0,1,2,\ldots$, is given by
\begin{equation}
\label{eqn:FullGD}
\myVec{s}^{(k+1)} \leftarrow \myVec{s}^{(k)} - \mu_k \nabla_{\myVec{s}} \mySet{L}_0(\myVec{s}^{(k)}),
\end{equation}
where $\mu_k$ is referred to as the {\em step size} of the {\em learning rate}. 

It is noted though that one must take caution to guarantee that we are still in the proximity of  $\myVec{s}^{(k)}$ where the approximation \eqref{eqn:Taylor} holds. A possible approach is to select the step size $\mu_k$ at each iteration using some form of line search, finding the one which maximizes the gap  $\mySet{L}_0(\myVec{s}^{(k)}) - \mySet{L}_0(\myVec{s}^{(k+1)})$ via, e.g., backtracking~\cite[Ch. 9.2]{boyd2004convex}.

\paragraph{Proximal Gradient Descent} 
Gradient descent relies on the ability to compute the gradients of $\mySet{L}_0(\cdot)$. Consider an objective function that can be decomposed into two functions as follows:
\begin{equation}
\label{eqn:Decompose1}
\mySet{L}_0(\myVec{s}) = g(\myVec{s})+h(\myVec{s}),
\end{equation}
where both $g(\cdot)$ and $h(\cdot)$ are convex, but only $g(\cdot)$ is necessarily differentiable. Such settings often represent regularized optimization, where $h(\cdot)$ represents the regularizing term. A common approach to implement a descent method which updates $\myVec{s}^{(k)}$ at each iteration $k$ is to approximate $g(\cdot)$ with its second-order Taylor series approximation. Nonetheless, the second-order derivatives, i.e., the Hessian matrix, is not explicitly computed (hence the corresponding method is still a first-order method). Instead, the Hessian matrix is approximated as $\frac{1}{\mu_k}\myMat{I}$ for some $\mu_k \geq 0$. This results in
\begin{equation}
\label{eqn:Taylor2}
g(\myVec{s}^{(k)} + \Delta \myVec{s}) \approx g(\myVec{s}^{(k)}) + \Delta \myVec{s}^T \nabla_{\myVec{s}} g(\myVec{s}^{(k)}) + \frac{1}{2\mu_k} \|\Delta \myVec{s}\|_2^2.
\end{equation}

Using the approximation in \eqref{eqn:Taylor2}, we aim at finding $\myVec{s}^{(k+1)} = \myVec{s}^{(k)} + \Delta \myVec{s}$ by minimizing
\begin{align}
\myVec{s}^{(k+1)} 
&= \mathop{\arg \min}_{\myVec{z}}  g(\myVec{s}^{(k)}) + ( \myVec{z} - \myVec{s}^{(k)})^T \nabla_{\myVec{s}} g(\myVec{s}^{(k)}) + \frac{1}{2\mu_k} \|\myVec{z} - \myVec{s}^{(k)}\|_2^2 +h({\myVec{z}}) \notag \\
&= \mathop{\arg \min}_{\myVec{z}}  \frac{1}{2\mu_k} \|\mu_k \nabla_{\myVec{s}} g(\myVec{s}^{(k)})\|_2^2  + ( \myVec{z} - \myVec{s}^{(k)})^T \nabla_{\myVec{s}} g(\myVec{s}^{(k)}) \notag \\
& \qquad \qquad  + \frac{1}{2\mu_k} \|\myVec{z} - \myVec{s}^{(k)}\|_2^2 +h({\myVec{z}}) +  g(\myVec{s}^{(k)}) -\frac{1}{2\mu_k} \|\mu_k \nabla_{\myVec{s}} g(\myVec{s}^{(k)})\|_2^2\notag \\
&= \mathop{\arg \min}_{\myVec{z}} \frac{1}{2\mu_k} \left\|\myVec{z} - \left(\myVec{s}^{(k)} - \mu_k \nabla_{\myVec{s}} g(\myVec{s}^{(k)})  \right) \right\|_2^2+h({\myVec{z}}).
\end{align}
Next, we define the {\em proximal mapping} as
\begin{equation}
\label{eqn:ProxDef}
{\rm prox}_{\phi}(\myVec{y})\triangleq \mathop{\arg \min}_{\myVec{z}}\frac{1}{2}\|\myVec{z}-\myVec{y}\|_2^2 + \phi(\myVec{z}).
\end{equation}
The resulting update equation, which is repeated over the iteration index $k=0,1,2,\ldots$, is given by
\begin{equation}
\label{eqn:ProxGD}
\myVec{s}^{(k+1)} \leftarrow  {\rm prox}_{\mu_k\cdot h}\left( \myVec{s}^{(k)} - \mu_k \nabla_{\myVec{s}} g(\myVec{s}^{(k)})\right),
\end{equation}
where we use ${\rm prox}_{\mu_k\cdot h}$ to represent the proximal mapping in \eqref{eqn:ProxDef} with $\phi(\myVec{z})\equiv \mu_k \cdot h(\myVec{z})$. 
Equation~\eqref{eqn:ProxGD} is the update equation of {\em proximal gradient descent}. Its key benefits stem from the following properties:
\begin{itemize}
	\item The proximal mapping can be computed analytically for many relevant $h(\cdot)$ functions, even if they are not differentiable.
	\item The proximal mapping is completely invariant of $g(\cdot)$ in \eqref{eqn:Decompose1}.
	\item The update equation in \eqref{eqn:ProxGD} coincides with conventional gradient descent as in \eqref{eqn:FullGD} when the term $h(\cdot)$ is constant. 
\end{itemize}

\begin{example}[ISTA]
	\label{exm:ISTA}
	Recall the $\ell_0$ regularized objective in \eqref{eqn:InvesObj1} used for describing the   super-resolution with sparse prior problem mentioned in Example~\ref{exm:InverseModel1}, where for simplicity we focus on the setting where $\myMat{\Psi} = \myMat{I}$, i.e., $\myVec{r}=\myVec{s}$. While this function is not convex, it can be relaxed into a convex formulation, known as the {\em LASSO} objective, by replacing the $\ell_0$ regularizer with an $\ell_1$ norm, namely:
	\begin{equation}
	\label{eqn:LASSO}
	f_{\rm LASSO}(\Input) = \mathop{\arg\min}\limits_{\myVec{s}} \frac{1}{2}\|\myVec{x}-\myMat{H}\myVec{s}\|^2 +\rho\|\myVec{s}\|_1.
	\end{equation}
	The objective in \eqref{eqn:LASSO} takes the form of \eqref{eqn:Decompose1} with $h(\myVec{s}) \equiv \rho\|\myVec{s}\|_1$ and $g(\myVec{s}) \equiv \frac{1}{2}\|\myVec{x}-\myMat{H}\myVec{s}\|^2$. In this case
	\begin{equation}
	\nabla_{\myVec{s}} g(\myVec{s}) = -\myMat{H}^T(\myVec{x}-\myMat{H}\myVec{s}).
	\end{equation}
	The proximal mapping is given by
	\begin{align}
	{\rm prox}_{\mu_k\cdot h}(\myVec{y}) = \mathop{\arg \min}_{\myVec{z}}\frac{1}{2}\|\myVec{z}-\myVec{y}\|_2^2 + \mu_k\rho\|\myVec{z}\|_1  &= \mySet{T}_{\mu_k\rho} (\myVec{y}),
	\end{align}
	where $\mySet{T}_{\beta}(\cdot)$ is the soft-thresholding operation applied element-wise,  given by
	\begin{equation}
	\mySet{T}_{\beta}(x) \triangleq {\rm sign}(x)\max(0,|x|-\beta). 
	\end{equation}
	This follows since $\left\|\myVec{y}-\myVec{s} \right\|_2^2   +  2\beta\| \csSignal \|_{1} = \sum (y_i -s_i)^2 +2\beta |s_i|$, and for  scalars $y_i$ and $s_i$, it holds that 
	\begin{align}
	\label{eqn:Derive1}
	\frac{d}{ds_i}\left((y_i-s_i)^2 + 2\beta|s_i|\right) 
	&= 2\left(s_i-y_i +\beta {\rm sign}(s_i)\right).
	\end{align}
	When compared to zero, \eqref{eqn:Derive1} yields $s_i = \mySet{T}_{\beta}(y_i)$. 
	The resulting formulation of the proximal gradient descent method is known as {\em \ac{ista}}~\cite{daubechies2004iterative}, and its update equation is given by
	\begin{equation}
	\label{eqn:ISTA}
	\myVec{s}^{(k+1)} \leftarrow  \mySet{T}_{\mu_k\rho}\left( \myVec{s}^{(k)} + \mu_k \myMat{H}^T(\myVec{x}-\myMat{H}\myVec{s}^{(k)}) \right).
	\end{equation}
\end{example}

\begin{example}[Projected Gradient Descent]
	\label{exm:PGD}
	Consider an optimization carried out over a closed set $\mySet{S}\subset \mySet{R}^n$. Such optimization problems can be re-formulated as
	\begin{equation*}
	\myVec{s}^* = \mathop{\arg \min}\limits_{\myVec{s} \in \mySet{S}}\mySet{L}_0(\myVec{s}) = \mathop{\arg \min}\limits_{\myVec{s}}\mySet{L}_0(\myVec{s}) + I_{\mySet{S}}(\myVec{s}), \qquad I_{\mySet{S}}(\myVec{s}) \triangleq \begin{cases}0 & \myVec{s}\in \mySet{S} \\ \infty & \myVec{s}\notin \mySet{S}\end{cases}.
	\end{equation*}
	In this case, the proximal mapping specializes into the projection operator denoted ${\Pi}_{\mySet{S}}$, since
	\begin{align}
	{\rm prox}_{\mu_k\cdot I_{\mySet{S}}}(\myVec{y}) &= \mathop{\arg \min}_{\myVec{z}}\frac{1}{2}\|\myVec{z}-\myVec{y}\|_2^2 + \mu_k I_{\mySet{S}}(\myVec{z}) \notag \\ &=  \mathop{\arg \min}_{\myVec{z} \in \mySet{S}}\|\myVec{z}-\myVec{y}\|_2^2 \equiv \Pi_{\mySet{S}}(\myVec{y}).
	\end{align}
	Consequently, the proximal gradient descent here coincides with projected gradient descent, i.e., 
	\begin{equation}
	\label{eqn:ProjGD}
	\myVec{s}^{(k+1)} \leftarrow  \Pi_{\mySet{S}}\left( \myVec{s}^{(k)} - \mu_k \nabla_{\myVec{s}} \mySet{L}_0(\myVec{s}^{(k)})\right).
	\end{equation}
\end{example}

\begin{figure}
	\centering
	\begin{subfigure}{0.45\textwidth}
		\centering
		{\includegraphics[width=\columnwidth]{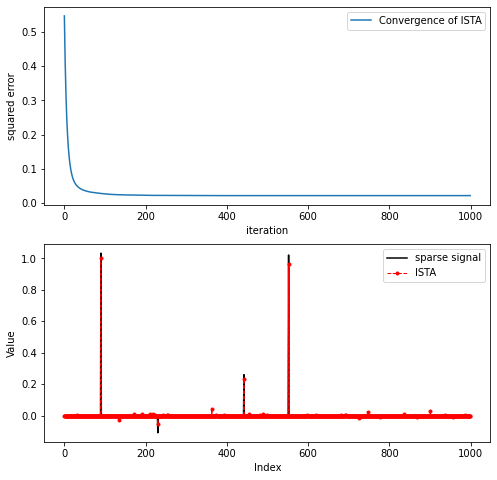}} 
		\caption{ \ac{ista} with step-size $\mu = 0.1$.
		}
		\label{fig:ISTA_mu01} 	
	\end{subfigure}
	$\quad$
	\begin{subfigure}{0.45\textwidth}
		\centering
		{\includegraphics[width=\columnwidth]{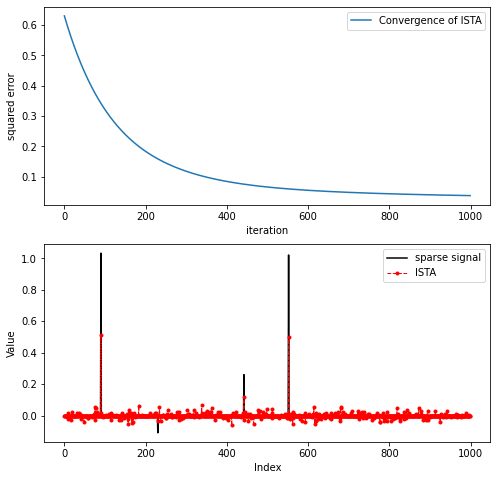}} 
		\caption{ \ac{ista} with step-size $\mu = 0.05$.
		}
		\label{fig:ISTA_mu005} 
	\end{subfigure}
	\caption{The convergence profile (upper figures) and recovered signal (lower figures) achieved by \ac{ista} for recovering a sparse high-dimensional signal from noisy low-dimensional observations.}
	\label{fig:ISTA}
\end{figure}

 First-order optimization procedures introduce additional hyperparameters, namely $\HypParam = \{\mu_k\}$, which are often fixed to a single step-size to facilitate tuning, i.e., $\HypParam = \mu$. The exact setting of these hyperparameters in fact affects the performance of the iterative algorithm, particularly when it is limited in the maximal number of iterations it can carry out. To illustrate this dependency, we evaluate \ac{ista} for the recovery of a signal with $n=1000$ entries, of which only $5$ are non-zero, from a noisy observations vector of comprised of $256$ measurements taken with a random Gaussian measurement matrix corrupted by noise with variance $\sigma^2 = 0.01$. We use two different step-sizes -- $\mu = 0.1$ (Fig.~\ref{fig:ISTA_mu01}) and $\mu=0.05$ (Fig.~\ref{fig:ISTA_mu005}) -- and limit the maximal number of \ac{ista} iterations to $1000$. Observing Fig.~\ref{fig:ISTA}, which depicts both the recovered signal as well as the convergence profile, i.e., the evolution of the squared error over the iterations, we note that the setting of the hyperparameters of the optimization procedure has a dominant effect on both the recovered signal and the convergence rate.

\subsection{Constrained Optimization}

Example \ref{exm:PGD} shows one approach to tackle constrained optimization, by casting the problem as an unconstrained objective which results in projected gradient descent as a first-order method. This is obviously not the only approach to tackle constrained optimization. The more conventional approach is based on Lagrange multipliers and duality. To describe these methods, let us first repeat the considered constrained optimization problem (with additional equality constraints):
\begin{eqnarray}
\label{eqn:OptProb1}
&{\rm minimize} \quad  &\mySet{L}_0(\myVec{s}),\\
&{\rm subject ~to}\quad  &\mySet{L}_i(\myVec{s}) \leq 0, \quad i=1,\ldots m, \notag\\
& &l_j(\myVec{s}) = 0, \quad j=1,\ldots r.\notag
\end{eqnarray}

The classic method of Lagrange multipliers converts the constrained optimization problem \eqref{eqn:OptProb1} into a single objective with additional auxiliary variables $\myVec{\eta} = [\eta_1,\ldots, \eta_m]$ and $\myVec{\mu}=[\mu_1,\ldots,\mu_r]$ with $\eta_i\geq 0$ and $\mu_j l_j(\myVec{s})=0$ for each $i\in 1,\ldots,m$ and $j\in 1,\ldots r$, respectively. The {\em Lagrangian} is defined as
\begin{equation}
\label{eqn:Lagrange1}
L(\myVec{s}, \myVec{\mu}, \myVec{\eta}) \triangleq \mySet{L}_0(\myVec{s}) +\sum_{i=1}^m \eta_i \mySet{L}_i(\myVec{s})  +\sum_{j=1}^r \mu_j l_j(\myVec{s}),
\end{equation}
and the {\em dual function} is
\begin{equation}
d(\myVec{\mu}, \myVec{\eta})\triangleq \min_{\myVec{s}}L(\myVec{s}, \myVec{\mu}, \myVec{\eta}). 
\end{equation}
The benefit of defining the dual function stems from the fact that for every non-negative $\eta_1,\ldots, \eta_m$ and for each $\myVec{s}$ satisfying the constraints in \eqref{eqn:OptProb1}, i.e., $\myVec{s}\in \mySet{S}$, it holds that 
\begin{align}
L(\myVec{s}, \myVec{\mu}, \myVec{\eta}) &= \mySet{L}_0(\myVec{s}) +\sum_{i=1}^m \eta_i \mySet{L}_i(\myVec{s})  +\sum_{j=1}^r \mu_j l_j(\myVec{s}) \notag \\
&\stackrel{(a)}{=} \mySet{L}_0(\myVec{s}) +\sum_{i=1}^m \eta_i \mySet{L}_i(\myVec{s})\stackrel{(b)}{\leq} \mySet{L}_0(\myVec{s}),
\end{align}
where $(a)$ and $(b)$ follow since for each $\myVec{s}\in\mySet{S}$ it holds that  $l_j(\myVec{s}) = 0$ and $\mySet{L}_i(\myVec{s}) \leq 0$, respectively.  
Consequently,
\begin{align}
d(\myVec{\mu}, \myVec{\eta}) &= \min_{\myVec{s}}L(\myVec{s}, \myVec{\mu}, \myVec{\eta}) 
\leq 
\min_{\myVec{s}} \mySet{L}_0(\myVec{s}) \leq \mySet{L}_0(\myVec{s}^*),\notag
\end{align}
i.e., for every non-negative $\eta_1,\ldots, \eta_m$, the dual function is a lower bound on the global minima of \eqref{eqn:OptProb1}. 
Under some conditions (Slater's condition, KKT, see \cite[Ch. 5]{boyd2004convex}), the maxima of the dual coincides with the global minima of the primal \eqref{eqn:OptProb1}, and thus solving \eqref{eqn:OptProb1} can be carried out by solving the dual problem 
\begin{eqnarray}
\label{eqn:DualProb1}
&{\rm maximize} \quad  & d(\myVec{\mu}, \myVec{\eta}),\\
&{\rm subject ~to}\quad  &\eta_i \geq 0, \quad i=1,\ldots m.
\end{eqnarray}

A common optimization algorithm which is based on duality is the \ac{admm} \cite{boyd2011distributed}. This algorithm aims at solving optimization problems where the objectives can be decomposed into two functions $g(\cdot)$ and $h(\cdot)$, with a constraint of the form
\begin{align}
\label{eqn:admm0}
\hat{\myVec{s}} &= \arg\mathop{\min}\limits_{\myVec{s}}\mathop{\min}\limits_{\myVec{v}} g(\myVec{s}) +   h(\myVec{v}) + \lambda \|\myVec{s} -\myVec{v}\|^2,  \\
&{\rm s.t. ~~~~~~~} \myMat{A}\myVec{v}+\myMat{B}\myVec{s} = \myVec{c}, \notag
\end{align}
for some fixed $\myMat{A},\myMat{B},\myVec{c}$. 
 \ac{admm} tackles the regularzied objective in \eqref{eqn:admm0} by formulating a dual problem as in \eqref{eqn:DualProb1}, which is solved by {\em alternating optimization}. 
 
 We next showcase a special case of \ac{admm} applied for solving regularized objectives, such as that considered in the context of super-resolution in Example~\ref{exm:InverseModel1}, for which \eqref{eqn:admm0} becomes  
\begin{align}
\hat{\myVec{s}}
&=\mathop{\arg\min}\limits_{\myVec{s}} \frac{1}{2}\|\myVec{x}-\myMat{H}\myVec{s}\|^2 +\phi(\myVec{s}).
\label{eqn:RegOpt4}
\end{align}
However, instead of solving \eqref{eqn:RegOpt4}, \ac{admm} decomposes the objective using variable splitting, formulating the objective as
\begin{align}
\label{eqn:admm1}
\hat{\myVec{s}} &= \mathop{\arg\min}\limits_{\myVec{s}}\mathop{\min}\limits_{\myVec{v}} \frac{1}{2}\|\myVec{x}-\csMatrix \myVec{s}\|^2 +   \phi(\myVec{v}) + \lambda \|\myVec{s} -\myVec{v}\|^2,  \\
&{\rm s.t. ~~~~~~~} \myVec{v}=\myVec{s}. \notag
\end{align}
The solution to \eqref{eqn:admm1} clearly coincides with that of \eqref{eqn:RegOpt4}. \ac{admm} tackles the regularzied objective in \eqref{eqn:admm1} by formulating a dual problem as in \eqref{eqn:DualProb1}, as detailed in the following example.

\begin{example}[ADMM]
	\label{exm:ADMM}
	Consider the constrained optimization problem in \eqref{eqn:admm1}. Since the formulation has only equality constraints, the Lagrangian is 
	\begin{align}
	L(\myVec{s}, \myVec{v}, \myVec{\mu}) 
	&= \frac{1}{2}\|\myVec{x}-\csMatrix \myVec{s}\|^2 +   \phi(\myVec{v}) + \lambda \|\myVec{s} -\myVec{v}\|^2 + \myVec{\mu}^T(\myVec{s} - \myVec{v}) \notag \\
	&\stackrel{(a)}{=} \frac{1}{2}\|\myVec{x}-\csMatrix \myVec{s}\|^2 +   \phi(\myVec{v}) + 
	\lambda\left\|\myVec{s} -\myVec{v} + \frac{1}{2\lambda}\myVec{\mu}\right\|^2 - \lambda\left\|\frac{1}{2\lambda} \myVec{\mu}\right\|^2,
	\label{eqn:AugLag}
	\end{align}
	where $(a)$ follows since for each $\myVec{y}$ (where here $\myVec{y} = \myVec{s}-\myVec{v}$) it holds that  $ \lambda\|\myVec{y} + \frac{1}{2\lambda}\myVec{\mu}\|^2 = \lambda \|\myVec{y}\|^2 + \myVec{\mu}^T\myVec{y} + \lambda\|\frac{1}{2\lambda} \myVec{\mu}\|^2$. Thus, by writing $\myVec{u} = \frac{1}{2\lambda}\myVec{\mu}$, we obtain the augmented Lagrangian
	\begin{align}
	\label{eqn:AugLag2}
	L(\myVec{s}, \myVec{v}, \myVec{u}) 
	&= \frac{1}{2}\|\myVec{x}-\csMatrix \myVec{s}\|^2 +   \phi(\myVec{v}) + 
	\lambda\|\myVec{s} -\myVec{v} + \myVec{u}\|^2 - \lambda\|\myVec{u}\|^2,
	\end{align}	
	and the dual is 
	\begin{equation}
	d(\myVec{u}) = \min_{\myVec{s}, \myVec{v}} L(\myVec{s}, \myVec{v}, \myVec{u}).
	\end{equation}
	
	As mentioned above, \ac{admm} solves the optimization problem in an alternating fashion. Namely, for each iteration $k$, it first optimizes $\myVec{s}$ to minimize the Lagrangian while keeping  $\myVec{v}$ and $\myVec{u}$ fixed, after which it optimizes  $\myVec{v}$ to minimize the Lagrangian, and then optimizes  $\myVec{u}$ to maximize the dual. 
	The resulting update equation for $\myVec{s}$ becomes
	\begin{align}
	\myVec{s}^{(k+1)} 
	&= \mathop{\arg\min}\limits_{\myVec{s}} L(\myVec{s}, \myVec{v}^{(k)}, \myVec{u}^{(k)}) \notag \\
	&= \mathop{\arg\min}\limits_{\myVec{s}}  \frac{1}{2}\|\myVec{x}-\csMatrix \myVec{s}\|^2 +  \lambda\|\myVec{s} -\myVec{v}^{(k)} + \myVec{u}^{(k)}\|^2.
	\end{align}
	Taking the gradient of  $L(\myVec{s}, \myVec{v}^{(k)}, \myVec{u}^{(k)})$ with respect to $\myVec{s}$ and comparing to zero yields
	\begin{equation}
	-\csMatrix^T(\myVec{x}-\csMatrix \myVec{s}^{(k+1)})  + 2\lambda(\myVec{s}^{(k+1)}  -\myVec{v}^{(k)} + \myVec{u}^{(k)}) = 0, 
	\end{equation}
	resulting in 
	\begin{equation}
	\myVec{s}^{(k+1)} = (\csMatrix^T\csMatrix  + 2\lambda \myMat{I})^{-1}(\csMatrix^T\myVec{x} + 2\lambda(\myVec{v}^{(k)} - \myVec{u}^{(k)})).
	\label{eqn:AdmmStp1}
	\end{equation}
	
	The update equation for $\myVec{v}$ is 
	\begin{align}
	\myVec{v}^{(k+1)} 
	&= \mathop{\arg\min}\limits_{\myVec{v}} L(\myVec{s}^{(k+1)}, \myVec{v}, \myVec{u}^{(k)}) \notag \\
	&= \mathop{\arg\min}\limits_{\myVec{v}} \phi(\myVec{v}) +  \lambda\|\myVec{s}^{(k+1)} -\myVec{v} + \myVec{u}^{(k)}\|^2 \notag \\
	&= \mathop{\arg\min}\limits_{\myVec{v}} \frac{1}{2\lambda}\phi(\myVec{v}) + \frac{1}{2} \| \myVec{v} -(\myVec{s}^{(k+1)} + \myVec{u}^{(k)})\|^2 \notag \\
	&= {\rm prox}_{\frac{1}{2\lambda}\cdot\phi}(\myVec{s}^{(k+1)} + \myVec{u}^{(k)}).
	\label{eqn:Prox1}
	\end{align}
	Finally, the auxiliary variable $\myVec{u}$, which should maximize the dual function, is updated via gradient ascent with step-size $\frac{\mu}{2\lambda}$, resulting in
	\begin{align}
	\myVec{u}^{(k+1)} 
	&=  \myVec{u}^{(k)} + \frac{\mu}{2\lambda}\nabla_{\myVec{u}=\myVec{u}^{(k)}} L(\myVec{s}^{(k+1)}, \myVec{v}^{(k+1)}, \myVec{u}) \notag \\ 
	&= \myVec{u}^{(k)} + \frac{\mu}{2\lambda}\nabla_{\myVec{u}=\myVec{u}^{(k)}} \left(\lambda\|\myVec{s}^{(k+1)} -\myVec{v}^{(k+1)} + \myVec{u}\|^2 - \lambda\|\myVec{u}\|^2\right) \notag \\
	&=\myVec{u}^{(k)} +\mu\left(\myVec{s}^{(k+1)} -\myVec{v}^{(k+1)}\right).
	\end{align}
	
	The resulting \ac{admm} optimization is summarized as Algorithm~\ref{alg:Algoadmm}.
	
	\begin{algorithm}  
		\caption{\ac{admm} for Problem~\eqref{eqn:admm1}}
		\label{alg:Algoadmm}
		\KwData{Fix   $\lambda, \mu>0$. Initialize  $\myVec{u}^{(0)}$,  $\myVec{v}^{(0)}$} 
		\For{$k=0,1,\ldots$}{
			Update $\myVec{s}^{(k+1)}$ via $
			\myVec{s}^{(k+1)} \leftarrow (\csMatrix^T\csMatrix  + 2\lambda \myMat{I})^{-1}(\csMatrix^T\myVec{x} + 2\lambda(\myVec{v}^{(k)} - \myVec{u}^{(k)}))$.\\
			\label{stp:prox}
			Update $\myVec{v}^{(k+1)}$ via $
			\myVec{v}^{(k+1)}  
			\leftarrow  {\rm prox}_{\frac{1}{2\lambda}\cdot\phi}(\myVec{s}^{(k+1)} + \myVec{u}^{(k)})$.\\
			Update $\myVec{u}^{(k+1)}$ via $
			\myVec{u}^{(k+1)}  
			\leftarrow \myVec{u}^{(k)} +\mu\left(\myVec{s}^{(k+1)} -\myVec{v}^{(k+1)}\right)$.
		}
		\KwOut{Estimate $\myVec{s}^{(k)}$.}
	\end{algorithm}

\end{example}

 In Example \ref{exm:ADMM},   the iterative solver introduces two hyperparameters, i.e., $\HypParam = [\lambda, \mu]$, where $\lambda$ is introduced in converting the objective in \eqref{eqn:RegOpt4} into \eqref{eqn:admm1}, while $\mu$ is used in the iterative minimization of \eqref{eqn:admm1}.

\section{Approximation and Heuristic Algorithms}
\label{subsec:MBApprox}
The model-based methods discussed in Sections~\ref{subsec:MBExplicit}-\ref{subsec:MBIterative} are derived directly as solvers, which are either explicit or iterative, to a closed-form optimization problem. Nonetheless, many important model-based methods, and particularly in signal processing related applications, are not obtained directly as solutions to an optimization problem. When tackling an optimization problem which is extremely challenging to solve in a computationally efficient manner (such as NP-hard problems), one often resorts to computationally inexpensive methods which are not guaranteed to minimize the objective. Common families of such techniques include approximation algorithms, e.g., greedy methods and dynamic programming, as well as heuristic approaches~\cite{williamson2011design}. Since signal processing applications are often applied in real-time on hardware-limited devices, computational efficiency plays a key role in their design, and thus such sub-optimal techniques are frequently used.

The term {\em approximation and heuristic algorithms} encompasses a broad range of diverse methods, whose main common aspect is the fact that they are not directly derived by solving a closed-form optimization problem. As a representative example of such algorithms in signal processing, we next detail the family of subspace methods for  tackling the \ac{doa} estimation problem  in Example~\ref{exm:DoaModel1}.

\begin{example}[Subspace Methods]
	\label{exm:Subspace}
	Consider the \ac{doa} estimation of narrowband sources formulated in Example~\ref{exm:DoaModel1} where the number of sources $d$ is smaller than the number of array elements $N$, the sources are non-coherent,  i.e., $\myMat{Y}$ is diagonal, and the noise is white, namely $\myMat{W} = \SigW\myMat{I}$ for some $\SigW > 0$. {The maximum likelihood estimator detailed in \eqref{eqn:DoAObj1} is typically computationally challenging to implement and relies on the assumption that the signals are temporally uncorrelated.} A popular alternative are subspace methods, which aim at recovering the \acp{doa} $\myVec{s}$ by dividing the covariance of  the observations $\Input_t$ into distinct {\em signal subspace} and {\em noise subspace}.
	
	In particular, under the above model assumptions, the covariance matrix of the observations $\Input_t$ in \eqref{eqn:DoAModel} is given by
	\begin{equation}
	\label{eqn:SubCov1}
	\E\{\Input_t\Input_t^H\} = \myMat{A}(\Label)\myMat{Y}\myMat{A}^H(\Label) + \SigW\myMat{I}.
	\end{equation}
	Note that $\myMat{A}(\Label)\myMat{Y}\myMat{A}^H(\Label)$ is an $N\times N$ matrix of rank $d < N$, and thus has $N-d$ eigenvectors corresponding to the zero eigenvalue, which are also the the eigenvectors corresponding to the $N-d$ least dominant eigenvalues of \eqref{eqn:SubCov1}. Let $\myVec{e}_n$ be such an eigenvector, i.e., $\myVec{e}_n^H\myMat{A}(\Label)\myMat{Y}\myMat{A}^H(\Label)\myVec{e}_n=0$. Since $\myMat{Y}$ is positive definite, it holds that $\myMat{A}^H(\Label)\myVec{e}_n=0$. Consequently, by letting $\myMat{E}_{\rm N}$ be the $(N-d)\times N$ matrix comprised of these $N-d$ least dominant eigenvectors, it holds that for each entry of $\myVec{s} = [s_1,\ldots, s_d]$: 
	\begin{equation}
	\label{eqn:SubspaceCore}
\myVec{a}^H(s_i)\myMat{E}_{\rm N}\myMat{E}_{\rm N}^H\myVec{a}
(s_i) = 
\big\| \myMat{E}_{\rm N}^H\myVec{a}({s_i}) \big\|^2 =  0, \qquad i\in 1,\ldots, d.
	\end{equation}
	Thus, subspace methods recover the \acp{doa} by seeking the steering vectors that are orthogonal to the noise subspace $\myMat{E}_{\rm N}$.	
\end{example}

The core equality of subspace methods in \eqref{eqn:SubspaceCore} is not derived from an optimization problem formulation, but is rather obtained from a set of arguments based on the understanding of the structure of the considered signals, which identifies the ability to decompose the input covariance into orthogonal signal and noise subspaces. This relationship gives rise to several classic \ac{doa} estimation methods, including the popular  \ac{music} algorithm~\cite{schmidt1986music}. 

\begin{example}[MUSIC]
	\label{exm:MUSIC}
		The \ac{music} algorithm exploits the subspace equality \eqref{eqn:SubspaceCore} to identify the \acp{doa} from the empirical estimate of the input covariance. To that aim, one first uses the $T$ snapshots to estimate the covariance in \eqref{eqn:SubCov1} as 
		\begin{equation}
		\label{eqn:MUSICCov}
		\myMat{C}_{\Input} = \frac{1}{T}\sum_{t=1}^T \Input_t \Input_t^H.
		\end{equation}
Then, its \acl{evd} is taken, from which the number of sources $\hat{d}$ is estimated as the number of dominant eigenvalues (e.g., by thresholding), while the eigenvectors corresponding to the remaining $N-\hat{d}$ eigenvalues are used to form the estimated noise subspace matrix $\hat{\myMat{E}}_{\rm N}$. 

The estimated noise subspace matrix $\hat{\myMat{E}}_{\rm N}$ is used to compute the  spatial spectrum, given by
		\begin{equation}\label{eqn:MUSICspectrum}
		{P}(\psi) = \frac{1}{\myVec{a}^H(\psi) \hat{\myMat{E}}_{\rm N}\hat{\myMat{E}}_{\rm N}^H \myVec{a}^H(\psi)}.
		\end{equation}
		The $\hat d$ dominant peaks of ${P}(\psi)$ are set as the estimated \ac{doa} angles $\hat{\Label}$.
\end{example}

An alternative subspace method is RootMUSIC~\cite{Barabell1983ImprovingTR}.
\begin{example}[RootMUSIC]
		\label{exm:RootMUSIC}
	RootMUSIC recovers the number of sources $\hat{d}$ and the  estimated noise subspace matrix $\hat{\myMat{E}}_{\rm N}$ in the same manner as MUSIC. However, instead of computing the spatial spectrum via~\eqref{eqn:MUSICspectrum}, it recovers the \acp{doa} from the roots of a polynomial formulation representing \eqref{eqn:SubspaceCore}. 
	
In particular, RootMUSIC formulates the Hermitian matrix $\myMat{E}=\hat{\myMat{E}}_{\rm N}\hat{\myMat{E}}_{\rm N}^\mathsf{H}$ using its diagonal sum coefficients ${e_n}$ via
\begin{equation}\label{eq:diagonal_coeff}
e_n = \sum_{i=0}^{N-1-n}[\myMat{E}]_{i,n+i},\quad  n \geq 0,
\end{equation}
where for $n < 0$, we set $e_n = e_{|n|}^\ast$. This allows to approximate \eqref{eqn:SubspaceCore}  as a polynomial equation of order $2N-2$ :
\begin{align} 
D(z)&= \sum_{i=0}^{N-1}\sum_{j=0}^{N-1}[\myVec{a}(\psi)]_i^\ast[\myMat{E}]_{ij}[\myVec{a}(\psi)]_j  
\notag  \\
&=\sum_{i=0}^{N-1}\sum_{j=0}^{N-1}{[\myMat{E}]_{ij}z^{i-j}}
=
\sum_{n=-(N-1)}^{N-1}{e_n z^n},
\label{eqn:polynomial}
\end{align}
where $z=e^{-j\pi \sin(\psi)}$. RootMUSIC  identifies the \acp{doa} from the roots of the polynomial \eqref{eqn:polynomial} and the roots map is viewed as the RootMUSIC spectrum. Since \eqref{eqn:polynomial} has $2N-2 > d$ roots (divided into symmetric pairs), while the roots corresponding to \acp{doa} should have unit magnitude, the $\hat{d}$ pairs of roots which are the closest to the unit circle are matched as the $\hat{d}$ sources \acp{doa} \cite{Barabell1983ImprovingTR}.
\end{example}

Examples~\ref{exm:MUSIC}-\ref{exm:RootMUSIC} showcase signal processing algorithms that are derived from revealing structures in the data that are identified based on domain knowledge and imposed assumptions, yet are not obtained by directly tackling their corresponding objective function. In fact, these \ac{doa} estimation algorithms do not rely on explicit knowledge of the objective parameters $\ObjParam$ in \eqref{eqn:DoAObj2}, but only on structures imposed on these parameters, i.e., that of non-coherent sources. Furthermore, they can operate with different number of snapshots $T$, where larger values of $T$ allow to better estimate the covariance via \eqref{eqn:MUSICCov}. In addition to estimating the desired $\Label$, the operation of the algorithms in Examples~\ref{exm:MUSIC}-\ref{exm:RootMUSIC} provides a visual interpretable representation of their decision via the \ac{music} spectrum in \eqref{eqn:MUSICspectrum} and the root map of RootMUSIC. 

An additional aspect showcased in the above examples is the dependency on reliable domain knowledge and faithful hardware representation. In particular, the ability to model the \ac{doa} estimation setup via \eqref{eqn:DoAModel} relies on the assumption that the sources are narrowband; the fact that the signal can be decomposed into signal and noise subspace depends on the assumption that the sources are non-coherent; the orthogonality of the steering vectors and the ability to compute them for each candidate angle holds when one possesses an array which is calibrated and its elements are indeed uniformly spaced with half-wavelength spacing. This set of structural assumptions is necessary for one to successfully recover \acp{doa} using subspace methods.

\begin{figure}
	\centering
	\begin{subfigure}{0.45\textwidth}
		\centering
		{\includegraphics[width=\columnwidth]{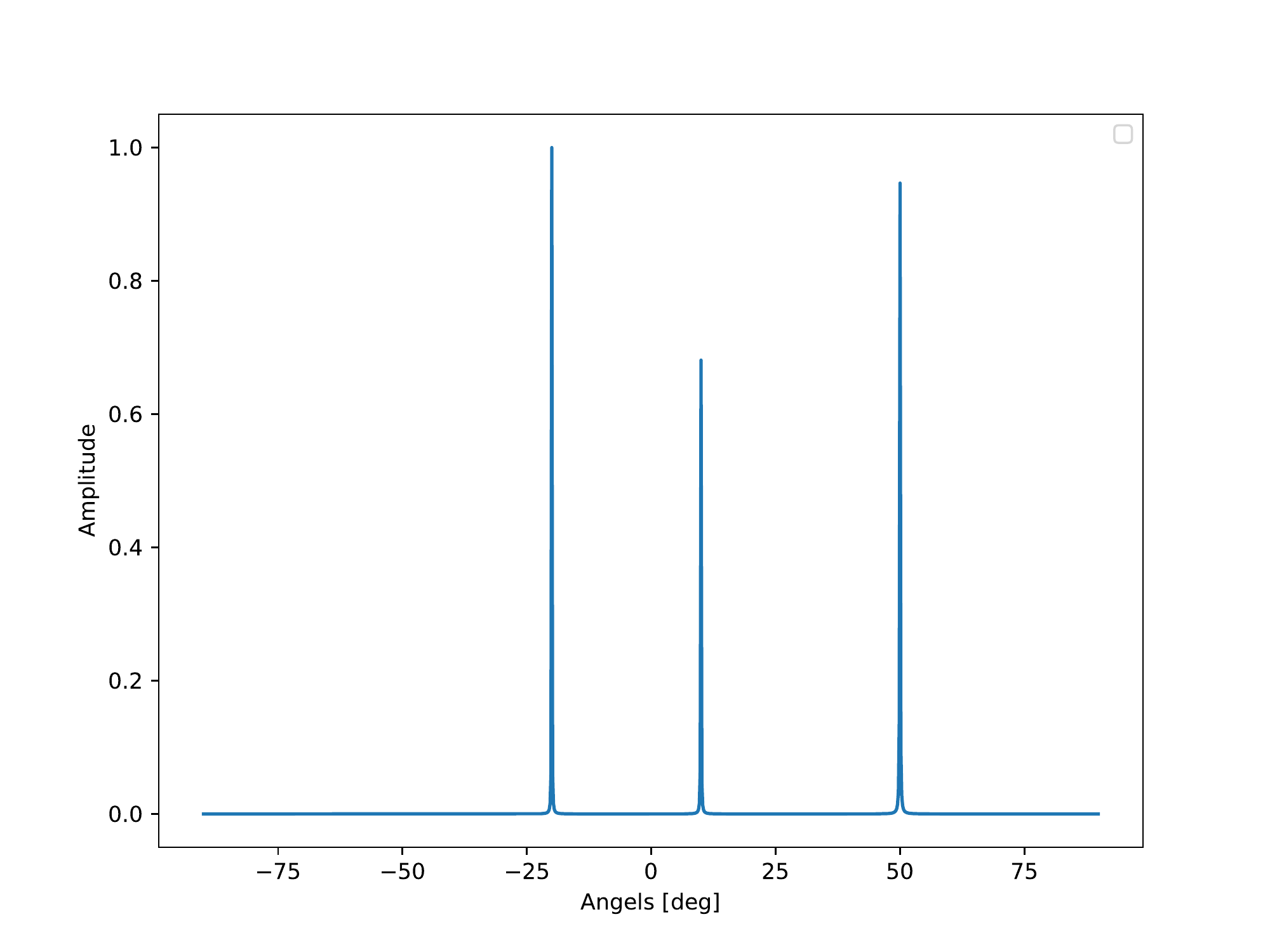}} 
		\caption{ \ac{music} spectrum, non-coherent sources.
		}
		\label{fig:MUSIC_nc} 	
	\end{subfigure}
	$\quad$
	\begin{subfigure}{0.45\textwidth}
		\centering
		{\includegraphics[width=\columnwidth]{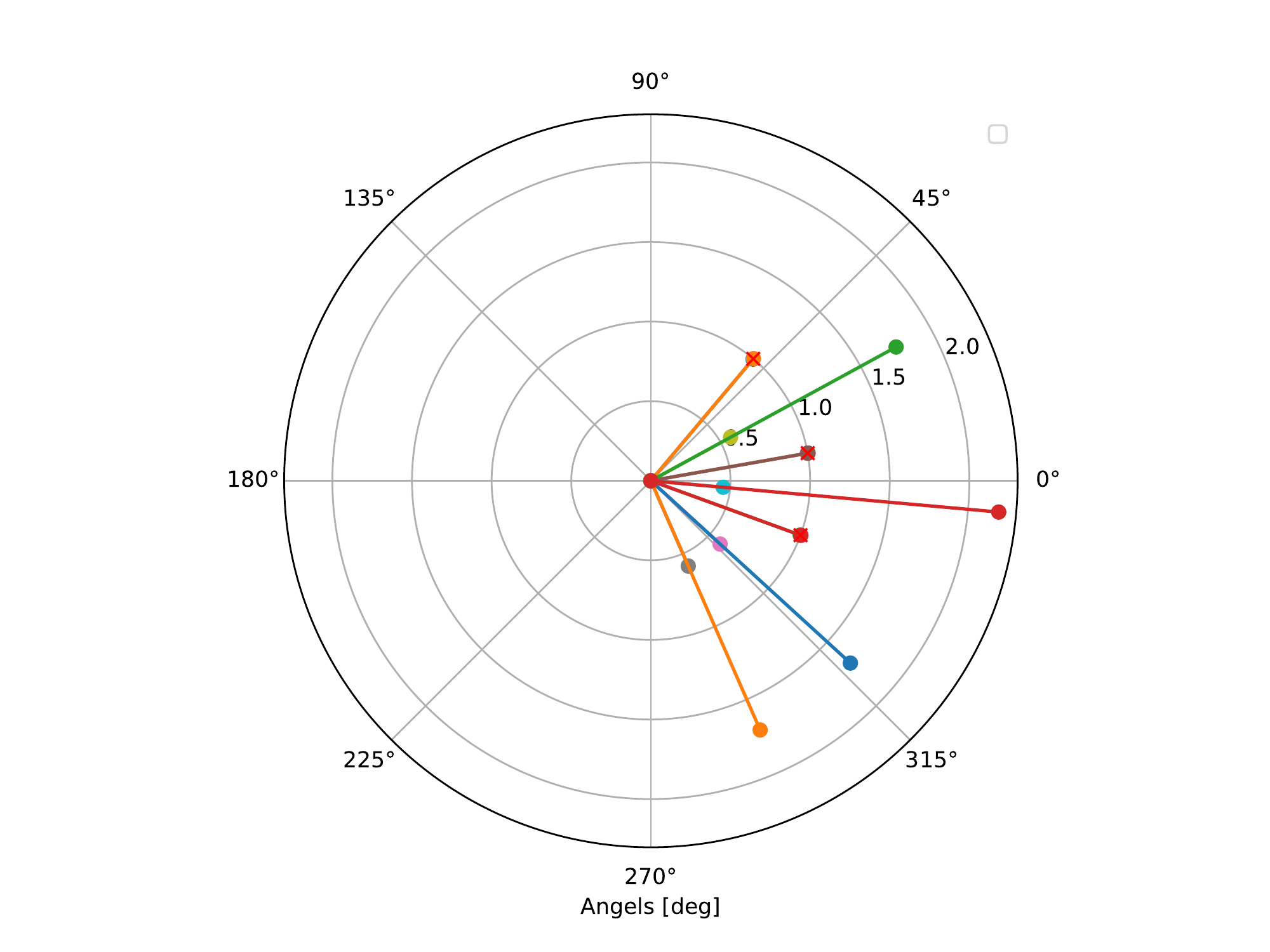}} 
		\caption{RootMUSIC spectrum, non-coherent sources.
		}
		\label{fig:RootMUSIC_nc} 
	\end{subfigure}\\
	\begin{subfigure}{0.45\textwidth}
	\centering
	{\includegraphics[width=\columnwidth]{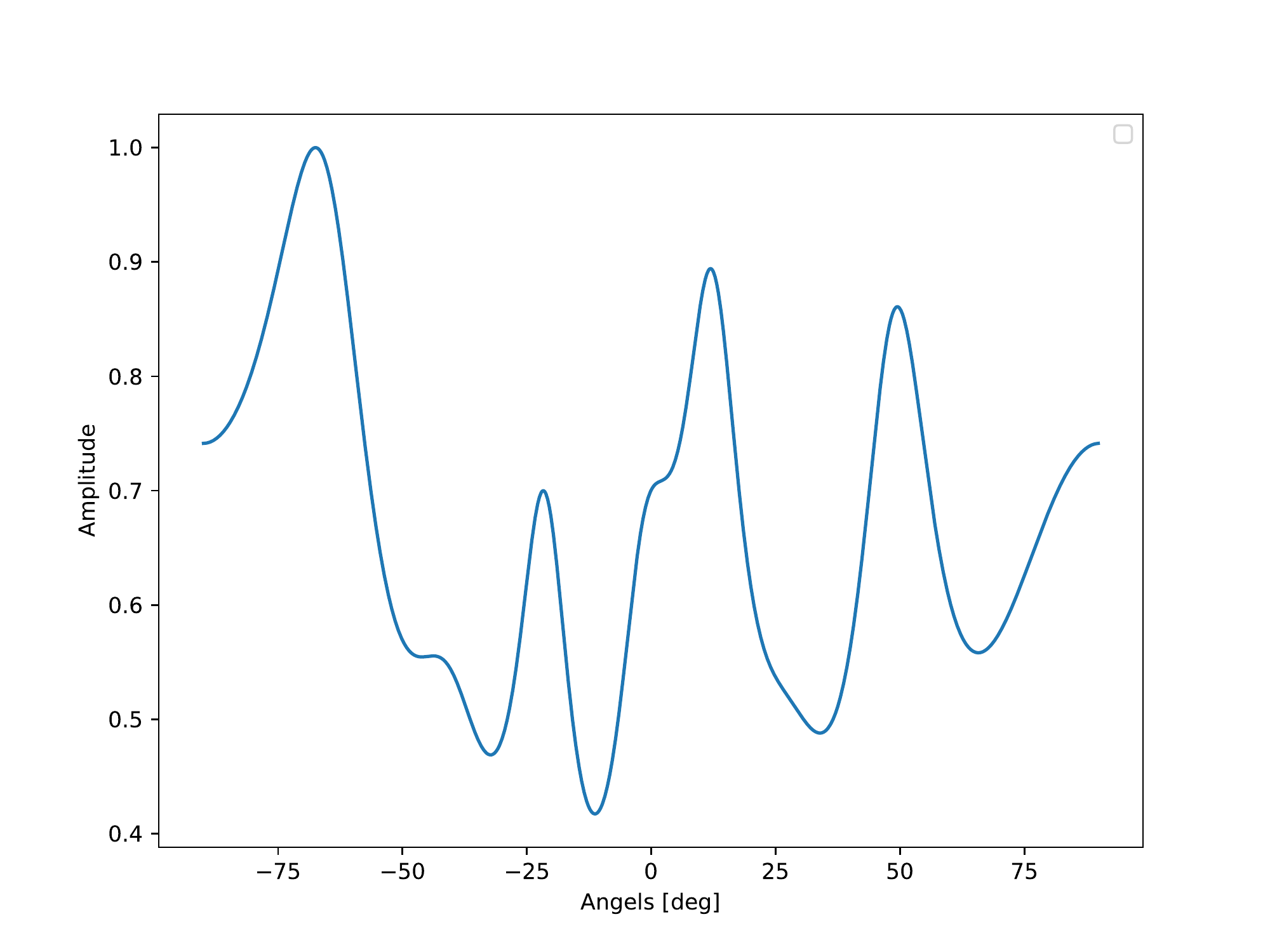}} 
	\caption{ \ac{music} spectrum, coherent sources. 
	}
	\label{fig:MUSIC_c} 	
\end{subfigure}
$\quad$
\begin{subfigure}{0.45\textwidth}
	\centering
	{\includegraphics[width=\columnwidth]{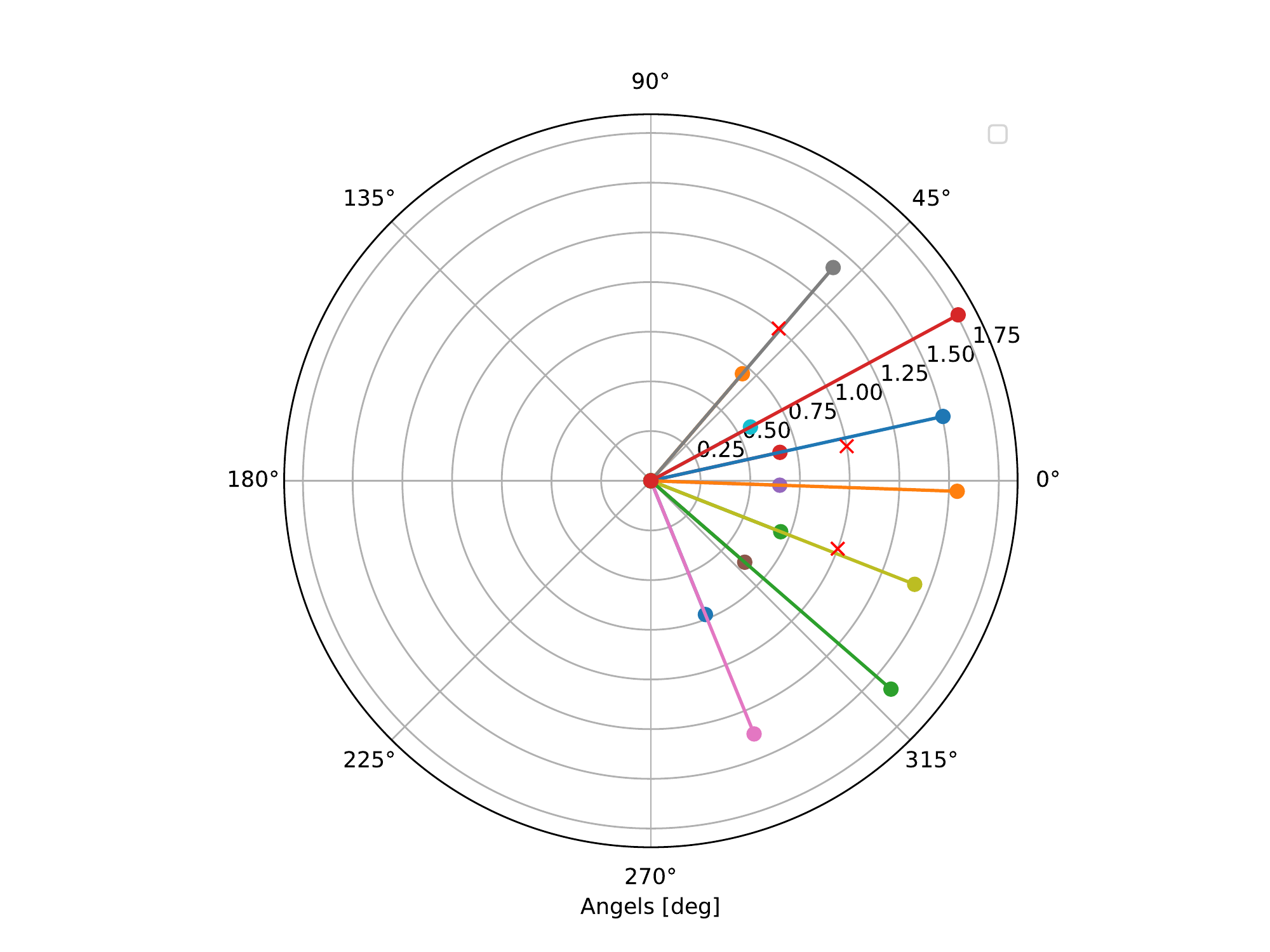}} 
	\caption{RootMUSIC spectrum, coherent sources. 
	}
	\label{fig:RootMUSIC_c} 
\end{subfigure}
	\caption{The spectrum obtained by MUSIC and RootMUSIC for recovering $d=3$ non-coherent (upper figures) and coherent (lower figures) sources located at angles $-22^\circ, 12^\circ, 50^\circ$.}
	\label{fig:Subspace}
\end{figure}

To showcase the operation of subspace methods and their dependence on the aforementioned assumptions, we evaluate both \ac{music} and RootMUSIC for recovering $d=3$ sources from $T=100$ snapshots taken by an array with $M=8$ half-wavelength spaced elements. When the sources are non-coherent (generated in an i.i.d. fashion, i.e., $\myMat{Y}=\myMat{I}$), the \ac{music} spectrum (Fig.~\ref{fig:MUSIC_nc}) exhibits clear peaks in the angles corresponding to the \acp{doa}, while RootMUSIC spectrum has roots lying on the unit circle on these angles (Fig.~\ref{fig:RootMUSIC_nc}). However, when the sources are coherent (which we simulate using the same waveforms, such that $\myMat{Y}$ is singular),  the resulting spectrum and root maps in Figs.~\ref{fig:MUSIC_c}-\ref{fig:RootMUSIC_c} no longer represent the true \acp{doa}.

\section{Summary}
\begin{tcolorbox}[width=\textwidth,colback={yellow}] 
\begin{itemize}
	\item The traditional model-based approach to design inference rules  imposes statistical or structural models on the underlying signals. 
	The  model is often a surrogate approximation adopted for simplicity and tractability. 
	\item 	Combining the loss measure and the  model yields a risk function, which can be used to formulate an objective which guides the inference procedure. 
	\item The objective function is typically parameterized, where the parameters often arise from the imposed model. Different setting of the objective parameters affect the optimization formulation,  representing different tasks.
	\item Once an objective is determined, it can lead to different inference mappings based on different approaches:
	\begin{itemize}
		\item In some cases, the optimization problem can be explicitly solved, and the decision rule is set as the corresponding solution.
		\item Often in practice, the optimization is tackled using iterative optimizers, and thus the inference rule is an iterative procedure, introducing additional parameters of the solver, referred to as hyperparameters.
		\item Alternatively, one can utilize algorithms that are not necessarily derived from the formulated problem, but rather based on exploiting structures in the  models using approximation and heuristic algorithms. 
	\end{itemize}
	\item Model-based approaches are highly sensitive to the validity of the  model  and the accuracy of its parameters. Even when the  model is fully known, the hyperparameters of the solver and objective parameters that are often tuned by hand can have a notable impact on  performance and runtime. 
\end{itemize}

\end{tcolorbox}

\chapter{Deep Learning}
\label{ch:Deep}
So far, we have discussed model-based decision making and optimization. These approaches rely on a formulation of the system objective which typically was obtained based on knowledge and approximations. 
While in many applications coming up with accurate and tractable models is difficult, we are often given access to data describing the setup. 
In this chapter, we start discussing \ac{ml}, and particularly deep learning, where decision making is carried out using \acp{dnn} whose operation is learned almost completely from data.

\section{Empirical Risk}
\label{sec:DeepObj}
 \ac{ml} systems learn their mapping from data. In a supervised setting, data is comprised of a training set consisting of $\Ntraining$ pairs of inputs and their corresponding labels, denoted $\Data = \{\Input^i, \Label^i\}_{i=1}^{\Ntraining}$. This data is referred to as the {\em training set}.
Since no mathematical model relating the input and the desired decision is imposed, the objective used for setting the decision rule $f(\cdot)$ as in \eqref{eqn:DecisionRule} is the {\em empirical risk}, given by 
\begin{equation}
\label{eqn:EmpRisk}
\mySet{L}_{\Data}(f) \triangleq \frac{1}{\Ntraining}\sum_{i=1}^{\Ntraining} l(f,\Input^i, \Label^i).
\end{equation}
While we focus our description in the sequel on supervised settings, \ac{ml} systems can also learn in an unsupervised manner. In such cases, the data set $\Data$ is comprised only of a set of examples $\{\Input^i\}_{i=1}^{\Ntraining} \subset \InputSpace$, and the loss measure $l$ is defined over $\mathcal{F}\times\InputSpace$, instead of over $\mathcal{F}\times \InputSpace \times \LabelSpace$.  
Since there is no label to predict, unsupervised \ac{ml} algorithms are often used to discover interesting patterns present in the given data.  Common tasks in this setting include clustering, anomaly detection, generative modeling, and compression.

\begin{figure}
	\centering
	\includegraphics[width=\columnwidth]{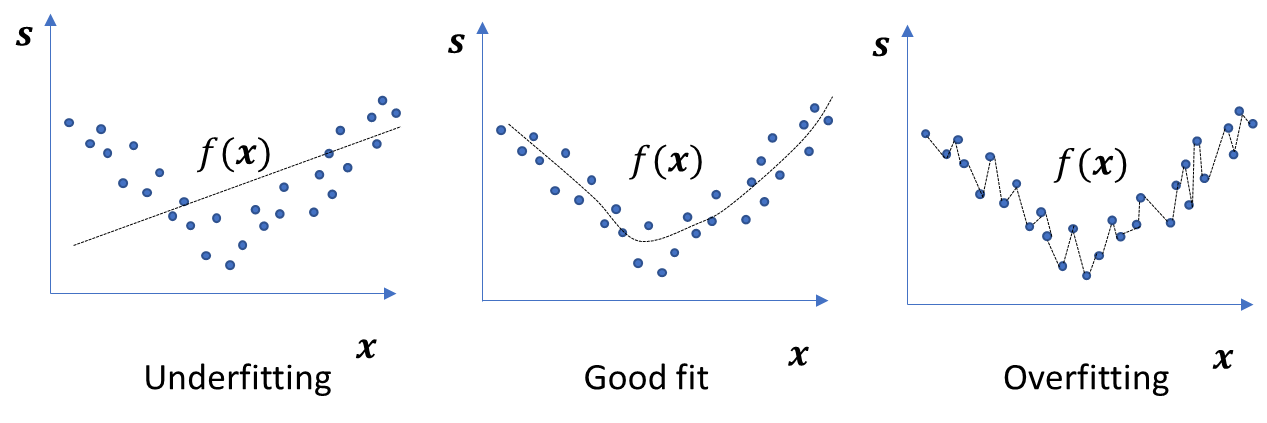}
	\caption{Illustration of overfitting versus underfitting for different structures imposed on the domain of feasible mappings.}
	\label{fig:OverfitvsUnderfit1}
\end{figure}

    The empirical risk in \eqref{eqn:EmpRisk} does not require any assumptions to be imposed on the relationship between the context $\Input$ and the desired decision $\Label$, and it thus 
allows to judge decisions solely based on their outcome and their ability to match the available data. 
As opposed to the model-based case, where decision mappings can sometimes be derived by directly solving the optimization problem arising from the risk formulation without initially imposing structure on the system, setting a decision rule based on \eqref{eqn:EmpRisk} necessitates restricting the domain of feasible mappings, also known as {\em inductive bias}. This stems from the fact that  one can usually form a decision rule which minimizes the empirical loss of \eqref{eqn:EmpRisk} by memorizing the data, i.e., overfit  \cite[Ch. 2]{shalev2014understanding}. The selection of the inductive bias is crucial to the generalization capabilities of the \ac{ml} model. As illustrated in Fig.~\ref{fig:OverfitvsUnderfit1}, allowing the system to implement arbitrary mappings can lead to overfitting, while restricting the feasible mappings to ones which may not necessarily suit the task is likely to result in underfitting.

\section{Neural Networks}
\label{sec:DeepNN}
  A leading strategy in \ac{ml}, upon which deep learning is based, is to assume a highly-expressive generic parametric model on the decision mapping, while incorporating optimization mechanisms and regularizing the empirical risk to avoid overfitting. In such cases, the decision box $f$ is dictated by a set of parameters denoted $\myVec{\theta}$, and thus the system mapping is written as $f_{\myVec{\theta}}$. In deep learning, $f_{\myVec{\theta}}$ is a \ac{dnn}, with $\myVec{\theta}$ being the network parameters.  Such highly-parametrized abstract systems can effectively approximate any Borel measurable mapping, as follows from the universal approximation theorem \cite[Ch. 6.4.1]{goodfellow2016deep}.

We next recall the basic formulation of \acp{dnn}, as well as some common architectures. 
\subsection{Basics of Neural Networks}

\begin{figure}
	\centering
	\includegraphics[width=0.5\linewidth]{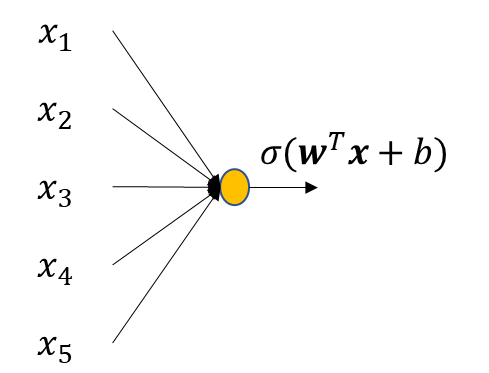}
	\caption{Neuron illustration, where the lines represent the weights $\myVec{w}$ and the  additive bias term $b$ while the node represents  the non-linear activation $\sigma(\cdot)$.}
	\label{fig:neuron}
\end{figure}

\paragraph{Artificial Neurons} 
The most basic building block of a neural network is the  {\em artificial neuron} (or merely {\em neuron}). It is a mapping $h_{\myVec{\theta}}:\mySet{R}^N \mapsto \mySet{R}$ which takes the form
\begin{equation}
\label{eqn:perceptron}
h_{\myVec{\theta}}(\Input) = 
\sigma(\myVec{w}^T \Input + b ), \qquad \myVec{\theta} = \{\myVec{w}, b\}.
\end{equation}
The neuron in \eqref{eqn:perceptron}, also illustrated in Fig.~\ref{fig:neuron}, is comprised of a parametric affine mapping (dictated by $\myVec{\theta}$) followed by some non-linear element-wise function $\sigma:\mySet{R} \mapsto \mySet{R}$ referred to as an {\em activation}. 

Stacking multiple neurons in parallel yields a {\em layer}. A layer with $M$ neurons can be written as $h_{\myVec{\theta}}:\mySet{R}^N \mapsto \mySet{R}^M$ and its mapping is given by
\begin{equation}
\label{eqn:layer}
h_{\myVec{\theta}}(\Input) = \sigma(\myMat{W}\Input+ \myVec{b}), \qquad \myVec{\theta} = \{\myMat{W}, \myVec{b}\},
\end{equation}
where the activation $\sigma(\cdot)$ is applied element-wise.

\paragraph{Activations} 
Activation functions are often fixed, i.e., their mapping is not parametric. 
Some notable examples of widely-used activation functions include:
\begin{example}[ReLU]
	The \ac{relu} is an extremely common  activation function, given by:
	\begin{equation}
	\sigma(x) = \max\{x,0\}.
	\end{equation}
\end{example}
\begin{example}[Leaky ReLU]
	A variation of the \ac{relu} activation includes an additional parameter $\alpha < 1$, typically set to $10^{-2}$, and is given by:
	\begin{equation}
	\sigma(x) = \max\{x,\alpha x\}.
	\end{equation}
\end{example}
\begin{example}[Sigmoid]
	Another common activation  is the sigmoid, which is given by:
	\begin{equation}
	\label{eqn:sigmoid}
	\sigma(x) = (1+\exp(-x))^{-1}.
	\end{equation}
\end{example}  
The importance of using activations stems from their ability to allow \acp{dnn} to realize non-affine mappings. 

\begin{figure}
	\centering
	\includegraphics[width=0.5\linewidth]{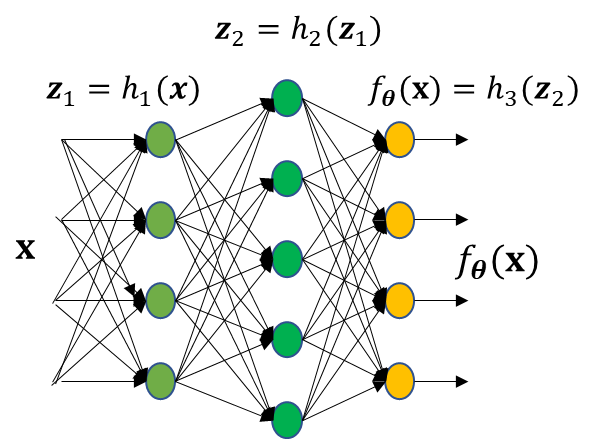}
	\caption{Mutli-layered perceptron illustration with $K=3$ layers. The lines represent the affine transformation (the  weights and the additive bias terms in each layer as in \eqref{eqn:DNNParams1}); the green nodes represent the non-linear activations $\sigma(\cdot)$, while the yellow nodes are the output layer.}
	\label{fig:mlop}
\end{figure}

\paragraph{Multi-Layered Perceptron}
While the layer mapping in \eqref{eqn:layer} may be limited in its ability to capture complex mappings, one can stack multiple layers to obtain a more flexible family of parameterized mappings. Such compositions are referred to as {\em multi-layered perceptrons}. Specifically, a \ac{dnn} $\dnnFunc$ consisting of $K$ layers $\{h_1, \ldots, h_K\}$ maps the input $\Input$ to the output $\hat{\Label} = \dnnFunc (\Input) = h_K \circ \cdots \circ h_1 (\Input)$, where $\circ$ denotes function composition. An illustration of a \ac{dnn} with $k=3$ layers (two hidden layers and one output layer) is illustrated in Fig.~\ref{fig:mlop}. 
Since each layer $h_k$ is itself a parametric function, the parameters of the entire network $\dnnFunc$ are the union of all of its layers' parameters, and thus $\dnnFunc$ denotes a \ac{dnn} with parameters $\myVec{\theta}$. In particular, by letting $\myMat{W}_k, \myVec{b}_k$ denote the configurable parameters of the $k$th layer $h_k(\cdot)$, the trainable parameters of the \ac{dnn} are written as
\begin{equation}
\label{eqn:DNNParams1}
\myVec{\theta} = \{\myMat{W}_k, \myVec{b}_k\}_{k=1}^{K}. 
\end{equation}
The \textit{architecture} of a \ac{dnn} refers to the specification of its layers $\{h_k\}_{k=1}^K$.

\paragraph{Output Layers} 
The choice of the output layer $h_K(\cdot)$ is tightly coupled with the task of the network, and more specifically, with the loss function $\mySet{L}$. In particular, the output layer dictates the possible outputs the parameteric mapping can realize. The following are commonly used output layers based on the system task:
\begin{itemize}
	\item {\bf Regression} tasks involve the estimation of a continuous-amplitude vector, e.g., $\LabelSpace = \mySet{R}^d$. In this case, the output must be allowed to take any value in $\mySet{R}^d$, and thus a common output layer is a {\em linear unit} of width $d$, i.e., 
	\begin{equation}
	h_k(\myVec{z}) = \myMat{W}_K\myVec{z} + \myVec{b}_K,
	\end{equation}
	where the number of rows of $\myMat{W}_K$ and dimension of $\myVec{b}_K$ is set to $d$.
	\item {\bf Detection} is a binary form of classification, i.e., $|\LabelSpace| = 2$. In classification tasks, one is typically interested in {\em soft outputs}, e.g., $\Pr(\Label| \Input)$, in which case the output is a probability vector over $\LabelSpace$. As $\LabelSpace$ is binary, a single output taking values in $[0,1]$, representing $\Pr(\Label=\Label_1| \Input)$, is sufficient. Thus, the typical output layer is a {\em sigmoid unit} as given in \eqref{eqn:sigmoid}, and the output layer is given by 
	\begin{equation}
	h_k(\myVec{z}) = \sigma(\myVec{w}_k^T\myVec{z} + b_k).
	\end{equation}
	\item {\bf Classification} in general allows any finite number of different labels, i.e. $|\LabelSpace| = d$, where $d$ is a positive integer not smaller than two. Here, to guarantee that the output is a probability vector over $\LabelSpace$,  classifiers typically employ the {\em softmax function} (e.g.\ on top of the output layer), given by:
	\begin{equation*}
	\textrm{Softmax}(\myVec{z}) = \left[\frac{\exp(z_1)}{\sum_{i=1}^{d} \exp(z_i)}, \ldots, \frac{\exp(z_d)}{\sum_{i=1}^{d} \exp(z_i)}\right].
	\end{equation*}
	The resulting output layer is given by 
	\begin{equation}
	h_K(\myVec{z}) = \textrm{Softmax}(\myMat{W}_K\myVec{z} + \myVec{b}_K),
	\end{equation}
	where the number of rows of $\myMat{W}_K$ and dimension of $\myVec{b}_K$ is set to $d$. 	Due to the exponentiation followed by normalization, the output of the softmax function is guaranteed to be a valid probability vector. 
\end{itemize}

\subsection{Common Architectures}
 Unlike model-based algorithms, which are specifically tailored to a given scenario, deep learning is model-agnostic. The unique characteristics of the specific scenario are encapsulated in 
the weights that are learned from data. The parametrized inference rule, e.g., the \ac{dnn} mapping, is generic and can be applied to a broad range of different problems. In particular, the multi-layered perception can combine any of its input samples in a parametric manner, and thus makes no assumption on the existence of some underlying structure in the data. 

While standard \ac{dnn} structures are model-agnostic and are commonly treated as black boxes, one can still incorporate some level of domain knowledge in the selection of the specific network architecture. We next review two common families of structured architectures: \acp{cnn} and \acp{rnn}.

\paragraph{Convolutional Neural Networks}
Convolutional layers are a special case of \eqref{eqn:layer}. However, they are specifically tailored to preserve the tensor representation of certain data types, e.g., images, while utilizing a reduced number of trainable parameters. The latter is achieved by connecting each neuron to only a subset of the input variables (e.g., pixels) and by reusing weights between different neurons of the same layer. The trainable parameters of convolutional layers are essentially a set of filters (typically two-dimensional), which slide over the spatial dimensions of the layer input.

Convolution operators are commonly studied in the signal processing literature in the context of time sequences, where one-dimensional filters are employed. In the case of causal finite response filtering, a one-dimensional convolution with kernel $K[i]$ of size $F$ applied to a time sequence $Z_1[n]$, yields a signal
\begin{equation}
\label{eqn:1DConv}
Z_2[n]=\sum_{i=0}^{F-1}Z_1[n-i]K[i].
\end{equation}
The operation in \eqref{eqn:1DConv} specializes the linear mapping of the perceptron by restricting the weights matrix to be Toeplitz. 

While \eqref{eqn:1DConv} corresponds to the traditional signal processing formulation of convolutions, and constitutes the basis of one-dimensional convolutional layers used in deep learning, the more common form of convolutional layers generalizes \eqref{eqn:1DConv} and considers tensor data, e.g., images, without reshaping them into vectors. 
\color{black} 
To formulate the operation of such convolutional layers, consider a layer with an input tensor $Z_1$ of dimensions $H_1 \times W_1 \times D_1$ (height $\times$ width $\times$ depth). The convolution layer is comprised of the following aspects: 
\begin{itemize}
	\item {\bf Convolution kernel} - the linear operation is carried out by a sliding kernel $K$ of dimensions $F \times F \times D_1 \times D_2$ (spatial extent *squared* $\times$ input depth $\times$ output depth).
	\item {\bf Bias} - as the transformation implemented is {\em affine} rather than {\em linear}, and thus a bias term is added. These biases are typically shared among all kernels mapped to a given output channel, and thus the trainable parameter $\myVec{b}$ is usually a $D_2 \times 1$ vector. 
	\item {\bf Zero padding} - since the input is of a finite spatial dimension (i.e., $H_1$ and $W_1$), convolving an $H_1\times W_1$ image with an $F \times F$ kernel yields an $(H_1 - F+1) \times (W_1 - F + 1)$ image. The natural approach to extend the dimensions is thus to zero-pad, i.e., add $P$ zeros around the image, resulting in the output being an $(H_1 - F+ 2P+ 1) \times (W_1 - F + 2P+1)$ image. 
	\item {\bf Stride} - in order to reduce the dimensionality of the output image, one can determine how the filter slides along the image, using the stride hyperparameter $S$. For $S=1$, the filter moves along the image pixel-by-pixel. Increasing $S$ implies that the filter skips $S-1$ pixels as it slides. Consequently, the output image is of dimensions $(\frac{H_1 - F+ 2P}{S}+ 1) \times (\frac{W_1 - F + 2P}{S}+1)$. 
\end{itemize}
To summarize, the output tensor $Z_2$ is of size $H_2 (=\frac{H_1 - F+ 2P}{S}+ 1) \times W_2 (= \frac{W_1 - F + 2P}{S}+1) \times D_2$, and its entries are computed as
\begin{align}
Z_2[n,m,l] = \sum_{i=0}^{F-1} \sum_{j=0}^{F-1}\sum_{d=0}^{D_1-1}&Z_1[n\cdot S+i - P,m\cdot S+j - P,d] \notag \\
&\cdot K[i,j,d,l] + b[l],
\label{eqn:conv4}
\end{align}
where $Z_1[n,m,d]$ is set to zero for $n \notin [0, H_1-1]$ and/or $n \notin [0, W_1-1]$.

\paragraph{Recurrent Neural Networks}
While \acp{cnn} can efficiently process spatial information in structured data such as images by combining neighboring features (e.g., pixels), \acp{rnn} are designed to  handle sequential information, i.e., time sequences. Time sequences comprise of a sequential order of samples. In this case, our data is a sequence of $T$ samples, denoted $\{\Input_t\}_{t=1}^T$. The {\em task} is to map the inputs into a label sequence  $\{\dnnLabel_t\}_{t=1}^T$. The sequential structure of the data implies that the inference of $\dnnLabel_t$ should not be made based on $\Input_t$ solely, but should also depend on  past inputs $\{\Input_j\}_{j=1}^{t-1}$.

\begin{figure}
	\centering
	\includegraphics[width=\linewidth]{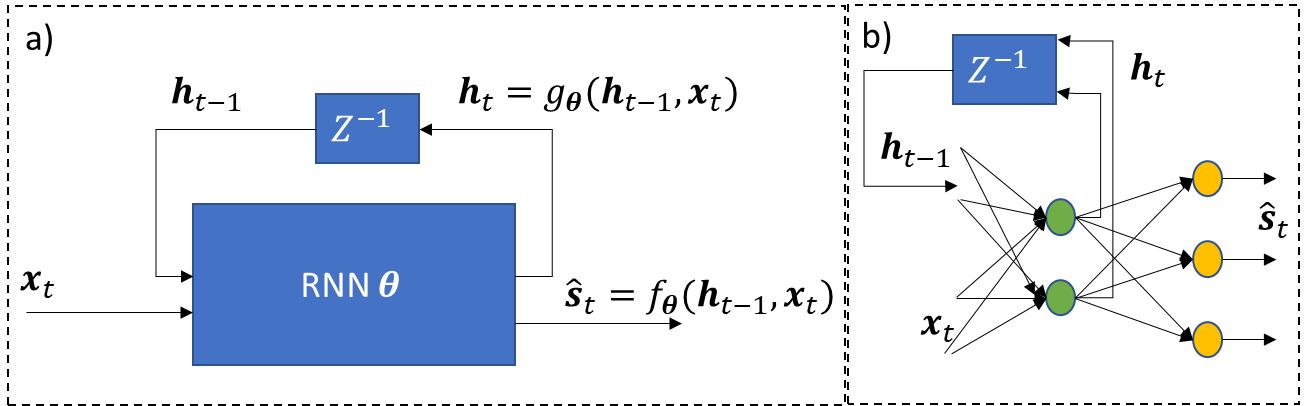}
	\caption{\acl{rnn} illustration: $(a)$ generic architecture; $(b)$ as a neural layer. The symbol $Z^{-1}$ represents a unit time delay. }
	\label{fig:RNNIllust1}
\end{figure}

Recurrent parametric models maintain an internal state vector, denoted $\myVec{h}_t$, representing the memory of the system. Now, the parametric mapping is given by:
\begin{equation}
\hat{\dnnLabel}_t= \dnnFunc(\Input_t, \myVec{h}_{t-1}),
\end{equation}
where the internal state also evolves via a learned parametric mapping 
\begin{equation}
\myVec{h}_{t} = g_{\dnnParam}(\Input_t, \myVec{h}_{t-1}).
\end{equation}
An illustration of such a generic \ac{rnn} is depicted in Fig.~\ref{fig:RNNIllust1}. The vanilla implementation of an \ac{rnn} is the single hidden layer model illustrated in Fig~\ref{fig:RNNIllust1}b, in which $\Input_t$ and the latent $\myVec{h}_{t-1}$ are first mapped into an updated hidden variable $\myVec{h}_t$ using a fully-connected layer, after which $\myVec{h}_t$ is used to generate the instantaneous output  $\hat{\dnnLabel}_t$ using another fully-connected output layer. In this case, we can write
\begin{equation}
\myVec{h}_{t} = g_{\dnnParam_1}(\Input_t, \myVec{h}_{t-1}); \qquad 
\hat{\dnnLabel}_t = f_{\dnnParam_2}(\myVec{h}_{t}); \quad
\dnnParam = [\dnnParam_1, \dnnParam_2].
\label{eqn:RNNParams}
\end{equation}

\section{Training}
\label{sec:DeepTrain}
Generally speaking, \ac{ml} methods and deep learning specifically operate without knowledge of a model relating the context and the desired decision. They rely on data in order to tune the parameters of the mapping, i.e., the network parameters $\myVec{\theta}$, to properly match the data. This is where optimization traditionally comes into play in the context of \ac{ml}. While in the previous chapter we discussed how optimization techniques are used for decision making, in deep learning they are typically applied to tune the parameters of the mapping. This procedure, where a data set $\mySet{D}$ is used to determine the parameters  $\myVec{\theta}$, is referred to as {\em training}, and is the conventional role of optimization in deep learning.

\subsection{Training Formulation}
As discussed in Section~\ref{sec:DeepObj}, \ac{ml} systems learn their mapping from data.   For a loss function $l(\cdot)$, a parametric model $\dnnFunc(\cdot)$, and labeled data set $\Data$, we wish to find $\dnnParam$ which minimizes the empirical risk of \eqref{eqn:EmpRisk}, which we henceforth write as $\mySet{L}_{\Data}(\dnnParam) \triangleq \mySet{L}_{\Data}(f_\dnnParam)$. 
Often, a regularizing term is added to reduce overfitting by imposing constraints on the values which those weights can take. This is achieved by  adding to the empirical risk $\mySet{L}_{\mySet{D}}(\dnnParam)$ (that depends on the training set $\mySet{D}$) a regularization term $\Phi(\dnnParam)$ which depends solely on the parameter vector $\dnnParam$. The resulting regularized loss is thus
\begin{equation}
\mySet{L}_{\Phi}(\dnnParam) = \mySet{L}_{\mySet{D}}(\dnnParam) + \lambda \Phi(\dnnParam),
\end{equation}
where $\lambda > 0$ is a hyperparameter representing the regularization strength, thus balancing the contribution of the summed terms. Consequently, training can be viewed as an optimization problem
\begin{equation}
\label{eqn:TrainOpt}
\dnnParam^* = \mathop{\arg \min}\limits_{\dnnParam} \mySet{L}_{\Phi}(\dnnParam).
\end{equation}

The optimization problem in \eqref{eqn:TrainOpt} is generally non-convex, indicating that actually finding $  \dnnParam^*$  is often infeasible. For this reason, we are usually just interested in finding a "good" set of the weights, and not the optimal one, i.e., finding a "good" local minima of $ \mySet{L}_{\Phi}$. Since neural networks are differentiable with respect to their weights and inputs (being comprised of a sequence of affine transformations and differentiable activations), a reasonable approach (and a widely adopted one) to search for such local minima is via first-order methods. Such methods involve the computation of the gradient \begin{align}
\nabla_{\dnnParam} \mySet{L}_{\Phi}(\dnnParam) 
&=   \nabla_{\dnnParam} \mySet{L}_{\Data}(\dnnParam) + \lambda  \nabla_{\dnnParam} \Phi(\dnnParam) \notag \\
&= \frac{1}{|\Data|} \sum_{\{\Input_t, \Label_t\} \in \Data} \nabla_{\dnnParam} l(\dnnFunc,\Input_t, \Label_t) + \lambda  \nabla_{\dnnParam} \Phi(\dnnParam).
\label{eqn:Gradient}
\end{align}

The main challenges associated with computing the gradients stem from the first term in \eqref{eqn:Gradient}, and particularly:
\begin{enumerate}
	\item The data set $\Data$ is typically very large, indicating that the summation in \eqref{eqn:Gradient} involves a massive amount of computations of $\nabla_{\dnnParam} l(\cdot)$ at all different training samples.
	\item Neural networks are given by a complex function of their parameters $\dnnParam$, indicating that computing each $\nabla_{\dnnParam} l(\cdot)$ may be challenging.
\end{enumerate}
Deep learning employs two key mechanism to overcome these challenges: the difficulty in computing the gradients is mitigated due to the sequential operation of neural networks via backporpagation, detailed in the following. The challenges associated with the data size are handled by replacing the gradients with a stochastic estimate, detailed in Subsection~\ref{subsec:StochasticGrad}.

\subsection{Gradient Computation}
\label{subsec:Gradient}
One of the main challenges in optimizing complex highly-parameterized models using gradient-based methods stems from the difficulty in computing the empirical risk gradient with respect to each parameter, i.e., $ \nabla_{\dnnParam} l(\dnnFunc,\Input^i, \Label^i)$ in \eqref{eqn:StochGrad}. In principle, the parameters comprising the entries of $\dnnParam$ may be highly coupled, making the computation of the gradient a difficult task.
Nonetheless, neural networks are not arbitrary complex models, but have a sequential structure comprised of a concatenation of layers, where each trainable parameter typically belongs to a single layer. This sequential structure facilitates the computation of the gradients using the  {\em backpropagation} process.

\paragraph{Backpropagation} 
The backpropagation method proposed   in  \cite{rumelhart1985learning} is based on the calculus chain rule. Suppose that one is given two multivariate functions such that $\myVec{y}=g(\myVec{x})$ and $l(\myVec{x})=f(\myVec{y})=f(g(\myVec{x}))$, where $f:\mySet{R}^n\mapsto \mySet{R}$ and $g:\mySet{R}^m\mapsto \mySet{R}^n$. By the chain rule, it holds that 
\begin{equation}
\frac{\partial l}{\partial x_i} = \sum_{j=1}^n \frac{\partial l}{\partial y_j} \frac{\partial y_j}{\partial x_i}.
\end{equation}

The formulation of the gradient computation via the chain rule is exploited to compute the gradients of the empirical risk of a multi-layered neural network with respect to its weights in a recursive manner. To see this, consider a neural network with $K$ layers, given by 
\begin{equation}
\label{eqn:nnKlayers}
\dnnFunc (\Input) = h_K \circ \cdots \circ h_1 (\Input),
\end{equation}
where each layer $h_k(\cdot)$ is comprised of a (non-parameterized) activation function $\sigma_k(\cdot)$ applied to an affine transformation with parameters $\dnnParam_k = (\myMat{W}_k, \myVec{b}_k)$. 
Define $\myVec{a}_k$ as the output to the $k$th layer, i.e., $\myVec{a}_{k} = h_k(\myVec{a}_{k-1})$, and $\myVec{z}_k$ as the affine transformation such that
\begin{equation*}
\myVec{z}_k = \myMat{W}_k\myVec{a}_{k-1} + \myVec{b}_k, \qquad \myVec{a}_{k} = \sigma_k(\myVec{z}_k).
\end{equation*}
Now, the empirical risk $l =  l(\dnnFunc,\Input, \Label)$ is a function of $\dnnFunc(\Input) = h_K \circ \cdots \circ h_{k+1}(\sigma_k(\myVec{z}_k))$. Consequently, we use the matrix version of the chain rule to obtain 
\begin{subequations}
	\label{eqn:Backprop1}
	\begin{align}
	\nabla_{\myVec{a}_{k-1}} l &=\myMat{W}_k^T \big(\nabla_{\myVec{z}_{k}} l\big), \label{eqn:Backprop1a}\\
	\nabla_{\myMat{W}_k}l &= \big(\nabla_{\myVec{z}_{k}}l \big)\myVec{a}_{k-1}^T,\label{eqn:Backprop1b}  \\ 
	\nabla_{\myVec{b}_k} l  &= \nabla_{\myVec{z}_{k}} l. \label{eqn:Backprop1c}
	\end{align}
	Furthermore, since $\myVec{a}_{k} = \sigma_k(\myVec{z}_k)$ it holds that
	\begin{equation}
	\nabla_{\myVec{z}_{k}} \mySet{L} = \nabla_{\myVec{a}_{k}} l \odot \sigma_k'(\myVec{z}_k), \label{eqn:Backprop1d}
	\end{equation}
\end{subequations}
where $\odot$ denotes the element-wise product, and $\sigma_k'(\cdot)$ is the element-wise derivative of the activation function $\sigma_k(\cdot)$. Equation \eqref{eqn:Backprop1} implies that the gradients of the empirical risk  with respect to $(\myMat{W}_k, \myVec{b}_k)$ can be obtained by first evaluating the outputs of each layer, i.e., the vectors $\{\myVec{z}_k, \myVec{a}_k\}$, also known as the {\em forward path}. Then, the gradients of the loss with respect to each layer's output can be computed recursively from the corresponding gradients of its subsequent layers via  \eqref{eqn:Backprop1a} and \eqref{eqn:Backprop1d}, while the desired weights gradients are obtained via \eqref{eqn:Backprop1b}-\eqref{eqn:Backprop1c}. This  computation starts with the gradient of the loss with respect to the \ac{dnn} output, i.e., $	\nabla_{\dnnFunc (\Input)} \mySet{L}$, which is dictated by the  loss function, i.e., how $\mySet{L}$ is computed for \ac{dnn} output $\dnnFunc (\Input)$. Then, this gradient is used to recursively update the gradients of the loss with respect to the parameters of the layers, going from the last layer ($h_K$) to the first one ($h_1$). 

\paragraph{Backpropagation Through Time}
Backpropagation via \eqref{eqn:Backprop1} builds upon the fact that neural networks are comprised of a sequential operation, allowing to compute the gradients of the loss with respect to the parameters of a given layer using solely the gradient of the loss with respect to its output. This operation has an underlying assumption that each layer has its own distinct weights. However, backpropagation can also be applied when weights are shared between layers. Since this is effectively the case in \acp{rnn} processing time sequences, the resulting adaptation of backporpagation is typically referred to as {\em backpropagation through time} \cite{sutskever2013training}; nonetheless, this procedure can be applied for any form of weight sharing between weights regardless of whether the network is an \ac{rnn} or if the data is a time sequence. 

To see how backpropagation through time operates, let us again consider a neural network with $K$ layers as in \eqref{eqn:nnKlayers} and use $\myVec{\alpha}_k$ to denote the features at the output of the $k$th layer. However, now there is some weight parameter $w_h$ that is shared by all layers, i.e., $w_h$ is an entry of $\dnnParam_k$ for each $k\in\{1,\ldots,K\}$. 
To compute the derivative of  the empirical risk with respect to $w_h$, we note that 
\begin{align}
\frac{\partial  l(\dnnFunc,\Input, \Label)}{\partial w_h} 
&=     \frac{\partial l(\dnnFunc,\Input, \Label) }{\partial\hat{\dnnLabel}} \frac{\partial \hat{\dnnLabel} \big(= \dnnFunc(\Input)\big)}{\partial \myVec{\alpha}_K}  \frac{\partial \myVec{\alpha}_K }{\partial w_h}.
\label{eqn:RNNBackprop}
\end{align}
Since the weight $w_h$ appears both in the $K$th layer as well as in its subsequent ones, it holds that
\begin{equation}
\frac{\partial \myVec{\alpha}_K }{\partial w_h}
=  \frac{\partial  h_K(\myVec{\alpha}_{K-1}; w_h)}{\partial w_h} + \frac{\partial  h_K(\myVec{\alpha}_{K-1}; w_h)}{\partial \myVec{\alpha}_{K-1}}  \frac{\partial \myVec{\alpha}_{K-1} }{\partial w_h}. \label{eqn:RNNBackprop2}
\end{equation}
We obtain a recursive equation relating $\frac{\partial \myVec{\alpha}_K }{\partial w_h}$ and $\frac{\partial \myVec{\alpha}_{K-1} }{\partial w_h}$. If $w_h$ would have been only a parameter of the $K$th layer, then  $\frac{\partial \myVec{\alpha}_{K-1} }{\partial w_h} = 0$, and we are left only with the first summand in \eqref{eqn:RNNBackprop}. When there is weight sharing, we can compute \eqref{eqn:RNNBackprop2} recursively $K$ times, eventually obtaining an expression for the gradients 
\begin{equation}
\label{eqn:RNNBackprop3}
\frac{\partial \myVec{\alpha}_K }{\partial w_h} = \frac{\partial  h_K(\myVec{\alpha}_{K-1}; w_h)}{\partial w_h} + 
\sum_{i=1}^{K-1}\left(\prod_{k=i+1}^{K}  \frac{\partial  h_k(\myVec{\alpha}_{k-1}; w_h)}{\partial \myVec{\alpha}_{k-1}} \right) \frac{\partial  h_i(\myVec{\alpha}_{i-1}; w_h)}{\partial w_h}.
\end{equation}
Plugging   \eqref{eqn:RNNBackprop3} into \eqref{eqn:RNNBackprop} yields the  expression for the desired gradients.

\subsection{Stochastic Gradients}
\label{subsec:StochasticGrad}
%
The method referred to as {\em (mini-batch) stochastic gradient descent} computes gradients   over a {\em random subset} of $\Data$, rather than the complete data set. At each iteration index $j$, a mini-batch comprised of $B$ samples, denoted $\Data_j$, is randomly drawn from $\Data$, and is used to compute the gradient. The gradient estimate is thus given by 
\begin{equation}
\nabla_{\dnnParam} \mySet{L}_{\Data_j}(\dnnParam) = \frac{1}{B} \sum_{\{\Input^i, \Label^i\} \in \Data_j} \nabla_{\dnnParam} l(\dnnFunc,\Input^i, \Label^i).
\label{eqn:StochGrad}
\end{equation}
When $\Data_j$ is drawn uniformly from all $B$-sized subsets of $\Data$ in an i.i.d. fashion,  \eqref{eqn:StochGrad} is a stochastic unbiased estimate of the true gradient $\nabla_{\dnnParam} \mySet{L}_{\Data}(\dnnParam)$. 

The usage of stochastic gradients for gradient descent based optimization of $\dnnParam$,  summarized as Algorithm~\ref{alg:AlgoSGD} below, requires $B$ gradient computations each time the parameters are updated, i.e., on each iteration. Since $B$ is typically much smaller then the data set size $|\Data|$, stochastic gradients  reduce the computational burden of training \acp{dnn}.

\begin{algorithm}  
	\caption{(Mini-batch) stochastic gradient descent}
	\label{alg:AlgoSGD}
	\KwData{Fix number of iterations $n$; step sizes $\{\mu_j\}$.}
	Initialize $\dnnParam_0$ randomly\;
	\For{$j=0,1,\ldots,n-1$}{
		Sample $B$ different samples uniformly from $\Data$ as $\Data_j$\;
		Estimate gradient as $\hat{\nabla}_{\dnnParam}\mySet{L}(\dnnParam_j) =  \nabla_{\dnnParam} \mySet{L}_{\Data_j}(\dnnParam)  + \lambda  \nabla_{\dnnParam} \Phi(\dnnParam_j)$\;
		Update parameters via		
		\begin{equation}
		\label{eqn:SGD2}
		\dnnParam_{j+1} \leftarrow \dnnParam_{j} - \mu_j \hat{\nabla}_{\dnnParam}\mySet{L}(\dnnParam_j).
		\end{equation} 
	}
	\KwOut{Trained parameters $\dnnParam_n$.}
\end{algorithm}

The term {\em stochastic gradient descent} is often used to refer to Algorithm~\ref{alg:AlgoSGD} with mini-batch size $B=1$, while for $B>1$ it is referred to as {\em mini-batch stochastic gradient descent}. Each time the training procedure goes over the entire data set, i.e., every $\lceil|\Data|/B\rceil$ iterations, are referred to as an {\em epoch}. 

\subsection{Update Rules}
The above techniques allow to compute an estimate of the gradient in \eqref{eqn:Gradient} with relatively feasible computational burden. 
One can now use first order methods for tuning the \ac{dnn} parameters $\dnnParam$, while replacing the gradient $\nabla_{\dnnParam} \mySet{L}_{\Phi}(\dnnParam)$ with its stochastic estimate $\hat{\nabla}_{\dnnParam} \mySet{L}(\dnnParam)$. 
The most straight-forward first-order optimizer which uses the stochastic gradients is stochastic gradient descent, detailed in Algorithm~\ref{alg:AlgoSGD}. 

There are a multitude of variants and extensions of \eqref{eqn:SGD2}, which are known to improve the learning characteristics. Here we review the leading approaches which can be divided into {\em momentum updates}, which introduce an additional additive term to \eqref{eqn:SGD2}, and {\em adaptive learning rates},  particularly the Adam optimizer \cite{kingma2014adam}, where \eqref{eqn:SGD2} is further scaled in a different manner for each weight. 

\paragraph{Momentum}
Momentum, as it name suggests, encourages the optimization path to follow its current direction. As the direction of the previous step is merely the difference $\dnnParam_j - \dnnParam_{j-1}$, momentum replaces the vanilla update rule \eqref{eqn:SGD2} with
\begin{equation}
\label{eqn:Momentum}
\dnnParam_{j+1} - \dnnParam_{j} = - \mu_j \hat{\nabla}_{\dnnParam} \mySet{L}(\dnnParam_j) + \beta_j\left(\dnnParam_j - \dnnParam_{j-1}\right). 
\end{equation}
The parameter $\beta_j$ in \eqref{eqn:Momentum} is referred to as the {\em damping factor}, and should take values in $(0,1)$. The setting of $\beta_j$ balances the contribution of the current direction (i.e., the momentum), and the current noisy stochastic gradient. In practice, the damping factor is typically set in the range $(0.9, 0.99)$, implying that momentum has a dominant impact on the optimization path.

\paragraph{Adam} 
Adaptive learning rate methods use a different learning rate for each weight. Here, the update equation \eqref{eqn:SGD2} becomes 
\begin{equation}
\dnnParam_{j+1} - \dnnParam_{j} = - \mu_j \myVec{v}_j \odot \hat{\nabla}_{\dnnParam} \mySet{L}(\dnnParam_j),
\label{eqn:Adaptive}
\end{equation}
where $\myVec{v}_j$ is the adaption vector, whose dimensions are equal to those of $\dnnParam$. 

Arguably the most widely-used adaptive update rule is the Adam method  \cite{kingma2014adam},  which scales the gradient of each element with an estimate of its root-mean-squared value accumulated over the learning iterations. Here, weights that receive high gradients will have their effective learning rate reduced, while weights that receive small or infrequent updates will have their effective learning rate increased.
This is achieved by maintaining an estimate of the (element-wise) sum-squared gradients in the vector
\begin{equation}
\tilde{\myVec{v}}_{j} = \alpha \tilde{\myVec{v}}_{j-1} + (1-\alpha) \big( \hat{\nabla}_{\dnnParam} \mySet{L}(\dnnParam_j)\big)^2,
\label{eqn:RMSProp}
\end{equation}
and a momentum term 
\begin{equation}
{\myVec{m}}_{j} = \alpha_1 {\myVec{m}}_{j-1} + (1-\alpha_1) \hat{\nabla}_{\dnnParam} \mySet{L}(\dnnParam_j).
\label{eqn:Adam}
\end{equation} 
The resulting  adaptation vector is set to
\begin{equation}
\label{eqn:Adaptation2}
[\myVec{v}_j]_i = \frac{[{\myVec{m}}_{j}]_i}{\sqrt{ [\tilde{\myVec{v}}_{j}]_i} + \epsilon},
\end{equation}
with $\alpha, \alpha_1, \epsilon$ being hyperparameters.  The full Adam algorithm proposed in \cite{kingma2014adam}, also includes a bias correction term not detailed here.

\newpage
\section{Summary}
\label{sec:DeepSum}

\begin{tcolorbox}[width=\textwidth,colback={yellow}] 
	\begin{itemize}
		\item \ac{ml} methods set inference rules without relying on mathematical modeling by using data to formulate an empirical estimate of the risk function as the design objective. This necessitates the need to impose a parametric model on the mapping, which in deep learning is that of \acp{dnn}. 
		\item The generic \ac{dnn} form is the multi-layered perceptron, which is a composition of parametric affine functions with intermediate non-linear activations, while the task affects the output layer. Such architectures are model-agnostic, and the specificities of the scenario which dictate the mapping are encapsulated in the parameters that are learned from data.
		\item \ac{dnn} architectures such as \acp{cnn} and \acp{rnn} are designed for structured data. Yet, like multi-layered perceptrons,  they are still invariant of the underlying statistical model, and are often viewed as black boxes, where one can confidently assign operational meaning only to their input and output. 
		\item Training, i.e., the tuning of the parameters of \acp{dnn} to match a given data set based on a specified loss measure, is typically based on stochastic variations of first-order optimizers combined with backpropagation to compute the gradients.
	\end{itemize}
	
\end{tcolorbox} 

\chapter{Model-Based Deep Learning}
\label{ch:MBDL}
The previous chapters focused on model-based optimization and deep learning, which are often viewed as fundamentally different approaches for setting inference rules. Nonetheless, both strategies typically use parametric mappings, i.e., the weights  $\myVec{\theta}$ of \acp{dnn} and the parameters $[\ObjParam,\HypParam]$ of model-based methods, whose setting is determined based on data and on knowledge of principled mathematical models. The core difference thus lies in the specificity and the parameterization of the inference rule type:  Model-based methods are {\em knowledge-centric}, e.g., rely on a characterization of an underlying  model. This knowledge allows model-based methods to employ inference mappings that are highly task-specific, and usually involve a limited amount of parameters that one can often set manually. Deep learning is {\em data-centric}, operating without specifying a statistical model on the data, and thus uses model-agnostic task-generic mappings that tend to be highly parametrized. 

\begin{sidewaysfigure}
	\centering
	\includegraphics[width=\linewidth]{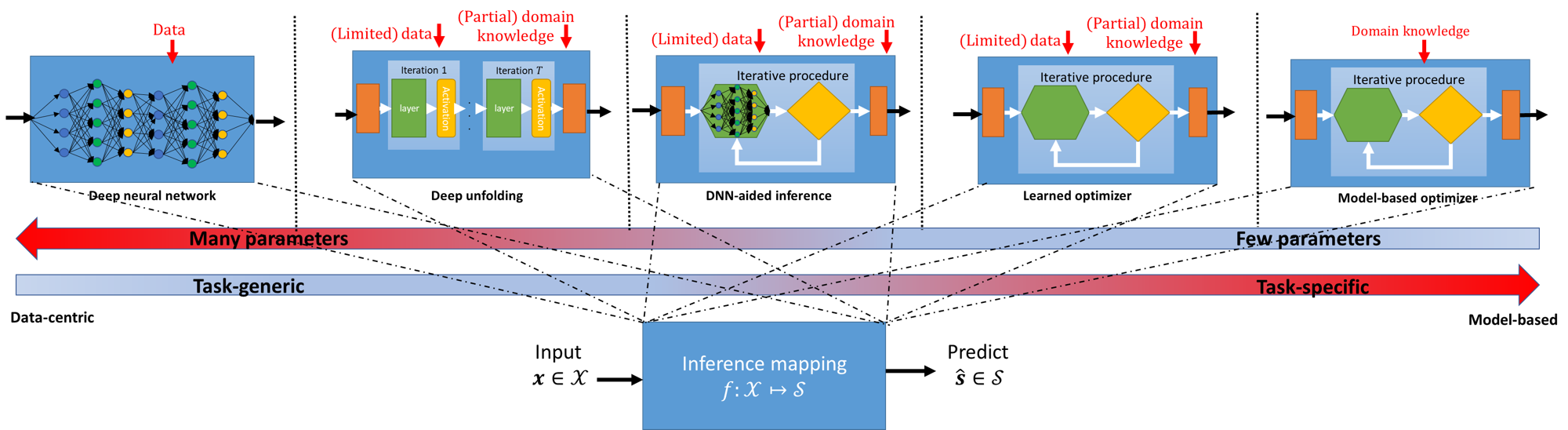}
	\caption{Continuous spectrum of specificity and parameterization with model-based methods and deep learning constituting the extreme edges of the spectrum.}
	\label{fig:Spectrum1}
\end{sidewaysfigure}

The identification of model-based methods and deep learning as two ends of a spectrum of specificity and parameterization indicates the presence of a continuum, as illustrated in Fig.~\ref{fig:Spectrum1}. In fact, many techniques lie in the middle ground, designing decision rules with different levels of specificity and parameterization by combining some balance of deep learning  with model-based optimization \cite{chen2021learning, monga2021algorithm, shlezinger2020model, shlezinger2022model}. This chapter is dedicated to exploring methodologies residing in this middle ground.

The learning of inference rules from data relies on three fundamental pillars: an {\em architecture}, dictating the family of mappings one can set; {\em data} which (in a supervised setting) comprises of pairs of inputs and their desired outputs; and a {\em learning algorithm} which uses the data to tune the architecture.
The core of model-based deep learning as the family of middle ground methodologies between model-based inference and deep learning lies mostly in its revisiting of  what type of inference rule is learned, namely, the {\em architecture}, and also how data is leveraged, i.e., the {\em learning algorithm}.

Specifically, model-based deep learning strategies are most suitable in the following settings:
\begin{itemize}
	\item {\bf Model deficiency}, where one has only some limited level of domain knowledge, e.g., a partial characterization of the underlying distribution $\Distribution$, as well as data.
	\item {\bf Algorithm deficiency}, where the task may be fully characterized mathematically, yet there is no efficient algorithm for tackling it.
\end{itemize}
In the context of the {\em learning algorithm}, the model deficient setting gives rise to a fundamental question: How to leverage the data to cope with the fact that the available domain knowledge is partial? One can envision two approaches -- a model-centric approach which uses the data to fit the missing domain knowledge, and a task-centric approach that aims to directly learn the inference mapping without estimating the missing model parameters. 	
We thus commence this chapter by discussing what it is that must be learned in such cases, relating these approaches to the notions of generative and discriminative learning \cite{ng2001discriminative, jebara2012machine} in Section~\ref{sec:GencDisc}.

The main bulk of the chapter is then dedicated to the {\em architecture} pillar, whose design via model-based deep learning can tackle both model and algorithm  deficiency.  We categorize systematic frameworks for designing decision mappings that are both knowledge and data-centric as a form of hybrid model-based deep learning optimization: The first strategy, coined {\em learned optimizers} \cite{agrawal2021learning} (Section~\ref{sec:LearnedOpt}), uses deep learning  automated tuning machinery to tune parameters of model-based optimization conventionally tuned by hand.  The second family of techniques, referred to as {\em deep unfolding} \cite{monga2021algorithm} (Section~\ref{sec:Unfolding}), converts iterative model-based optimizers with a fixed number of iterations into a \ac{dnn}. We conclude with methods which augment model-based optimization with dedicated deep learning tools as a form of \ac{dnn}-aided optimization. These include the replacement of a complex prior (coined {\em \ac{dnn}-aided priors}, see Section~\ref{sec:DNNpriors}), and the incorporation of \acp{dnn} for carrying out internal computations (referred to as {\em \ac{dnn}-aided inference}, see Section~\ref{sec:DNNinference}) \cite{shlezinger2020model}. An illustration of this division is depicted in Fig.~\ref{fig:Spectrum1}. 

%
%
%
%
%
%

\section{Learning to Cope with Model Deficiency}
\label{sec:GencDisc}
As discussed above, model deficiency refers to settings where there is both a (possibly limited) data set $\Data$ and partial characterization of the model $\Distribution$. Broadly speaking and following the terminology of \cite{ng2001discriminative}, one can consider two main approaches for designing inference rules in such settings, where one has to cope with the fact that the available domain knowledge is incomplete. These approaches arise from the notions of {\em generative learning} and {\em discriminative learning}. We next recall these \ac{ml} paradigms in the context of data-driven inference with partial domain knowledge, after which we exemplify the differences and interplay between these approaches using a dedicated example where both learning paradigms are amenable to tractable analysis.

\color{black}

\paragraph{Generative Learning}
The generative approach aims at designing the decision rule $f$ to minimize the generalization error $\mySet{L}_{ \Distribution}(f)$ in \eqref{eqn:Risk}, i.e., as in model-based inference. The available data $\Data$ is thus used to estimate the data generating distribution $\Distribution$. Given the estimated distribution, denoted by $\hat{\Distribution}_{\Data}$, one then seeks the inference rule which minimizes the risk function with respect to $\hat{\Distribution}_{\Data}$, i.e., 
\begin{equation}
\label{eqn:MBOpt}
f^* = \mathop{\arg \min}\limits_{f\in \mySet{F}} \mySet{L}_{\hat{\Distribution}_{\Data}}(f).
\end{equation}
 
\paragraph{Discriminative Learning}
The discriminative approach bypasses the need to fit the underlying distribution, and uses data 
to directly tune the inference rule based on the empirical risk $\mySet{L}_{\Data}(f)$ in \eqref{eqn:EmpRisk}.  
As such, discriminative learning encompasses conventional supervised training of \acp{dnn} discussed in  Chapter~\ref{ch:Deep}. However, it can also accommodate partial domain knowledge as a form of {\em model-based discriminative learning}.
In particular, while the formulation in Chapter~\ref{ch:Deep} is model-agnostic and allows the inference rule to take a broad range of mappings parametrized as abstract neural networks, model-based discriminative learning leverages the available domain knowledge to determine what structure the inference rule takes. Namely, the feasible set of inference rules  is now constrained to a pre-determined parameterized structure, denoted $\mySet{F}_{\Distribution}$, that is known to be suitable for the problem at hand, and the learning task becomes finding
\begin{equation}
\label{eqn:DiscOpt}
\dnnParam^*= \mathop{\arg \min}\limits_{f_{\dnnParam}\in \mySet{F}_{\Distribution}} \mySet{L}_{\Data}(f_{\dnnParam}).
\end{equation}
The additional constraint in \eqref{eqn:DiscOpt} boosts the specificity and reduces the parameterization of the inference mapping compared with black-box model agnostic end-to-end learning.

\begin{figure}
	\centering
\includegraphics[width=\linewidth]{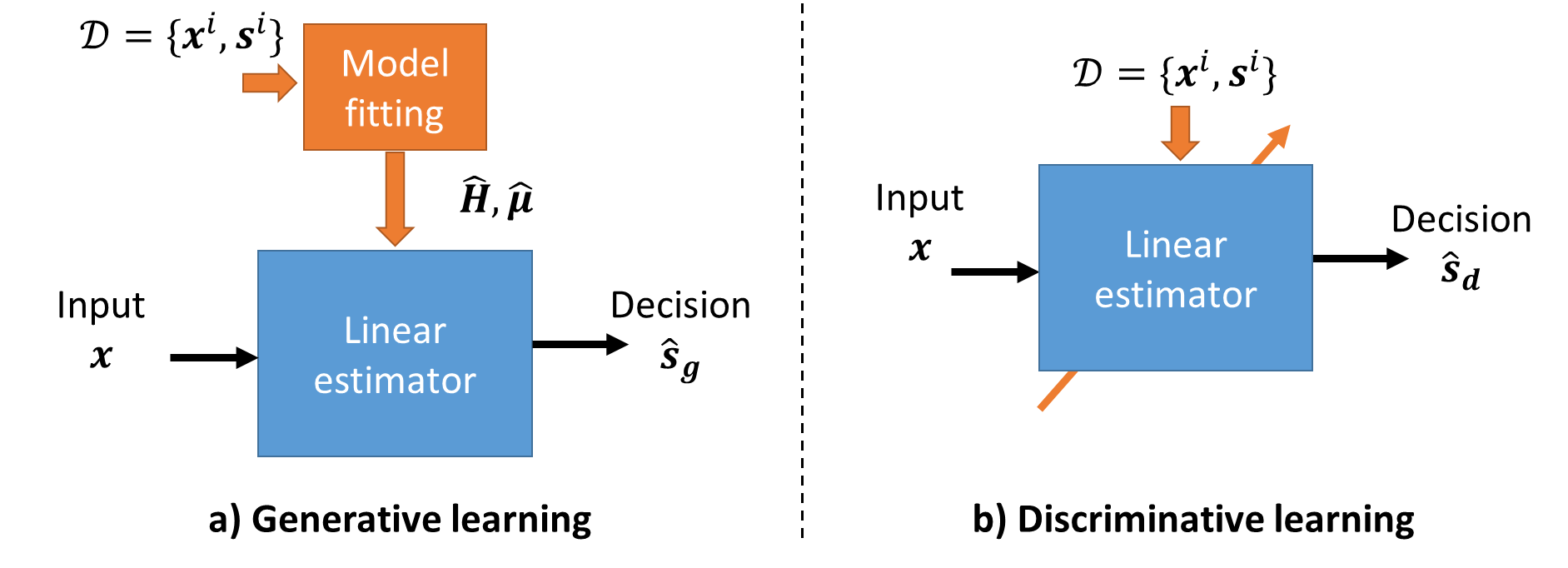}
\caption{Illustration of the linear estimation of Example~\ref{exm:DiscVsGen} obtained via generative learning $(a)$ and discriminative learning $(b)$.}
\label{fig:Linear1}
\end{figure}

\smallskip
To exemplify the differences and interplay between the generative and discriminative approach for hybrid model-based/data-driven inference in an analytically tractable manner, we present the following example, taken from \cite{shlezinger2022discriminative}, and illustrated in Fig.~\ref{fig:Linear1}.

\begin{example}
	\label{exm:DiscVsGen}
	We consider an estimation task with the $\ell_2$ loss given in \eqref{eqn:L2Loss}, where the risk function becomes the \ac{mse} as in~\eqref{eqn:MMSE}.
In this setting, one has prior knowledge that the target $\Label$ admits a Gaussian distribution with mean $\myVec{\mu}_\Label$ and covariance $\myMat{C}_{\Label\Label}$, i.e. 
	$\Label\sim\mathcal{N}(\myVec{\mu}_\Label,\myMat{C}_{\Label\Label})$,
	and that the measurements follow a linear model 
	\begin{equation}
	\label{eqn:IORel}
	\Input = \myMat{H}\Label + \myVec{w}, \quad \myVec{w}\sim\mathcal{N}(\myVec{\mu},\sigma^2\myMat{I}). 
	\end{equation}
 In \eqref{eqn:IORel}, $\myVec{w}$ is a Gaussian noise with mean $\myVec{\mu}$ and covariance matrix $\sigma^2\myMat{I}$, and is assumed to be  independent of  $\Label$. 
Consequently, the observed $\Input$ and the desired $\Label$ obey a jointly Gaussian distribution $\Distribution$. 
	The available domain knowledge is partial in the sense that  the parameters $\myMat{H}$ and $\myVec{\mu}$ are unknown. Nonetheless, we are given access to a data set $\Data =\{\Input^i, \Label^i\}_{i=1}^{\Ntraining}$ comprised of i.i.d. samples drawn from $\Distribution$. 
	
	To formulate the generative and discriminative estimators for the considered linear Gaussian setting, we define the sample first-order and second-order moments of $(\Input,\Label)$ computed from $\Data$ as 
	\begin{align*}
	\bar{\Input}&= \frac{1}{\Ntraining}\sum_{i=1}^{\Ntraining} \Input^i,~~~ \bar{\Label}=\frac{1}{\Ntraining}\sum_{i=1}^{\Ntraining} \Label^i, \\
	 \hat{\myMat{C}}_{\Label\Input} &= \frac{1}{\Ntraining} \sum_{i=1}^{\Ntraining} (\Label^i - \bar{\Label})(\Input^i -\bar{\Input})^T, \\
	  \hat{\myMat{C}}_{\Label\Label} &=\frac{1}{\Ntraining} \sum_{i=1}^{\Ntraining} (\Label^i - \bar{\Label})(\Label^i - \bar{\Label})^T, \\
	  	 \hat{\myMat{C}}_{\Input\Input} &= \frac{1}{\Ntraining} \sum_{i=1}^{\Ntraining} (\Input^i - \bar{\Input})(\Input^i - \bar{\Input})^T.
	\end{align*}
	
	The {\bf  generative approach} uses data to estimate the missing domain knowledge parameters, i.e., it uses $\Data$ to estimate the matrix $\myMat{H}$ and  the noise mean $\myVec{\mu}$. Since these parameters are considered to be deterministic and unknown, they are fitted to the data using the maximum likelihood rule. Letting $\Distribution(\Input,\Label;\myMat{H}, \myMat{\mu})$ be the joint distribution of $\Input$ and $\Label$ for given $\myMat{H}$ and $\myVec{\mu}$, the log-likelihood can be written as
	\begin{eqnarray}
	\label{eqn:loglike1}
	\log \Distribution(\Input^i,\Label^i;\myMat{H},\myVec{\mu})=
	\log \Distribution(\Input^i|\Label^i;\myMat{H},\myVec{\mu})+ \log \Distribution(\Label^i)\nonumber\\
	={\text{const}}
	-\frac{1}{\sigma^2} \|\Input^i-\myMat{H}\Label^i -\myVec{\mu}\|_2^2.
	\end{eqnarray}
	In \eqref{eqn:loglike1}, ${\text{const}}$ denotes a constant term, which is not a function of the unknown parameters $\myMat{H}$ and $\myVec{\mu}$. 
	 Since $\Data$ is comprised of  i.i.d. samples drawn from the jointly Gaussian generative distribution $\Distribution$, the maximum likelihood estimates are obtained from \eqref{eqn:loglike1} as
	\begin{equation}
	\label{eqn:loglike}
	\hat{\myMat{H}},\hat{\myVec{\mu}} = \mathop{\arg\max}\limits_{\myMat{H}, \myVec{\mu}} \sum_{i=1}^{\Ntraining}\|\Input^i-\myMat{H}\Label^i -\myVec{\mu}\|_2^2.
	\end{equation}
	The solutions to \eqref{eqn:loglike} are given by \cite[Section 3.3]{theodoridis2020machine}
	\begin{equation}
	\label{eqn:hat_H}
	\hat{\myVec{\mu}} = \bar{\Input}-
	\hat{\myMat{H}}\bar{\Label}, \quad 
	\hat{\myMat{H}} = \hat{\myMat{C}}_{\Input\Label} \hat{\myMat{C}}_{\Label\Label}^{-1},
	\end{equation}
	where 	$ \hat{\myMat{C}}_{\Input\Label}= \hat{\myMat{C}}_{\Label\Input}^T$. 
	
	The estimates $\hat{\myMat{H}}$ and $\hat{\myVec{\mu}}$ are then used to form the  estimator in \eqref{eqn:MBOpt}, as illustrated in Fig.~\ref{fig:Linear1}(a). The estimated $\hat{\Distribution}_{\Data}$ is obtained from the linear model 
	\begin{equation}
	\label{eqn:IORel_est}
	\Input = \hat{\myMat{H}}\Label + \tilde{\myVec{w}}, \quad \tilde{\myVec{w}}\sim\mathcal{N}(\hat{\myVec{\mu}},\sigma^2\myMat{I}),
	\end{equation} 
	where we recall that 	$\Label\sim\mathcal{N}(\myVec{\mu}_\Label,\myMat{C}_{\Label\Label})$. 
	Since the estimated distribution,  $\hat{\Distribution}_{\Data}$,
	is a jointly Gaussian distribution,
	the solution of  \eqref{eqn:MBOpt} is given by the conditional expectation \eqref{eqn:MMSE}, i.e.,
	\begin{align}
	\hat{\Label}_g=f^*( \Input) &=
	\E_{(\Input,\Label)\sim \hat{\Distribution}_{\Data}} \{\Label | \Input\} \notag \\ 
	&\stackrel{(a)}{=} {\myVec{\mu}}_\Label \! + \!{\myMat{C}}_{\Label \Label}\hat{\myMat{H}}^T ( \hat{\myMat{H}} {\myMat{C}}_{\Label \Label}\hat{\myMat{H}}^T \! + \!\sigma^2\myMat{I})^{-1} ( \Input \! - \!\bar{\Input}\! - \!\hat{\myMat{H}}({\myVec{\mu}}_\Label\! - \!\bar{\Label})) \notag \\
	&\stackrel{(b)}{=}
	{\myVec{\mu}}_\Label \! + \!( \hat{\myMat{H}}^T\hat{\myMat{H}}\! + \!\sigma^2 {\myMat{C}}_{\Label \Label}^{-1})^{-1} \hat{\myMat{H}}^T( \Input \! - \!\bar{\Input}\! - \!\hat{\myMat{H}}({\myVec{\mu}}_\Label\! - \!\bar{\Label})).
	\label{eqn:MBOpt_linear2_5}
	\end{align}
	Here, $(a)$ follows from the estimated jointly Gaussian model \eqref{eqn:IORel_est}, and  $(b)$ is obtained using the  matrix inversion lemma (assuming that  $\hat{\myMat{H}}^T\hat{\myMat{H}}\! + \!\sigma^2 {\myMat{C}}_{\Label \Label}^{-1}$ is invertible). 
	\\
	
	The {\bf discriminative approach} leverages the partial domain knowledge regarding the underlying joint Gaussianity to set the structure of the estimator, namely, that it should take a linear form. This implies that $\mySet{F}_{\Distribution}$ is the set of parameterized mappings of the form
	\begin{equation}
	\label{f_linear}
f_{\myVec{\theta}}(\Input)=\myMat{A}\Input+\myVec{b}, \quad \myVec{\theta}=\{\myMat{A},\myVec{b}\}.
	\end{equation}
For the considered parametric model, the available data is used to directly identify the parameters which minimize the empirical risk, as illustrated in Fig.~\ref{fig:Linear1}(b), i.e., 
\begin{align} 
\myVec{\theta}^* 
&=  \mathop{\arg \min}\limits_{f_{\myVec{\theta}} \in \mySet{F}_{\Distribution}}  \frac{1}{\Ntraining}\sum_{i=1}^{\Ntraining} 
\|\Label^i - f_{\myVec{\theta}}(\Input^i)\|_2^2 \notag \\
&= \mathop{\arg \min}\limits_{\myMat{A},\myVec{b}} \frac{1}{\Ntraining}\sum_{i=1}^{\Ntraining} 
\|\Label^i - \myMat{A}\Input^i-\myVec{b}\|_2^2,
\label{eqn:DDOpt_linear2}
\end{align}
which is solved by $\myVec{b}^* = \bar{\Label}  -  \myMat{A}^*\bar{\Input}$ with $\myMat{A}^* = \hat{\myMat{C}}_{\Label\Input} \hat{\myMat{C}}_{\Input\Input}^{-1}$. The resulting discriminative learned estimator is given by 
\begin{equation} \label{Monte_Carlo_LMMSE}
\hat{\Label}_d= f_{\myVec{\theta}^*}(\Input)=
\hat{\myMat{C}}_{\Label\Input} \hat{\myMat{C}}_{\Input\Input}^{-1}(\Input-\bar{\Input})+\bar{\Label}.
\end{equation}
The estimator is in fact the sample linear \ac{mmse} estimate, obtained by plugging the sample-mean and sample covariance matrices  into  the linear \ac{mmse} estimator. 
	
\end{example}

Example~\ref{exm:DiscVsGen}, which considers a simple yet common scenario of linear estimation with a partially-known measurement model, provides closed-form expressions for the suitable estimators attained via generative learning and via discriminative learning. The resulting estimators in \eqref{eqn:MBOpt_linear2_5} and in \eqref{Monte_Carlo_LMMSE} are generally different. Yet, they can be shown to coincide given sufficient number of data samples, i.e.,  they both approach the linear \ac{mmse} estimator when $\Ntraining\rightarrow\infty$. Nonetheless, in the presence of model-mismatch, e.g., when the true $\Distribution$ is not a linear one, but is given by  $\Input = g(\myMat{H},\Label) + \myVec{w}$ for some non-linear function $g(\cdot)$, the learned estimators differ also for $\Ntraining\rightarrow\infty$. In this case, the discriminative estimator is  preferred as it approaches the linear \ac{mmse} estimator regardless of the underlying model, while the generative learning approach, which is based on a mismatched statistical model, yields a  linear estimate that differs from the LMMSE estimator, and is thus sub-optimal at  $\Ntraining\rightarrow\infty$.

\begin{figure}
	\centering
		\centering
	\begin{subfigure}{0.45\textwidth}
		\centering
		{\includegraphics[width=\columnwidth]{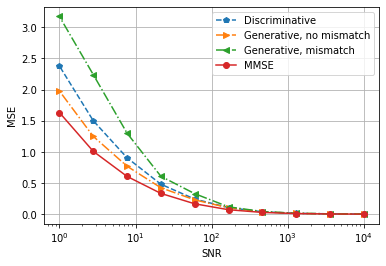}} 
		\caption{ \ac{mse} vs. \ac{snr} $1/\sigma^2$; $\Ntraining = 100$.
		}
		\label{fig:GenVsDiscSimSNR} 	
	\end{subfigure}
	$\quad$
	\begin{subfigure}{0.45\textwidth}
		\centering
		{\includegraphics[width=\columnwidth]{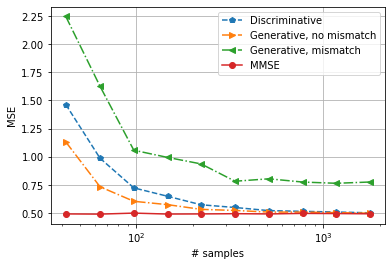}} 
		\caption{\ac{mse} vs. number of samples $\Ntraining$; $ \sigma = 0.3$.
		}
		\label{fig:GenVsDiscSimNt} 
	\end{subfigure} 
	\caption{\ac{mse} of discriminative versus generative learning of a linear estimator in a jointly Gaussian model with and without model mismatch, compared with the \ac{mmse}.}
	\label{fig:GenVsDiscSim}
\end{figure}

This behavior is empirically observed in Fig.~\ref{fig:GenVsDiscSim}. There, we compare the generative and the discriminative learning estimators for signals in $\mySet{R}^{30}$, when the target covariance has exponentially decaying off-diagonal entries, i.e.,  $[\myMat{C}_{\Label\Label}]_{i,j}=e^{-\frac{|i-j|}{5}}$. While the discriminative approach is invariant of this setting, for the generative estimator we consider both the case in which $\myMat{C}_{\Label\Label}$ is accurately known as well as a mismatched case where it is approximated as the identity matrix.  Observing Fig.~\ref{fig:GenVsDiscSimSNR}, we note that in the high \ac{snr} regime, all estimators achieve performance within a minor gap of the MMSE. However, in lower \ac{snr} values, the generative estimator, which fully knows $\myMat{C}_{\Label\Label}$, outperforms the discriminative approach due to its ability to incorporate the prior knowledge of $\sigma^2$ and of the statistical moments of $\Label$. Nonetheless, in the presence of small mismatches in  $\myMat{C}_{\Label\Label}$, the discriminative approach yields improved MSE, indicating on its ability to better cope with modeling mismatches compared with generative learning. In Fig.~\ref{fig:GenVsDiscSimNt} we observe that the effect of model mismatch does not vanish when the number of samples increases, and the mismatched generative model remains within a notable gap from the MMSE, while both the discriminative learning estimator and the non-mismatched generative one approach the MMSE as $\Ntraining$ grows. 
\\

The above noted improved robustness of discriminative learning to model mismatch is exploited in model-based deep learning methodologies, detailed in the sequel. A repeated rationale which is shared among many of these methodologies leverages the available domain knowledge to identify a suitable model-based algorithm for the problem at hand, and then converts this algorithm into a {\em trainable discriminative architecture}, where data is used to directly tune the selected solver. 
 This rationale is useful for not only for the model  deficit case, as discussed here, but also for tackling algorithm deficiency. There, one can parameterize an efficient model-based method that exists for a surrogate related model, and train it in a discriminative manner to be applicable for the problem at hand. 
 
 Consequently, discriminative learning dominates the training of architectures designed via learned optimization (Section~\ref{sec:LearnedOpt}) and deep unfolding (Section~\ref{sec:Unfolding}). Generative learning approaches are utilized in some forms of model-based deep learning via \ac{dnn}-aided priors and \ac{dnn}-aided inference, as discussed in Sections~\ref{sec:DNNpriors}-\ref{sec:DNNinference}. These allow to integrate a pre-trained architecture into an existing model-based method, and can thus benefit from both the generative and discriminative approaches. 
\color{black}


\section{Learned Optimization}
\label{sec:LearnedOpt}
Recall that in Chapter~\ref{ch:MB}, when we discussed decision making via model-based optimization, we identified two main parameters of such solvers:
\begin{itemize}
	\item {\em Objective parameters $\ObjParam$:} These are parameters  used in the formulation of the resulting optimization problem, which the decision making system is designed to solve.
	They typically arise from the model imposed on the task. However, some objective parameters, e.g., regularization coefficients, do not stem directly from the system model, and are selected so that a solver based on the objective achieves satisfactory results.  
	 In some cases (e.g., the Kalman filter example discussed in Example~\ref{exm:LQGPoliciy}), one can directly solve the resulting problem, i.e., mathematically derive $\arg\min_{f}\mySet{L}_{\ObjParam}(f)$.  As we discussed, the model-based objective $\mySet{L}_{\ObjParam}$ is often a simplified approximated surrogate of the true system objective, and thus modifying $\ObjParam$ can better fit the solution to the actual problem being solved (and not just the one we can formulate).
	\item {\em Hyperparameters $\HypParam$:} Often in practice, even simplified surrogate objectives do not give rise to a direct solution (such as the sparse recovery example discussed in Example~\ref{exm:ISTA}), but can be tackled using iterative solvers. Iterative solvers are associated with their own parameters (e.g., step-sizes and initialization), which are hyperparameters of the problems. 
\end{itemize}

Tuning of the above parameters is traditionally a cumbersome task, involving manual configuration over extensive simulations. Learned optimization is a method to utilize deep learning techniques to facilitate this aspect of optimization.
In particular, learned optimizers use conventional model-based methods for decision making, while tuning the parameters and hyperparameters of classic solvers via automated deep learning training \cite{agrawal2021learning}. This form of model-based deep learning leverages data to optimize the optimizer. While learned optimization bypasses the traditional daunting effort of manually fitting the decision rule parameters, it involves the introduction of new hyperparameters of the training procedure that must to be configured (typically by hand). 

Learned optimization effectively converts an optimizer into an \ac{ml} model. Since automated tuning of \ac{ml} models is typically carried out using gradient-based first-order methods, a key requirement is for the optimizer to be differentiable, namely, that one can compute the gradient of its decision with respect to the parameter being optimized.

\subsection{Explicit Solvers}
Learned optimization focuses on optimizing parameters conventionally tuned manually; these are parameters whose value does not follow from prior knowledge of the problem being solved, and thus their modification affects only the solver, and not the problem being solved. For model-based optimizers based on explicit solutions, the parameters available are only those of the objective $\ObjParam$. Nonetheless, some of these parameters stem from the fact the objective is inherently surrogate to the actual problem being solved, and thus require tuning, as shown in the following example, based on the running example of tracking of dynamic systems.

\begin{example}
	\label{exm:BPKalman}
	Consider a dynamic system characterized by a linear Gaussian state-space model as in \eqref{eqn:ssmodel1}, namely
	\begin{subequations}
		\label{eqn:ssmodel1a}
		\begin{align}
		\myVec{s}_{t+1} &= \myMat{F}\myVec{s}_t+ \myVec{v}_t, \qquad \myVec{v}_t\sim\mathcal{N}(0, \myMat{V}) \\
		\myVec{x}_t &= \myMat{H}\myVec{s}_{t} +\myVec{w}_t, \qquad \myVec{w}_t\sim\mathcal{N}(0, \myMat{W}).
		\end{align}
	\end{subequations}
	For the filtering task (\ac{mse} estimation of $\myVec{s}_t$ from $\{\myVec{x}_\tau\}_{\tau\leq t}$), we know that the optimal solution is the Kalman filter, given by the update equations:
	\begin{align}
	\hat{\Label}_{t|t} = f(\{\myVec{x}_\tau\}_{\tau\leq t}) &= \myMat{F}\hat{\Label}_{t-1|t-1} +\myMat{K}_t(\Input_t - \myMat{H}\myMat{F}\hat{\Label}_{t-1|t-1}),
	\label{eqn:KF1}
	\end{align}
	where the Kalman gain is updated via
	\begin{align}
	\myMat{K}_t &= \myMat{\Sigma}_{t|t-1} \myMat{H}^T\big(\myMat{H}\myMat{\Sigma}_{t|t-1}\myMat{H}^T + \myMat{W} \big)^{-1},
	\label{eqn:KGain}
	\end{align}
	and the second-order moments are updated as
	\begin{subequations}
		\label{eqn:MomUp}
		\begin{align}
		\myMat{\Sigma}_{t|t-1} &= \myMat{F}\myMat{\Sigma}_{t-1|t-1}\myMat{F}^T + \myMat{V}, \\
		\myMat{\Sigma}_{t|t} &= \myMat{\Sigma}_{t|t-1} - \myMat{K}_t\myMat{H}\myMat{\Sigma}_{t|t-1}.
		\end{align}
	\end{subequations}
	
	In such  settings, the linear mappings $\myMat{F}, \myMat{H}$ often arise from understanding the physics of the problem. Nonetheless, in practice, one typically does not have a concrete stochastic model for the noise signals, which are often introduced as a way to capture stochasticity, and thus $\myMat{V}$ and $\myMat{W}$ are often tuned by hand. 
	
	Given a data set comprised of $\Ntraining$ trajectories of $T$ observations along with their corresponding states and actions, i.e., $\mySet{D}=\{\{\myVec{x}_t^i, \myVec{s}_t^i\}_{t=1}^T\}_{i=1}^{\Ntraining}$, one may set the trainable parameters to be $\myVec{\theta}=[\myMat{V},\myMat{W}]$, and write the Kalman filter output \eqref{eqn:KF1} for a given setting as $f_{\myVec{\theta}}(\{\myVec{x}_\tau\}_{\tau\leq t})$
	and optimize them via 
	\begin{equation}
	\label{eqn:SGD1}
	\myVec{\theta}_{j+1}= \myVec{\theta}_{j} - \eta_j \nabla_{\myVec{\theta}} \mySet{L}_{\mySet{D}_j}(f_{\myVec{\theta}_j}).
	\end{equation} 
	In \eqref{eqn:SGD1}, $\mySet{D}_j$ is a randomly selected mini-batch of $\mySet{D}$, and the loss is
	\begin{equation}
	\mySet{L}_{\mySet{D}}(f_{\myVec{\theta}}) = \frac{1}{T\cdot|\mySet{D}|}\sum_{i=1}^{|\mySet{D}|}\sum_{t=1}^{T} \|\myVec{s}_t^i - f_{\myVec{\theta}}(\{\myVec{x}^i_\tau\}_{\tau\leq t})\|^2_2 + \phi(\myVec{\theta}),
	\end{equation}
	where $\phi(\cdot)$ is a regularizing term (e.g., a scalar multiple of the $\ell_2$ or $\ell_1$ norm). 
	
	In order to optimize $\myVec{\theta}=[\myMat{V},\myMat{W}]$ using \eqref{eqn:SGD1}, thus fitting the Kalman filter to data, one must be able to compute the gradient of $ \hat{\Label}_{t|t}$ with respect to both $\myMat{V}$ and $\myMat{W}$.  To show that these gradients can indeed be computed, let us focus on a scalar case (though we keep the multivariate notations, e.g.,  $\myMat{V}$ and $\myMat{W}$, for brevity), where both  $\Input_t$ and $\Label_t$ are scalars. Here
	\begin{align}
	\frac{\partial \hat{\Label}_{t|t} }{\partial \myMat{V}} 
	&= \myMat{F}\frac{\partial \hat{\Label}_{t-1|t-1} }{\partial \myMat{V}}+\frac{\partial \myMat{K}_t}{\partial \myMat{V}}(\Input_t - \myMat{H}\myMat{F}\hat{\Label}_{t-1|t-1}) - \myMat{K}_t \myMat{H}\myMat{F}\frac{\partial\hat{\Label}_{t-1|t-1}}{\partial \myMat{V}},
	\label{eqn:KGRec2}
	\end{align}
	where, by \eqref{eqn:KGain}, we have that
	\begin{align}
	\frac{\partial \myMat{K}_t}{\partial \myMat{V}} &= \frac{\partial \myMat{K}_t}{\partial  \myMat{\Sigma}_{t|t-1}}  \frac{\partial \myMat{\Sigma}_{t|t-1}}{\partial \myMat{V}},
	\label{eqn:KGRec2a}
	\end{align}
	in which $\frac{\partial \myMat{K}_t}{\partial  \myMat{\Sigma}_{t|t-1}}$ is obtained from~\eqref{eqn:KGain}  and 
	\begin{align}
	\frac{\partial \myMat{\Sigma}_{t|t-1}}{\partial \myMat{V}} &=\myMat{F}        \frac{\partial \myMat{\Sigma}_{t-1|t-1}}{\partial \myMat{V}}\myMat{F}^T + \myMat{I}.
	\label{eqn:KGRel1}
	\end{align}
	Now, by \eqref{eqn:MomUp} we have that
	\begin{align}
	\frac{\partial \myMat{\Sigma}_{t|t}}{\partial \myMat{V}} 
	&= \frac{\partial \myMat{\Sigma}_{t|t-1}}{\partial \myMat{V}} - \frac{\partial \myMat{K}_t}{\partial \myMat{V}}\myMat{H}\myMat{\Sigma}_{t|t-1} -\myMat{K}_t \myMat{H}\frac{\partial \myMat{\Sigma}_{t|t-1}}{\partial \myMat{V}} \notag \\
	&= \frac{\partial \myMat{\Sigma}_{t|t-1}}{\partial \myMat{V}} - \frac{\partial \myMat{K}_t}{\partial  \myMat{\Sigma}_{t|t-1}}  \frac{\partial \myMat{\Sigma}_{t|t-1}}{\partial \myMat{V}}\myMat{H}\myMat{\Sigma}_{t|t-1} -\myMat{K}_t \myMat{H}\frac{\partial \myMat{\Sigma}_{t|t-1}}{\partial \myMat{V}} \notag \\
	&\stackrel{(a)}{=}\left(\myMat{I} - \frac{\partial \myMat{K}_t}{\partial  \myMat{\Sigma}_{t|t-1}}  \myMat{H}\myMat{\Sigma}_{t|t-1} -\myMat{K}_t \myMat{H}\right) \frac{\partial \myMat{\Sigma}_{t|t-1}}{\partial \myMat{V}} \notag \\
	&\stackrel{(b)}{=}\left(\myMat{I} - \frac{\partial \myMat{K}_t}{\partial  \myMat{\Sigma}_{t|t-1}}  \myMat{H}\myMat{\Sigma}_{t|t-1} -\myMat{K}_t \myMat{H}\right)\left(\myMat{F}        \frac{\partial \myMat{\Sigma}_{t-1|t-1}}{\partial \myMat{V}}\myMat{F}^T + \myMat{I}\right),
	\label{eqn:RecRel1}
	\end{align}
	where $(a)$ holds since we assumed scalar quantities for brevity, and $(b)$ is due to \eqref{eqn:KGRel1}. 
	
	Note that \eqref{eqn:RecRel1} formulates the computation of the gradient $\frac{\partial \myMat{\Sigma}_{t|t}}{\partial \myMat{V}} $ as a recursive relationship, implying that it can be computed recursively as a form of  backpropagation through time. Then, the computed value is used to set $\frac{\partial \myMat{K}_t}{\partial \myMat{V}}$ in \eqref{eqn:KGRec2a}, which is substituted into the gradient expression in \eqref{eqn:KGRec2}. Note that once $\frac{\partial \myMat{K}_t}{\partial \myMat{V}}$  is obtained,  \eqref{eqn:KGRec2} forms an additional recursive relationship.  While the above derivations are stated for $\myMat{V}$, one can formulate similar recursions for $\myMat{W}$. Once these gradients are computed, they can be used to optimize $\myVec{\theta}$ from data via \eqref{eqn:SGD1}.
	
\end{example}

\begin{figure}
	\centering
	\begin{subfigure}{0.42\textwidth}
		\centering
		{\includegraphics[width=\columnwidth]{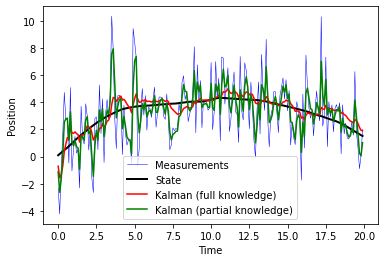}} 
		\caption{ Kalman filter with  mismatched compared with full  knowledge.
		}
		\label{fig:KalmanLONotLearned} 	
	\end{subfigure}
	$\quad$
	\begin{subfigure}{0.42\textwidth}
		\centering
		{\includegraphics[width=\columnwidth]{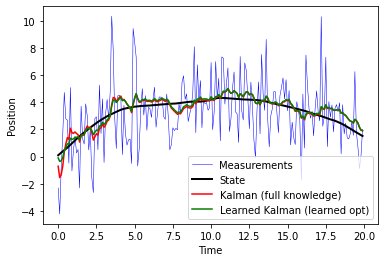}} 
		\caption{ Kalman filter with learned  parameters compared with full  knowledge.
		}
		\label{fig:KalmanLOLearned} 
	\end{subfigure}
	\caption{Tracking a dynamic system from noisy observations using the model-based Kalman filter with mismatches and with objective parameters learned via learned optimization.}
	\label{fig:KalmanLO}
\end{figure}

To illustrate the ability to tune objective parameters by learning through an explicit solver via learned optimization, we next simulate a Kalman filter for tracking a two-dimensional state vector from a noisy observations of the first entry, as in the study reported in Fig.~\ref{fig:Kalman}. Here, a mismatched model corresponds to inaccurate knowledge of the state evolution noise covariance $\myMat{V}$, which degrades accuracy as shown in Fig.~\ref{fig:KalmanLONotLearned}. Then, by treating $\myMat{V}$ as a trainable parameter and converting the Kalman filter into a discriminative model,  one can learn its setting by backpropagating through the algorithm, yielding performance that approaches that achieved with full domain knowledge, as illustrated in Fig.~\ref{fig:KalmanLOLearned}, where we used a data set comprised of $\Ntraining=20$ trajectories.

\subsection{Iterative Optimizers}


When the inference rule being learned is an iterative solver, one can use data to automatically tune the hyperparameters $\HypParam$. As opposed to  the objective parameters $\ObjParam$, they often have only a minor effect on the solution when the algorithm is allowed to run to convergence, and so are of secondary importance.   But when the iterative algorithm is stopped after a predefined number of iterations, they affect the decisions, and therefore also the design objective. Additional parameters that can be automatically tuned from data are objective parameters that do not follow from the underlying statistical model, e.g., regularization coefficients, which are conventionally tuned manually and not based on some principled modeling.

The main challenge with applying the same approach used in Example~\ref{exm:BPKalman} for iterative solvers stems from the challenge in taking their gradients. This is due to the fact that the output is not an explicit function of the input and the hyperparameters, but an implicit one, i.e., there is no closed-form expression for the inference rule output. Moreover, the number of iterations can vary between different inputs. Nonetheless, one can still compute the gradients by letting the algorithm run until it converges, and then taking the gradient of the loss computed over a data set, as shown in the following example:

\begin{example}
	\label{exm:BPISTA}
	Recall the \ac{ista}, introduced in Example~\ref{exm:ISTA}, as an iterative solver for the convex  LASSO problem  in \eqref{eqn:LASSO}, i.e.,
	\begin{equation}
	\label{eqn:LASSO2}
	\hat{\myVec{s}} = \mathop{\arg\min}\limits_{\myVec{s}} \frac{1}{2}\|\myVec{x}-\myMat{H}\myVec{s}\|^2 +\rho\|\myVec{s}\|_1,
	\end{equation}
	which is obtained as an $\ell_1$ convex relaxation of the super-resolution with sparse prior problem. The iterations of \ac{ista} are given by \eqref{eqn:ISTA}, namely,
	\begin{equation}
	\label{eqn:ISTA2}
	\myVec{s}^{(k+1)} \leftarrow  \mySet{T}_{\beta=\mu\rho}\left( \myVec{s}^{(k)} + \mu \myMat{H}^T(\myVec{x}-\myMat{H}\myVec{s}^{(k)}) \right),
	\end{equation}
	where $\mySet{T}_{\beta}(\cdot)$ is the soft-thresholding operation applied element-wise:
	\begin{equation}
	\mySet{T}_{\beta}(x) \triangleq {\rm sign}(x)\max(0,|x|-\beta). 
	\end{equation}

	Now, given a labeled data set $\Data = \{\Input^i, \Label^i\}_{i=1}^{\Ntraining}$, we are interested in optimizing two hyperparameters of the algorithm: the step size $\mu$ (which is a parameter of the iterative solver) and the soft threshold $\beta = \mu \rho$ (which is a parameter of the objective that does not follow from the underlying model), i.e., $\myVec{\theta}=[\mu, \beta]$. 
	
By letting $\hat{\myVec{s}}^i = f(\Input^i;\myVec{\theta})$  denote the converged output of \ac{ista} applied to $\Input_t$ with setting $\myVec{\theta}$, 
 the empirical risk is given by
	\begin{equation}
	\mySet{L}_{\mySet{D}}(\myVec{\theta}) = \frac{1}{|\mySet{D}|}\sum_{(\Input^i, \Label^i)\in \mySet{D}}\|{\myVec{s}}^i - f(\Input^i;\myVec{\theta})\|_2^2.
	\label{eqn:ISTAEmpLoss}
	\end{equation}
	Now, in order to use \eqref{eqn:ISTAEmpLoss} to optimize $\myVec{\theta}$ using first-order methods, we need to be able to compute the gradient
	\begin{align}
	\nabla_{\myVec{\theta}} f(\Input^i;\myVec{\theta}) = \Big[\frac{\partial\myVec{s}^{(K^i)} }{\partial \mu}, \quad \frac{\partial\myVec{s}^{(K^i)} }{\partial \beta}  \Big]^T,
	\label{eqn:ISTAgrad}
	\end{align}
	where $K^i$ is the number of iterations required to achieve convergence.
	Essentially, we will treat $K_t$ as fixed after convergence is achieved, and set $\myVec{\theta}$ to get the best performance out of these iterations. 
	
	To compute \eqref{eqn:ISTAgrad}, we note that 
	\begin{align}
	\frac{\partial}{\partial \beta} \mySet{T}_{\beta}(x)  &=  -{\rm sign}(x)\myVec{1}_{\beta < |x|}, \qquad 
	\frac{\partial}{\partial x} \mySet{T}_{\beta}(x)  = \myVec{1}_{\beta < |x|},
	\end{align}
	where $\myVec{1}_{(\cdot)}$ is the indicator function.
	Using this and \eqref{eqn:ISTA2}, we obtain the (element-wise) recursion
	\begin{align}
	\frac{\partial\myVec{s}^{(k+1)} }{\partial \mu} &= \myVec{1}_{\beta < |\myVec{s}^{(k)} + \mu \myMat{H}^T(\myVec{x}-\myMat{H}\myVec{s}^{(k)})|} \odot 
	\frac{\partial \left( \myVec{s}^{(k)} + \mu \myMat{H}^T(\myVec{x}-\myMat{H}\myVec{s}^{(k)})  \right)}{\partial \mu} \notag \\
	&= \myVec{1}_{\beta < |\myVec{s}^{(k)} + \mu \myMat{H}^T(\myVec{x}-\myMat{H}\myVec{s}^{(k)})|} \notag \\
	&\qquad  \odot 
	 \left(\myMat{H}^T\left(\myVec{x}-\myMat{H}\myVec{s}^{(k)}\right) + \left(\left(\myMat{I}-\mu\myMat{H}^T\myMat{H}\right) \frac{\partial\myVec{s}^{(k)} }{\partial \mu} \right)\right).
	\end{align}
	Similarly, we can compute the gradient with respect to $\beta$ via
	\begin{align}
	\frac{\partial\myVec{s}^{(k+1)} }{\partial \beta} =& -{\rm sign}\big(\myVec{s}^{(k)} + \mu \myMat{H}^T(\myVec{x}-\myMat{H}\myVec{s}^{(k)})\big)\myVec{1}_{\beta < |\myVec{s}^{(k)} + \mu \myMat{H}^T(\myVec{x}-\myMat{H}\myVec{s}^{(k)})|}  \notag \\
	& + \mySet{T}_{\beta}\left( \myVec{s}^{(k)} + \mu \myMat{H}^T(\myVec{x}-\myMat{H}\myVec{s}^{(k)}) \right)\odot \left(\left(\myMat{I}-\mu\myMat{H}^T\myMat{H} \right)\frac{\partial\myVec{s}^{(k)} }{\partial \beta} \right).
	\end{align}
	The above expressions for the gradients are used to update the hyperparameters based on the empirical risk~\eqref{eqn:ISTAEmpLoss}. 	Once the hyperparameters are learned from data, inference follows using the conventional \ac{ista}. 
\end{example}

Learned optimization of the hyperparameters of iterative solvers replace the often tedious manual tuning of these hyperparameters with an automated data-driven pipeline. While Example~\ref{exm:BPISTA} focuses on \ac{ista}, a similar derivation can be applied to alternative optimizers such as \ac{admm}, as we show next.

\begin{example}
	\label{exm:1ParADMM}
	Let us consider  the \ac{admm} optimizer, which aims at solving
	\begin{equation}
	\label{eqn:Recovery1ab}
	f_{\rm MAP}(\Input) = \mathop{\arg\min}\limits_{\myVec{s}} \frac{1}{2}\|\myVec{x}-\myMat{H}\myVec{s}\|^2 +\sigma^2\phi(\myVec{s}),
	\end{equation}
	via the iterative steps summarized as Algorithm~\ref{alg:Algoadmm}. This iterative algorithm has two main hyperparameters, i.e., $\HypParam =\{\lambda,\mu\}$. Learned optimization allows to  leverage a labeled data set $\Data = \{\Input^i, \Label^i\}_{i=1}^{\Ntraining}$ to best tune these hyperparameters from data, i.e., to set the learned parameters to be $\myVec{\theta}=\HypParam$. These parameters are tuned in an automated fashion based on the following optimization problem
	\begin{equation}
	\myVec{\theta}^* = \mathop{\arg \min}_{\myVec{\theta}=\{\lambda, \mu\}} \frac{1}{\Ntraining}\sum_{i=1}^{\Ntraining}\|f_{\myVec{\theta}}(\myVec{x}^i) - \myVec{s}^i\|^2,
	\label{eqn:LearnedAdmmObj}
	\end{equation}
	where $f_{\myVec{\theta}}(\cdot)$ is \ac{admm} with hyperparameters $\myVec{\theta}$.
	
	To tackle \eqref{eqn:LearnedAdmmObj} via first-order methods and automated deep learning optimization, we need to show that the iterative algorithm  is differentiable with respect to $\myVec{\theta}$. As  seen in Section~\ref{sec:LearnedOpt}, this can be done by treating the algorithm as a discriminative learning model while using  a variant of backpropagation through time when each of the iteration steps is differentiable. In Algorithm~\ref{alg:Algoadmm}, the only computation which has to be specifically examined is the proximal mapping
	\begin{equation}
	\label{eqn:ProxDefa}
	{\rm prox}_{\lambda \phi}(\myVec{y})\triangleq \mathop{\arg \min}_{\myVec{z}}\frac{1}{2}\|\myVec{z}-\myVec{y}\|_2^2 + \lambda \phi(\myVec{z}).
	\end{equation}
	Consequently, in order to backpropagate through \ac{admm}, thus using deep learning tool to tune the hyperparameters of \ac{admm}, one has to be able to compute the gradient of \eqref{eqn:ProxDefa} with respect to $\myVec{\theta}=\HypParam$. We next show that this can be done, focusing on the hyperparameter $\lambda \in \HypParam$ which explicitly appears in the formulation of the proximal mapping.
	
	When $\phi(\cdot)$ is differentiable, and the proximal mapping is unique, this derivative  can be approximated using its definition
	\begin{equation}
	\frac{\partial{\rm prox}_{\lambda \phi}(\myVec{y})}{\partial \lambda} = \frac{1}{\epsilon} \left(\partial{\rm prox}_{(\lambda + \epsilon) \phi}(\myVec{y}) - \partial{\rm prox}_{\lambda \phi}(\myVec{y})\right).
	\end{equation}
	Now, we write $\myVec{z}^* =  {\rm prox}_{\lambda \phi}(\myVec{y}) $ 
	and $\myVec{z}^* + \Delta \myVec{z} = {\rm prox}_{(\lambda + \epsilon) \phi}(\myVec{y})$. Next, we note that at $\myVec{z}^* + \Delta \myVec{z} $, the gradient of the proximal mapping argument must be zero, being a global minima of the right hand side of \eqref{eqn:ProxDefa}, and thus
	\begin{align}
	&\nabla_{\myVec{z}}\frac{1}{2}\|\myVec{z}-\myVec{y}\|_2^2 +  (\lambda + \epsilon)\cdot \phi(\myVec{z}) \Big|_{\myVec{z} = \myVec{z}^* + \Delta \myVec{z}}\notag \\ &  \qquad =
	(\lambda + \epsilon) \cdot \nabla \phi(\myVec{z}^* + \Delta \myVec{z}) + \myVec{z}^* + \Delta \myVec{z}-\myVec{y} = 0.
	\label{eqn:DerProx}
	\end{align}
	
	Next, we approximate $  \nabla \phi(\myVec{z}^* + \Delta \myVec{z})$ using its Taylor series approximation around $\myVec{z}^*$, such that
	\begin{align}
	\nabla \phi(\myVec{z}^* + \Delta \myVec{z}) \approx \nabla \phi(\myVec{z}^* ) + \myMat{H}_{\phi}(\myVec{z}^*)\Delta \myVec{z},
	\end{align}
	with $\myMat{H}_{\phi}$ being the Hessian matrix. 
	Substituting this into \eqref{eqn:DerProx} yields
	\begin{align}
	&(\lambda + \epsilon) \cdot(\nabla \phi(\myVec{z}^* ) + \myMat{H}_{\phi}(\myVec{z}^*)\Delta \myVec{z}) + \myVec{z}^* + \Delta \myVec{z}-\myVec{y} = 0 \notag \\
	&\Rightarrow \left(\lambda \nabla \phi(\myVec{z}^* ) + \myVec{z}^* -\myVec{y}  \right) +\epsilon \nabla \phi(\myVec{z}^* ) +  (\lambda + \epsilon)\myMat{H}_{\phi}(\myVec{z}^*)\Delta \myVec{z} + \Delta \myVec{z} = 0 \notag \\
	&\stackrel{(a)}{\Rightarrow} \epsilon \nabla \phi(\myVec{z}^* ) +  (\lambda + \epsilon)\myMat{H}_{\phi}(\myVec{z}^*)\Delta \myVec{z} + \Delta \myVec{z} = 0, 
	\end{align}
	where $(a)$ follows since $\myVec{z}^* =  {\rm prox}_{\lambda \phi}(\myVec{y})$. 
	Consequently, 
	\begin{align}
	\Delta \myVec{z} = -\epsilon\left(\myMat{I} +  (\lambda + \epsilon)\myMat{H}_{\phi}(\myVec{z}^*) \right)^{-1} \nabla \phi(\myVec{z}^* ),
	\end{align}
	implying that 
	\begin{equation}
	\frac{\partial{\rm prox}_{\lambda \phi}(\myVec{y})}{\partial \lambda} =  -\left(\myMat{I} +  \lambda \myMat{H}_{\phi}(\myVec{z}^*) \right)^{-1} \nabla \phi(\myVec{z}^* ),
	\end{equation}
	whose computation depends on the differentiable function $\phi(\cdot)$. 
		Once the hyperparameters are learned from data, inference follows using the conventional \ac{admm}, i.e., Algorithm~\ref{alg:Algoadmm}. 
		
		The gains of hyperparameter tuning via learned optimization ability are illustrated in Fig.~\ref{fig:LearnedADMM}. There, learned optimization of the hyperaparameters of \ac{admm}  is applied to recover a $200 \times 1$ sparse vector from $150$ noisy compressed observations, obtained using a Gaussian measurement matrix. In particular, \ac{admm} runs until either convergence is achieved (difference of less than $10^{-2}$ between consecutive estimates) or until $1000$ iterations are exhausted. The hyperparameters are learned using the Adam optimizer~\cite{kingma2014adam} with $60$ epochs over a data set of $1000$ labeled samples. The results in Fig.~\ref{fig:LearnedADMM} demonstrate the ability of learned optimization to notably facilitate hyperparameter tuning. 
		
	
	\begin{figure}
		\centering
		\includegraphics[width=0.7\columnwidth]{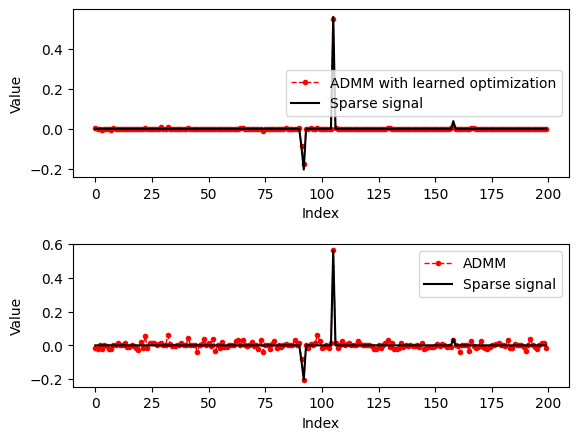}
		\caption{Sparse signal recovered using ADMM with hyperparameters optimized via learned optimization (upper figure) compared with ADMM with manually tuned hyperparemetrs (lower figure).}
		\label{fig:LearnedADMM}
	\end{figure}

\end{example}

\section{Deep Unfolding}
\label{sec:Unfolding}

A relatively common and well-established methodology for combining model-based methods and deep learning is that of deep unfolding, also referred to as {\em deep unrolling} \cite{monga2021algorithm}. Originally proposed in 2010 by Greger and LeCun for sparse recovery \cite{gregor2010learning}, deep unfolding converts iterative optimizers into trainable \acp{dnn}.

\begin{figure}
	\centering
	\includegraphics[width=0.8\columnwidth]{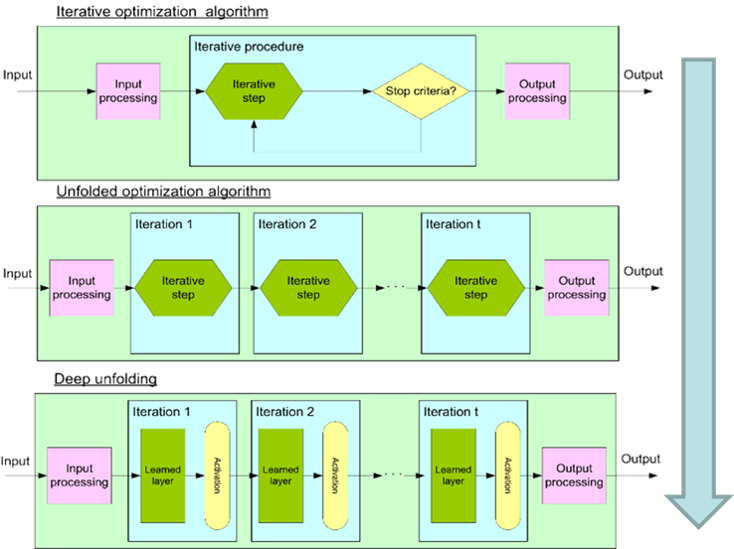}
	\caption{Deep unfolding outline illustration.}
	\label{fig:Unfolding1}
\end{figure}

As the name suggests, the method relies on unfolding an iterative algorithm into a sequential procedure with a fixed number of iterations. Then, each iteration is treated as a layer, with its trainable parameters $\myVec{\theta}$ being either only the hyperparameters $\HypParam$, or also the objective parameters $\ObjParam$.    An illustration of this approach is depicted in Fig.~\ref{fig:Unfolding1}.

Converting an iterative optimizer into a \ac{dnn} facilitates optimizing different parameters for each iteration, being transformed into trainable parameters of different layers. This is achieved by training the decision box end-to-end, i.e., by evaluating the system output based on data. For instance, letting $K$ be the number of unfolded iterations, deep unfolding can learn iteration-dependent hyperparameters $\{\HypParam_k\}_{k=1}^K$ and  objective parameters $\{\ObjParam_k\}_{k=1}^K$. This increases the parameterization and abstractness compared with learned optimization of iterative solvers, which typically reuses the learned hyperparameters and runs until convergence (as in model-based optimizers). Nonetheless, for every setting of $\{\ObjParam_k\}_{k=1}^K$ and $\{\HypParam_k\}_{k=1}^K$, a deep unfolded system effectively carries out its decision using $K$ iterations of some principled iterative solver known to be suitable for the problem.

The above rationale specializes into three forms of deep unfolding, which notably vary in their specificity versus parameterization tradeoff illustrated in Fig.~\ref{fig:Spectrum1}. We refer to these approaches, ordered from the most task-specific to the most parametrized one, as {\em Unfolded Hyperparameters}, {\em Unfolded Objective Parameters}, and {\em Unfolded Abstracted Optimizers}.

\subsection{Learning Unfolded Hyperparameters}
Deep unfolded networks can be designed to improve upon model-based optimization in two main criteria: convergence speed and model abstractness. The former is achieved by the fact that the resulting system operates with a fixed number of iterations, which can be set to be much smaller compared with that usually required to achieve convergence. This is combined with the natural ability of deep unfolding to learn iteration-dependent hyperparameters to enable accurate decisions to be achieved within this predefined number of iterations, as exemplified next: 

\begin{example}
	\label{exm:UnfADMM}
	Let us consider again the \ac{admm} optimizer as in Example~\ref{exm:1ParADMM}. We are interested in being able to accurately carry out \ac{admm} with a low and fixed number of iterations by leveraging a labeled data set $\Data = \{\Input^i, \Label^i\}_{i=1}^{\Ntraining}$. 
	
	A deep unfolded \ac{admm} is obtained by setting the decision to be $\hat{\myVec{s}} = \myVec{s}_K$ for some fixed $K$, and allowing each iteration to use hyperparameters $[\lambda_k, \mu_k]$, that are stacked into the trainable parameters vector $\myVec{\theta}$. These hyperparameters are learned from data via solving the optimization problem
	\begin{equation}
	\myVec{\theta}^* = \mathop{\arg \min}_{\myVec{\theta}=\{\lambda_k, \mu_k\}_{k=1}^K} \frac{1}{\Ntraining}\sum_{i=1}^{\Ntraining}\|f_{\myVec{\theta}}(\myVec{x}^i) - \myVec{s}^i\|^2.
	\label{eqn:UnfAdmmObj}
	\end{equation}
	Since, the iterative algorithm is differentiable with respect to $\myVec{\theta}$, as shown in Example~\ref{exm:1ParADMM}, the per-iteration hyperparameters can be learned based on   \eqref{eqn:UnfAdmmObj} via first-order methods and automated deep learning optimization.

	\begin{figure}
		\centering
		\begin{subfigure}{0.7\textwidth}
			\centering
			{\includegraphics[width=\columnwidth]{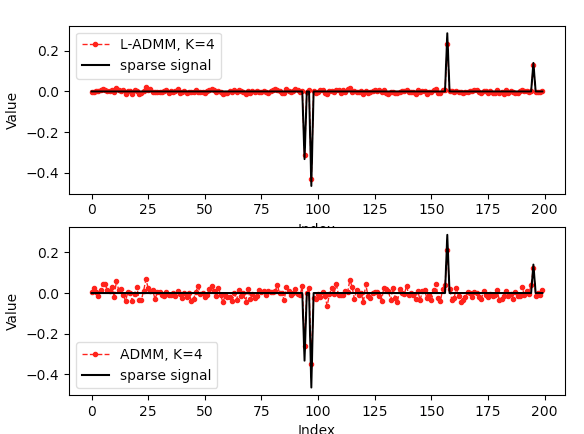}} 
			\caption{ Unfolded \ac{admm} (L-ADMM) versus conventional \ac{admm} with $K=4$ iterations.
			}
			\label{fig:L_ADMM_K4} 	
		\end{subfigure}
		$\quad$
		\begin{subfigure}{0.7\textwidth}
			\centering
			{\includegraphics[width=\columnwidth]{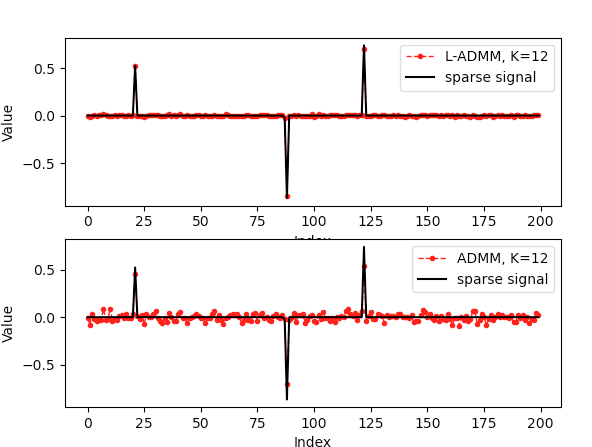}} 
			\caption{Unfolded \ac{admm} (L-ADMM) versus conventional \ac{admm} with $K=12$ iterations.
			}
			\label{fig:L_ADMM_K12} 
		\end{subfigure}
		\caption{Comparing unfolded \ac{admm} with conventional \ac{admm} with the same number of iterations for recovering a $200\times 1$ sparse signal from $150$ compressed noisy observations. The unofolded \ac{admm} is trained using $1000$ labeled pairs.}
		\label{fig:L_ADMM}
	\end{figure}
	
	Once the hyperparameters are learned from data, inference is carried out using $K$ \ac{admm} iterations with the per-iteration configured settings. The resulting architecture, as compared with the original \ac{admm} and its version with learned optimization, is illustrated in Fig.~\ref{fig:ADMMComp2}. Unfolding \ac{admm} and learning iteration-dependent hyperparameters notably improves performance compared to conventional \ac{admm} with fixed hyperparaemeters, as visualized in Fig.~\ref{fig:L_ADMM}. There, both unfolding into $K=4$ as well as $K=12$ layers are contrasted with conventional \ac{admm} with the same number of iterations, showcasing the notable gains of deep unfolding with learned hyperparameters in improving performance withing a fixed number of iterations. 
\end{example}


\subsection{Learning Unfolded Objective Parameters}
Example~\ref{exm:UnfADMM} preserves the operation of \ac{admm} with $K$ iterations, as only the hyperparameters are learned. Nonetheless, one can also transform iterative solvers into more abstract \acp{dnn} by also tuning the objective of each iteration. Here, the trainable architecture  jointly learns the hyperparameters and the objective parameters $\ObjParam$ per each iteration. This can be viewed as if each iteration  follows a different objective, such that the output of the system after $K$ such iterations most accurately matches the desired value. 
As a result, the trained \ac{dnn} can realize a larger family of abstract mappings and is likely to deviate from the model-based optimizer, which serves here as a principled initialization for the system, rather than its fixed structure.

\begin{figure}
	\centering
	\includegraphics[width=\linewidth]{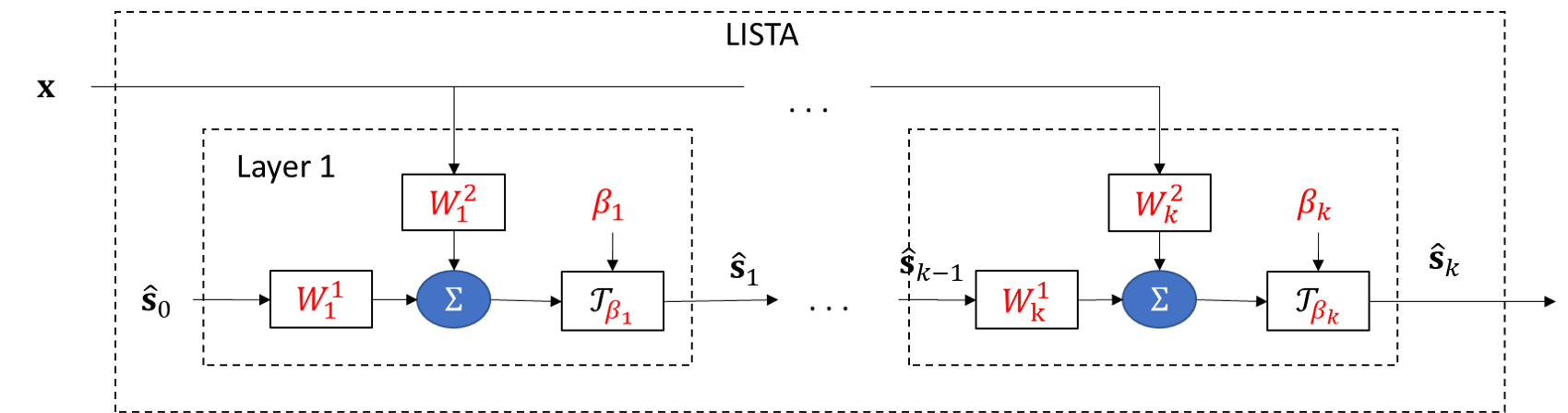}
	\caption{Learned iterative soft-thresholding algorithm architecture.}
	\label{fig:LISTA}
\end{figure}

\begin{example}
	\label{exm:LISTA}
	Consider again the \ac{ista} optimizer, which solves the convex  LASSO problem in \eqref{eqn:LASSO2} 
	via the iterative update equations in \eqref{eqn:ISTA2}. 	
	The  learned \ac{ista} (LISTA) \ac{dnn} architecture, which is the first introduction of deep unfolding methodology \cite{gregor2010learning}, unfolds \ac{ista} by fixing $K$ iterations and replacing the update step in \eqref{eqn:ISTA} with 
	\begin{equation}
	\label{eqn:LISTA}
	\myVec{s}^{(k+1)} = \mySet{T}_{\beta_k}\left( \myMat{W}_k^1\myVec{x} +  \myMat{W}_k^2  \myVec{s}^{(k)} \right).
	\end{equation} 
	Note that for $\myMat{W}_k^1 = \mu\csMatrix^T$, $\myMat{W}_k^2 = \myMat{I} -\mu\csMatrix^T\csMatrix$,  and $\beta_k=\mu \rho$, \eqref{eqn:LISTA} coincides with the model-based \ac{ista}.  
	The trainable parameters $\myVec{\theta}=\big[\{\myMat{W}_k^1, \myMat{W}_k^2, \beta_k\}_{k=1}^K\big]$ are learned from data via end-to-end training, building upon the ability to backpropagate through \ac{ista} iterations discussed in Example~\ref{exm:BPISTA}.  An illustration of the unfolded LISTA is given in Fig.~\ref{fig:LISTA}.
\end{example}

\begin{figure}
	\centering
	\includegraphics[width=0.6\linewidth]{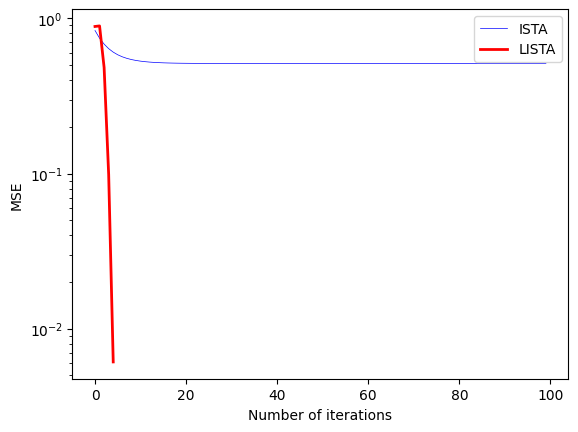}
	\caption{\ac{mse} performance of LISTA and \ac{ista} versus number of iterations $K$.}
	\label{fig:LISTA_LOSS}
\end{figure}

While in Example~\ref{exm:LISTA} each layer has different parameters, one can enforce the parameters to be equal across layers; since the objective parameters are  optimized, doing so still preserves the ability of deep unfolded networks to jointly match the solver and its objective to the data. Fig.~\ref{fig:LISTA_LOSS} compares \ac{ista} to  LISTA, in which the objective parameters are shared between different iterations, i.e., the trainable parameters are $\myVec{\theta}=\big[\myMat{W}^1, \myMat{W}^2, \{ \beta_k\}_{k=1}^K\big]$. In Fig.~\ref{fig:LISTA_LOSS}, LISTA is trained for $K\in \{1,\ldots,13\}$ iterations, and is contrasted with \ac{ista} (where the objective is not learned but set via $\myMat{W}^1 = \mu\csMatrix^T$, $\myMat{W}^2 = \myMat{I} -\mu\csMatrix^T\csMatrix$) with the same number of iterations for recovering a $200\times 1$ $4$-sparse signal from $150$ compressed noisy measurements, where the \ac{mse} is averaged over $1000$ realizations. The results in Fig.~\ref{fig:LISTA_LOSS} showcase the ability of deep unfolding with learned objective parameters to notably improve performance over conventional model-based optimization by jointly learning a surrogate objective along with dedicated hyperparameters. 

{For additional examples on deep unfolding for jointly learning the objective and the hyperparameters see examples in \cite{solomon2019deep, li2020efficient, dardikman2020learned}, and the surveys in \cite{sahel2022deep,monga2021algorithm}.}

\subsection{Learning Unfolded Abstracted Optimizers}
The third variant of the deep unfolding methodology uses an iterative optimizer to form an abstract \ac{dnn} whose operation is inspired by the model-based optimizer rather than preserving or even specializing it. An example for such an unfolding of an iterative optimization algorithm, which introduces further abstraction beyond reparameterization of the model-based iterations, is the DetNet architecture proposed in \cite{samuel2019learning}. This algorithm is obtained by unfolding the projected gradient descent iterations for   detecting binary vectors in linear settings with Gaussian noise. This corresponds to, e.g., symbol detection in \ac{mimo} communications, where the constraint follows from the discrete nature of digital symbols. 

\begin{figure*}
	\centering
	\includegraphics[width=\linewidth]{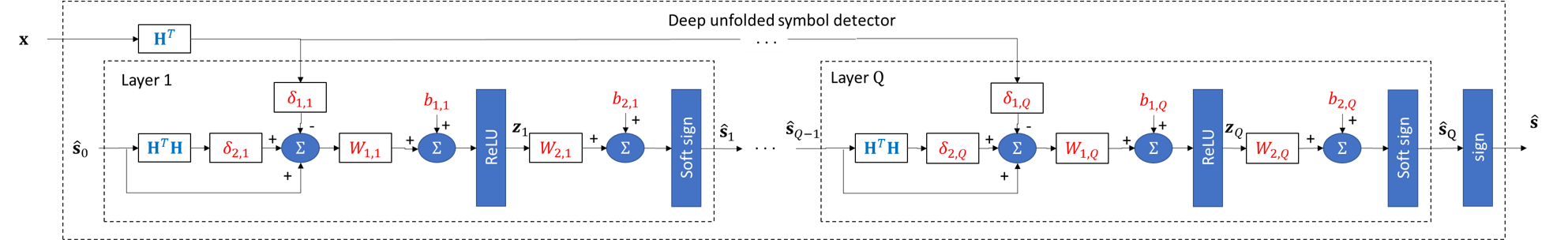} 
	\caption{DetNet illustration. Parameters in red fonts are learned in training, while those in blue fonts are externally provided.} 
	\label{fig:DetNet}
\end{figure*}

\begin{example}
	\label{exm:DetNet}
	Consider the optimization problem formulated by
	\begin{equation}
	\label{eqn:DetNetObj}
	\hat{\myVec{s}} = \mathop{\arg \min}\limits_{\myVec{s}\in\{\pm 1\}^\Nusers} \|\Input-\myMat{H}\myVec{s}\|^2,
	\end{equation}
	where $\myMat{H}$ is a known matrix.  
	While directly solving \eqref{eqn:DetNetObj} involves an exhaustive search over the $2^\Nusers$ possible  combinations, it can be tackled with affordable computational complexity using the iterative projected gradient descent algorithm. Let $\mySet{P}_{\mySet{S}}(\cdot)$ denote the projection operator into $\mySet{S}$, which for $\mathcal{S}=\{\pm 1\}^K$ is the element-wise sign function. Projected gradient descent iterates via
	\begin{align}
	\hat{\myVec{s}}^{(k + 1)} 
	&= \mySet{P}_{\mySet{S}}\left(\hat{\myVec{s}}^{(k)} - \eta_k \left.\frac{\partial \|\Input-\myMat{H}\myVec{s}\|^2}{\partial \myVec{s}}\right|_{\myVec{s} = \hat{\myVec{s}}^{(k)}} \right) \notag \\
	&=  \mySet{P}_{\mySet{S}}\left(\hat{\myVec{s}}^{(k)} - \eta_k\myMat{H}^T\Input + \eta_k\myMat{H}^T\myMat{H}\hat{\myVec{s}}^{(k)} \right)
	\label{eqn:ProjGrad}
	\end{align}
	where $\eta_k$ denotes the step size at iteration $k$, and $\hat{\myVec{s}}_0$ is set to some initial guess.

	DetNet builds upon the observation that \eqref{eqn:ProjGrad} consists of two stages: gradient descent computation, i.e., gradient step $\hat{\myVec{s}}^{(k)} - \eta_k\myMat{H}^T\Input + \eta_k\myMat{H}^T\myMat{H}\hat{\myVec{s}}^{(k)}$, and projection, namely, applying $ \mySet{P}_{\mySet{S}}(\cdot)$. Therefore, each unfolded iteration is represented as two sub-layers: The first sub-layer learns to compute the gradient descent stage by treating the step-size as a learned parameter and applying a \ac{fc} layer with ReLU activation to the obtained value. For iteration index $k$, this results in 
	\begin{equation*}
	\myVec{z}^{(k)} \!=\! {\rm ReLU}\left(\myMat{W}_{1,k}\left((\myMat{I}\! +\! \delta_{2,k}\myMat{H}^T\myMat{H})\hat{\myVec{s}}^{(k-1)} \!-\! \delta_{1,k}\myMat{H}^T\Input  \right) \!+\! \myVec{b}_{1,k}  \right)
	\end{equation*}
	in which $\{\myMat{W}_{1,k}, \myVec{b}_{1,k}, \delta_{1,k}, \delta_{2,k}\}$ are learnable parameters. 
	The second sub-layer learns the projection operator by approximating the sign operation with a soft sign activation proceeded by an \ac{fc} layer, leading to
	\begin{equation}
	\hat{\myVec{s}}^{(k)} = {\rm soft~sign}\left( \myMat{W}_{2,k} \myVec{z}^{(k)} +\myVec{b}_{2,k} \right),
	\label{eqn:Layer2}
	\end{equation}
	where the soft sign operation is applied element-wise and is given by ${\rm soft~sign}(x)=\frac{x}{1+|x|}$. 
	Here, the learnable parameters are $\{\myMat{W}_{2,k}, \myVec{b}_{2,k}\}$. 
	The resulting deep network is depicted in Fig. \ref{fig:DetNet}, in which the output after $K$ iterations is used as the estimated symbol vector by taking the sign of each element.  
	
	Let  $\myVec{\theta} = \{(\myMat{W}_{1,k}, \myMat{W}_{2,k}, \myVec{b}_{1,k}, \myVec{b}_{2,k}, \delta_{1,k}, \delta_{2,k})\}_{k=1}^Q$ be the trainable parameters of DetNet\footnote{The formulation of DetNet in \cite{samuel2019learning} includes an additional sub-layer in each iteration intended to further lift its input into higher dimensions and introduce additional trainable parameters, as well as reweighing of the outputs of subsequent layers. As these operations do not follow directly from unfolding  projected gradient descent, they are not included in the description here.}.  To tune $\myVec{\theta}$, the overall network is trained end-to-end  to minimize the empirical weighted $\ell_2$ norm loss over its intermediate layers, given by 
	\begin{equation}
	\label{eqn:LossDetNet}
	\mySet{L}(\myVec{\theta}) = \frac{1}{\Ntraining}\sum_{i=1}^{\Ntraining}\sum_{k=1}^{K} \log(k)\|\myVec{s}^i - \hat{\myVec{s}}^{(k)}(\Input^i; \myVec{\theta}) \|^2
	\end{equation}
	where $\hat{\myVec{s}}_k(\Input^i; \myVec{\theta})$ is the output of the $k$th layer of DetNet with parameters $\myVec{\theta}$ and input $\Input^i$. This loss measure accounts for the interpretable nature of the unfolded network, in which the output of each layer is a further refined estimate of $\myVec{s}$.
	
\end{example}

\section{DNN-Aided Priors}
\label{sec:DNNpriors}
\begin{figure}
	\centering
	\includegraphics[width=0.9\columnwidth]{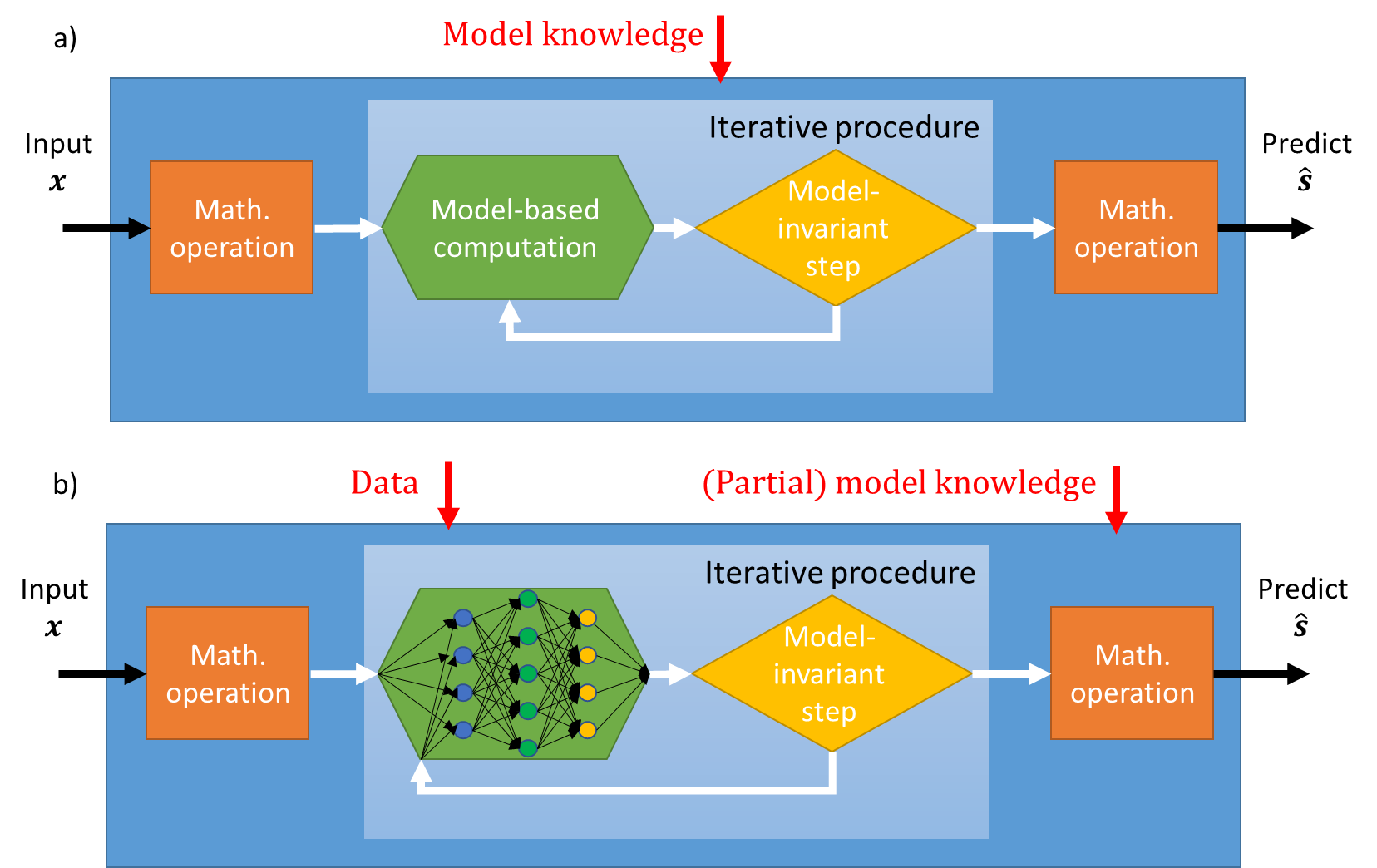}
	\caption{DNN-aided inference illustration: $a)$ a model-based algorithm comprised of multiple iterations with intermediate model-based computations; $b)$ A data-driven implementation of the algorithm, where the specific model-based computations are replaced with dedicated learned deep models.}
	\label{fig:DNN-Aided1}
\end{figure}

In the previous section we introduced the strategy of deep unfolding for combining principled mathematical models with data-driven \acp{dnn} \cite{gregor2010learning}. In unfolded networks, the resultant inference system is a deep network whose architecture imitates the operation of a model-based iterative optimizer, and we discussed three different unfolding approaches which arise from the inherent parameterization of iterative algorithms.  The next sections introduce \ac{dnn}-aided inference, which is a family of model-based deep learning algorithms in which \acp{dnn} are incorporated into model-based methods. Here, inference is carried out using a traditional model-based method, while some of the intermediate computations are augmented by \acp{dnn}, as illustrated in Fig.~\ref{fig:DNN-Aided1}.
 
The main motivation of \ac{dnn}-aided inference is to exploit the established benefits of model-based methods, in terms of performance, complexity, and suitability for the problem at hand. Deep learning is incorporated to mitigate sensitivity to inaccurate model knowledge, facilitate operation in complex environments, limit computational burden, and enable application of the algorithm in new domains. In this section, we  focus on \ac{dnn}-aided inference for facilitating optimization over complex and intractable signal domains (coined {\em structure-agnostic \ac{dnn}-aided inference} in \cite{shlezinger2020model}). 

While model-based deep learning is mostly about converting model-based algorithms into trainable architectures that can be learned end-to-end, namely, via discriminative learning, the family of \ac{dnn}-aided priors scheme is typically concerned with generative learning, i.e., learning the statistical model rather than the inference task. However, as opposed to conventional generative learning as exemplified in Section~\ref{sec:GencDisc}, where learning is used to fill in missing parameters of an imposed system model, here the rationale is to exploit the abstractness of \acp{dnn} to implicitly capture the underlying statistics, without having to explicitly model it.


	\ac{dnn}-aided priors is thus a family of methods which use deep learning tools to implicitly learn structures and statistical properties of the signal of interest, in a manner that is amenable to model-based optimization. These inference systems are particularly relevant for various inverse problems in signal processing, including denoising, sparse recovery, deconvolution, and super resolution \cite{ongie2020deep}. 	Tackling such problems typically involves imposing some structure on the signal domain. This prior knowledge is then incorporated into a model-based optimization procedure, which recovers the desired signal with provable performance guarantees. 
	
	 To formulate this, we consider optimization problems which take on  the form 
\begin{equation}
\label{eqn:regOpt0}
\hat{\myVec{s}} = \mathop{\arg\max}\limits_{\myVec{s}} \mySet{L}(\myVec{s}) = \mathop{\arg\max}\limits_{\myVec{s}}  g_{\myVec{x}}(\myVec{s}) +\phi(\myVec{s}).
\end{equation}
Note that \eqref{eqn:regOpt0} is stated as an unconstrained optimization whose objective is comprised of two terms -- a {\em data matching} term   $g_{\myVec{x}}(\cdot)$, which depends on the observations $\myVec{x}$, and a {\em regularization term}, or {\em prior}, denoted $\phi(\cdot)$, that is invariant of $\myVec{x}$. 

Optimization problems of the form \eqref{eqn:regOpt0} are tackled by various forms of deep unfolding, as discussed in Section~\ref{sec:Unfolding}. There, the optimization problem is known and $\phi(\cdot)$ is given (up to perhaps some missing parameters such as regularization coefficients);  model-based deep learning is thus employed to realize an effective solver,  e.g., to cope with {\em algorithm deficiency.} \ac{dnn}-aided priors tackle a complementary challenge, which  stems from the difficulty in characterizing the prior term $\phi(\cdot)$, i.e., {\em model deficiency}. 
For instance, how does one capture the marginal distribution of natural images? or of human speech? 

%

We next discuss two approaches for designing inference rules via \ac{dnn}-aided priors to cope with the challenge of characterizing prior terms for various types of data. The first, termed {\em plug-and-play networks}, augments iterative optimizers such as \ac{admm} with a \ac{dnn} to bypass the need to express a proximal mapping. The second approach, coined {\em deep priors}, employs deep generative networks for capturing the signal prior in a manner that is amenable to applying gradient-based iterative optimizers for the resulting optimization problem.

\subsection{Plug-and-Play Networks}
Plug-and-play networks utilize deep denoisers as learned proximal mappings, which are key ingredients in many iterative optimization methods. Here, one uses \acp{dnn} to carry out an optimization procedure which relies on a regularized objective without having to compute it, i.e., without expressing the desired signal domain in tractable form. 
We focus on the combination of plug-and-play networks with \ac{admm} optimization, though this methodology can also be combined with other optimizers such a \ac{ista}. We exemplify it for tackling the regularized objective of \eqref{eqn:RegOpt4}, i.e.,
\begin{align}
\hat{\myVec{s}}
&=\mathop{\arg\min}\limits_{\myVec{s}} \frac{1}{2}\|\myVec{x}-\myMat{H}\myVec{s}\|^2 +\phi(\myVec{s}).
\label{eqn:RegOpt4a}
\end{align}

\paragraph{Plug-and-Play \ac{admm}} 
The \ac{admm} algorithm \cite{boyd2011distributed} was introduced in Example~\ref{exm:ADMM} and formulated in Algorithm~\ref{alg:Algoadmm} as an iterative optimizer seeking the saddle point of the primal-dual formulation of the optimization problem in \eqref{eqn:RegOpt4}.
The key challenge in implementing Algorithm~\ref{alg:Algoadmm} stems from the computation of the proximal mapping in Step \ref{stp:prox}. There are two main difficulties associated with this step: 
\begin{enumerate}
	\item Explicit knowledge of the prior is required, which is often not available.
	\item Even when one has a good approximation of $\phi(\cdot)$, computing the proximal mapping  may still be extremely challenging to carry out analytically. 
\end{enumerate}

\begin{sidewaysfigure}
	\centering
	\includegraphics[width=\linewidth]{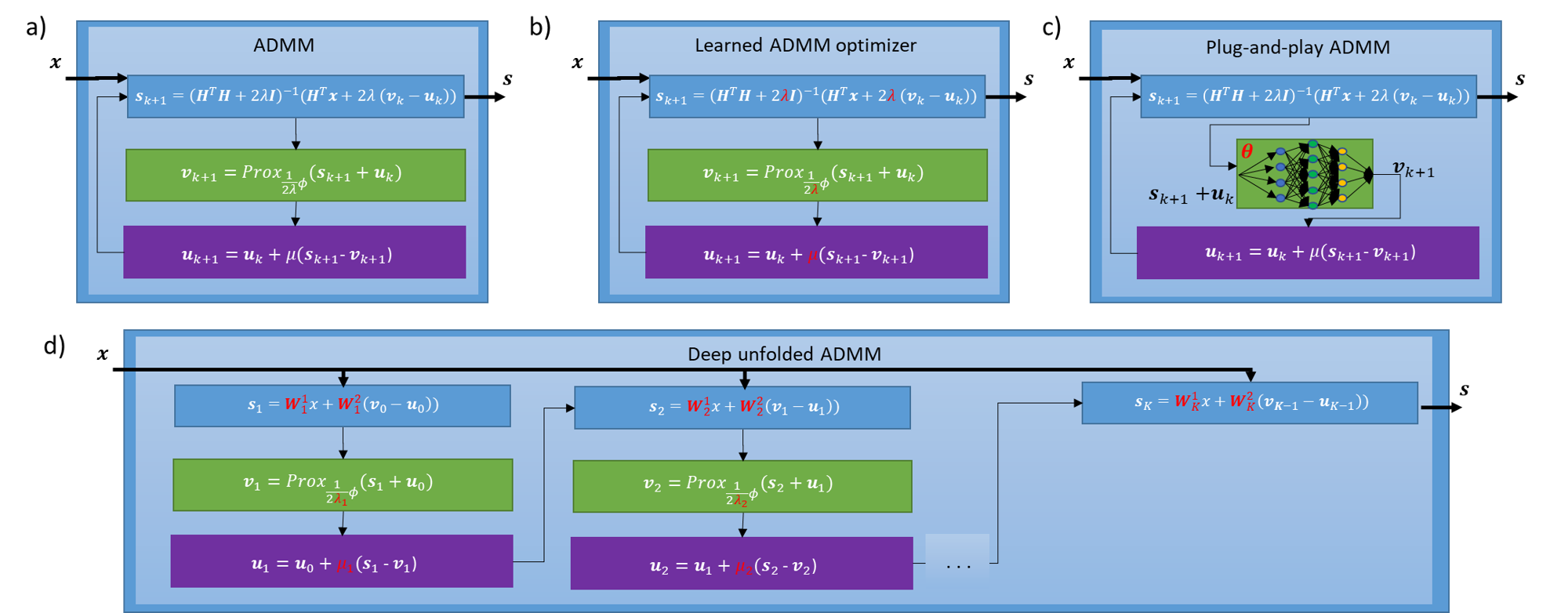}
	\caption{An illustration of the model-based deep learning strategies arising from the \ac{admm} optimizer (Algorithm~\ref{alg:Algoadmm}), where variables in \textcolor{red}{red fonts} represent trainable parameters: $a)$ the model-based optimizer; $b)$ a learned \ac{admm} optimizer; $c)$ plug-and-play \ac{admm}; and $d)$ deep unfolded \ac{admm}.}
	\label{fig:ADMMComp2} 
\end{sidewaysfigure}

Nonetheless, the proximal mapping is invariant of the task. In particular, as we show in the following example, it is the solution to a denoiser  for signals with prior $\phi(\cdot)$.
\begin{example}
	\label{exm:denoise}
	Consider a {\em denoising} task of recovering $\myVec{s}$ from its noisy measurements of the form 
	\begin{equation}
	\myVec{x} = \myVec{s} + \myVec{w},
	\end{equation}
	where $\myVec{w} \sim \mathcal{N}(\myVec{0}, \sigma^2 \myMat{I})$. In this case, the \ac{map} rule of Example~\ref{exm:InverseModel0} can be applied, where the log conditional probability becomes
	\begin{align}
	-\log p(\myVec{x}|\myVec{s}) &=    -\log (2\pi)^{-k/2} {\rm det} (\sigma^2 \myMat{I})^{-1/2} \exp\{- \frac{1}{2\sigma^2}\|\myVec{x}-\myVec{s}\|^2 \} \notag \\
	&= \frac{1}{2\sigma^2}\|\myVec{x}-\myVec{s}\|^2  +{\rm const}.
	\end{align}
	Consequently, the \ac{map} rule of \eqref{eqn:Recovery1} for the denoising task is given by
	\begin{align}
	\hat{\myVec{s}}_{\rm MAP}   
	&=\mathop{\arg\min}\limits_{\myVec{s}} \frac{1}{2}\|\myVec{x}-\myVec{s}\|^2 +\sigma^2\phi(\myVec{s}) \notag \\
	&= {\rm prox}_{\sigma^2\cdot \phi}(\myVec{x}).
	\label{eqn:RegOpt3}
	\end{align}
 	It is noted that even when the noise is not known to obey a Gaussian distribution, modeling the denoiser task as \eqref{eqn:RegOpt3} is still reasonable being the least-squares objective. 
\end{example}
Denoisers are common \ac{dnn} models, which can be trained in an unsupervised manner (e.g., autoencoders) and are known to operate reliably on signal domains with intractable priors (e.g., natural images). One can thus implement Algorithm~\ref{alg:Algoadmm} without having to specify the prior $\phi(\cdot)$ by replacing Step~\ref{stp:prox} with a \ac{dnn} denoiser, as illustrated in Fig.~\ref{fig:ADMMComp2}.


The term {\em plug-and-play} \cite{ahmad2020plug} is used to describe decision mappings as in the above example where pre-trained models are plugged into model-based optimizers without further tuning, as illustrated in Fig.~\ref{fig:ADMMComp2}(c). Nonetheless, this methodology can also incorporate deep learning into the optimization procedure by, e.g., unfolding  the iterative optimization steps into a large \ac{dnn} whose trainable parameters are those of the smaller networks augmenting each iteration, as in \cite{gilton2019neumann,wei2022deep}. This approach allows to benefit from both the ability of deep learning to implicitly represent complex domains, as well as the inference speed reduction  of deep unfolding along with its robustness to uncertainty and errors in the model parameters assumed to be known. Nonetheless, the fact that the iterative optimization must be learned from data in addition to the prior on $\mySet{S}$ implies that typically larger amounts of labeled data are required to train the system, compared to using the model-based optimizer.

\subsection{Deep Priors}
The plug-and-play approach avoids having to actually compute the prior $\phi(\cdot)$ by going directly to the optimization algorithm which builds upon the regularized objective, and specifically selecting an optimizer where the incorporation of $\phi(\cdot)$ can be replaced with a standard denoiser (due to, e.g., variable splitting as in \ac{admm}). An alternative approach to augment model-based solvers with pre-trained \acp{dnn} is the usage of {\em deep priors} \cite{bora2017compressed}.
As opposed to plug-and-play networks, which focus on the {\em solver} that is based on the regularized optimization problem~\eqref{eqn:regOpt0}, e.g., on \ac{admm},  deep priors use \acp{dnn} to {\em replace the optimization problem with a surrogate one}. Here, deep learning tools are particularly employed to yield a surrogate optimization problem which is amenable to tackling using gradient based approaches, while being a reliable approximation of the true (possibly intractable) optimization problem.

\paragraph{Generative Priors}
Consider again the super-resolution problem detailed in Example~\ref{exm:InverseModel0}. However,  let us assume that we have access to a bijective mapping from some latent space $\mySet{Z}$ to the signal space $\mySet{S}$, denoted $G:\mySet{Z}\mapsto \mySet{S}$. In this case, assuming discrete-valued random variables, the prior term $\phi(\myVec{s})$ becomes
\begin{align}
\phi(\myVec{s}) &= -\log p(\myVec{s}) + {\rm const} \notag \\
&= -\log p(\myVec{z} = G^{-1}(\myVec{s})) + {\rm const} := \tilde{\phi}(\myVec{z})|_{\myVec{z} = G^{-1}(\myVec{s})}.
\label{eqn:deepPrior1}
\end{align}
We can now write the regularized optimization problem as 
\begin{align}
\hat{\myVec{s}} 
&= G(\hat{\myVec{z}}), \quad \text{where} \notag \\
\hat{\myVec{z}} &=\mathop{\arg\min}\limits_{\myVec{z}} g_{\myVec{x}}\left(G(\myVec{z})\right) +\tilde{\phi}(\myVec{z}).
\label{eqn:RegOpt3alt0}
\end{align}
For the \ac{map} rule in \eqref{eqn:Recovery1}, this becomes
\begin{align}
\hat{\myVec{s}}_{\rm MAP}  
&= G(\myVec{z}_{\rm MAP}), \quad \text{where} \notag \\
\myVec{z}_{\rm MAP} &=\mathop{\arg\min}\limits_{\myVec{z}} \frac{1}{2}\|\myVec{x}-\myMat{H}G(\myVec{z})\|^2 +\sigma^2\tilde{\phi}(\myVec{z}).
\label{eqn:RegOpt3alt}
\end{align}
The rationale of generative priors is thus to find a mapping $G(\cdot)$  with which the surrogate optimization problem \eqref{eqn:RegOpt3alt0} is a faithful approximation of the true optimization problem~\eqref{eqn:regOpt0}, e.g., one can employ \eqref{eqn:RegOpt3alt} to seek the \ac{map} rule \eqref{eqn:Recovery1}. Specifically, when using deep learning for generative priors, the learned prior is differentiable, facilitating tackling the surrogate problem via gradient based methods, as detailed next, where we focus our description on the approximation of the \ac{map} rule \eqref{eqn:Recovery1}.

\paragraph{Deep Generative Priors}
When one has access to a mapping $G(\cdot)$ for which the latent prior $\tilde{\phi}(\cdot)$ is computable and differentiable, the surrogate regularized objective \eqref{eqn:RegOpt3alt} can be used to recover the \ac{map} estimate $\hat{\myVec{s}}_{\rm MAP}$. The learning of such mappings $G(\cdot)$ that map from a known distribution into a complex signal distribution is what is carried out by deep generative models, such as \acp{gan}. Exploiting this property for regularized optimization was proposed in \cite{bora2017compressed}. Here, that latent variable $\myVec{z}$ takes a zero-mean Gaussian distribution with i.i.d. entries, and thus there exists $\lambda >0$ for which
\begin{equation}
\sigma^2\tilde{\phi}(\myVec{z}) = \lambda \|\myVec{z}\|^2 + {\rm const}.
\end{equation}

Deep generative priors use a pre-trained \ac{dnn}-based prior $G_{\myVec{\theta}}$, typically a \ac{gan} trained over the domain of interest in an unsupervised manner. Consequently, the design of \ac{dnn}-aided priors commences with training a deep generative model for the considered data type. Once the trained $G_{\myVec{\theta}}$ is available, one can formulate the surrogate optimization problem as
\begin{equation}
\myVec{z}_{\rm MAP} =\mathop{\arg\min}\limits_{\myVec{z}} \frac{1}{2}\|\myVec{x}-\myMat{H}G_{\myVec{\theta}}(\myVec{z})\|^2 + \lambda \|\myVec{z}\|^2 := \mathop{\arg\min}\limits_{\myVec{z}}\mySet{L}(\myVec{z}).
\label{eqn:csLoss}
\end{equation}
Even though the exact formulation of $G_{\myVec{\theta}}$ may be highly complex, one can tackle \eqref{eqn:csLoss} via gradient-based optimization, building upon the fact that \acp{dnn} allow simple computation of gradients via backpropagation, e.g.
\begin{equation}
\myVec{z}_{k+1} =  \myVec{z}_{k} -\eta_k \nabla_{\myVec{z}=\myVec{z}_k}\mySet{L}(\myVec{z}).
\label{eqn:gradStep}
\end{equation}
Note that the gradient is computed in \eqref{eqn:gradStep} not with respect to the weights (as done in conventional \ac{dnn} training), but with respect to the input of the network.

\begin{figure}
	\centering
	\includegraphics[width=0.7\columnwidth]{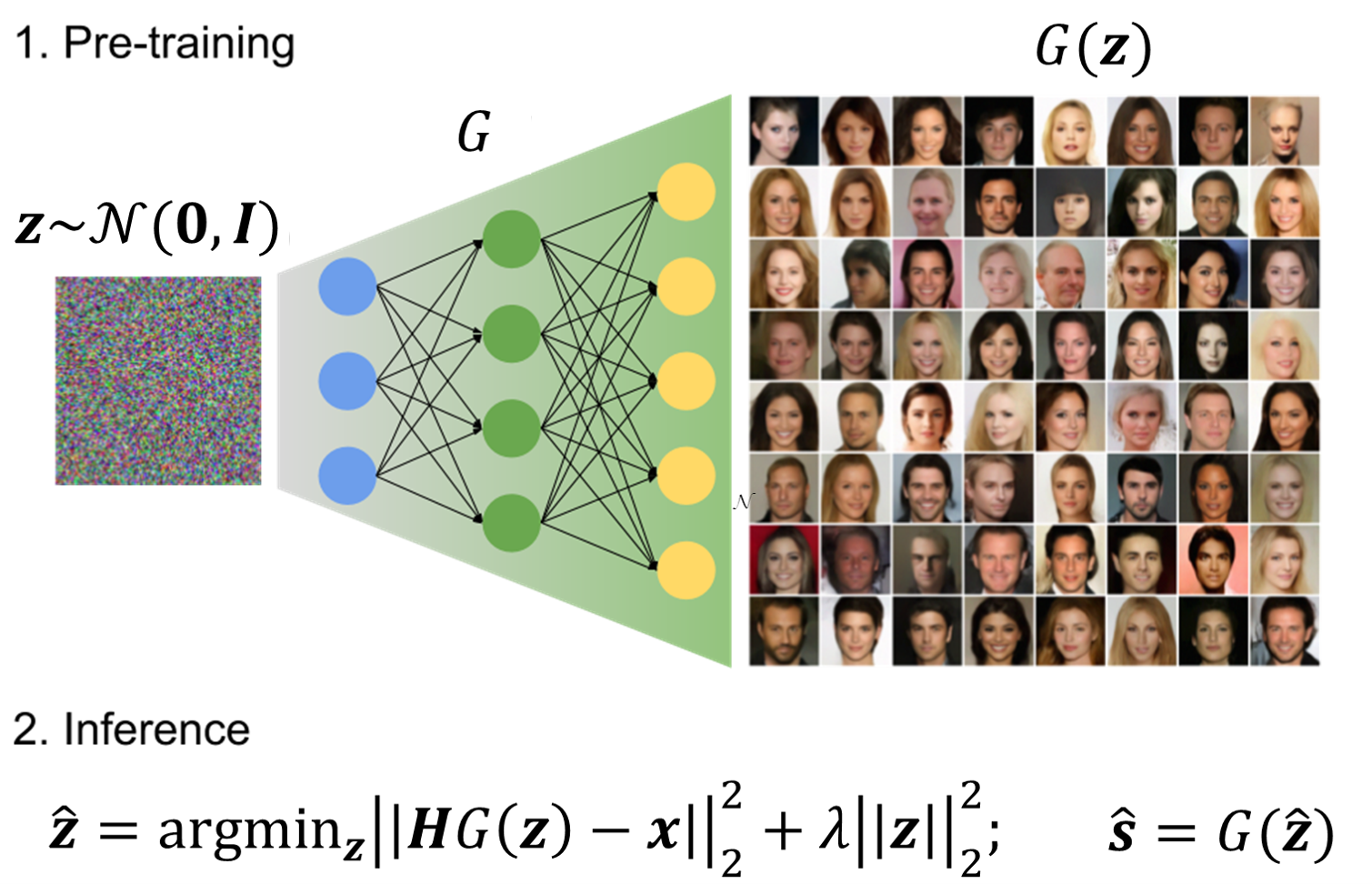}
	\caption{High-level overview of tackling   \eqref{eqn:RegOpt3} with a \ac{dnn}-based prior. The generator network $G$ is pre-trained to map Gaussian latent variables to plausible signals in the target domain. Then signal recovery is done by finding a point in the range of $G$ that minimizes the reconstruction error via gradient-based optimization over the latent variable.} 
	\label{fig:CSGM_flow}	 
\end{figure}

In summary, deep generative priors replace the constrained optimization over the complex input signal with tractable optimization over the latent variable $\csLatent$, which follows a known simple distribution. This is achieved using a pre-trained \ac{dnn}-based prior $G_{\myVec{\theta}}$ to map it into the domain of interest. Inference is performed by minimizing $\mySet{L}$ in the latent space of $G_{\myVec{\theta}}$ via, e.g., \eqref{eqn:gradStep}. {An illustration of the system operation is depicted in Fig.~\ref{fig:CSGM_flow}}.

\section{DNN-Aided Inference}
\label{sec:DNNinference}
 The last strategy we review combines conventional \ac{dnn} architectures with model-based optimization to enable the latter to operate reliably in complex domains. The rationale here is to preserve the  objective and structure of a model-based decision mapping suitable for the problem at hand based on the available domain knowledge, while augmenting computations that rely on approximations and missing domain knowledge with model-agnostic \acp{dnn}. \ac{dnn}-aided inference thus aims at benefiting from the best of both worlds by accounting in a principled manner for the available domain knowledge while using deep learning to cope with the elusive aspects of the problem description. 
 
 Unlike the strategies discussed in the previous sections, which are relatively systematic and can be viewed as recipe-style methodologies, \ac{dnn}-aided inference accommodates a broad family of different techniques for augmenting model-based optimizers with \acp{dnn}. We next discuss two representative \ac{dnn}-aided inference approaches, with the first replacing internal computations of a model-based method with \acp{dnn}, while the latter adds an external \ac{dnn} to correct internal computations. We exemplify these approaches  based on our running examples.{Additional examples from other domains in signal processing and communications can be found in, e.g., \cite{shlezinger2019viterbinet, shlezinger2020inference, luijten2020adaptive}.}

\subsection{Replacing Model-Based Computations with \acp{dnn}}
Often, model-based algorithms can be preferred based on some structural knowledge of the underlying model, despite the fact that its subtleties may (and often are) unknown. 
In such cases, one can exploit domain knowledge of statistical structures to carry out model-based inference methods in a data-driven fashion. These hybrid systems thus 
utilize deep learning not for the overall inference task, but for robustifying and relaxing the model-dependence of established model-based inference algorithms designed specifically for the structure induced by the specific  problem being solved. 

%

We next demonstrate how this rationale is translated into hybrid model-based/data-driven algorithms. We consider both the running example of tracking of dynamic systems,  based on \cite{revach2021kalmannet}, and that of \ac{doa} estimation, which is based on \cite{Shmuel2023deep}.

\begin{example}
	\label{exm:KNet}
	Consider a dynamic system characterized by a state-space model
	\begin{subequations}
		\label{eqn:ssmodela}
		\begin{align}
		\label{eqn:ssmodel1b}
		\myVec{s}_{t+1} &= f(\myVec{s}_t)+ \myVec{v}_t, \\
		\myVec{x}_t &=  h(\myVec{s}_{t}) +\myVec{w}_t.
		\label{eqn:ssmodel2}
		\end{align}
	\end{subequations}
	Here, $f(\cdot)$ and $h(\cdot)$ are possibly non-linear mappings, and the noise sequences $\myVec{v}_t, \myVec{w}_t$ are zero-mean and i.i.d. in time. 
	However, we do not have a concrete stochastic model for the noise signals, which are often introduced as a way to capture stochasticity, and thus $\myMat{V}$ and $\myMat{W}$ are often tuned by hand. 
	Yet, we are given a data set comprised of $\Ntraining$ trajectories of $T$ observations along with their corresponding states and actions, i.e., $\mySet{D}=\{\{\myVec{x}_t^i, \myVec{s}_t^i\}_{t=1}^T\}_{i=1}^{\Ntraining}$.
	We focus on the filtering task, i.e., \ac{mse} estimation of $\myVec{s}_t$ from $\{\myVec{x}_\tau\}_{\tau\leq t}$.
	
	Clearly, if the noise is Gaussian with known moments and the functions  $f(\cdot)$ and $h(\cdot)$ are linear, then the Kalman filter achieves the minimal \ac{mse}. Even when $f(\cdot)$ and $h(\cdot)$ are non-linear, one can still use variants of the Kalman filter, such as the extended Kalman filter, in which the prediction stage is obtained by
	\begin{subequations}
		\label{eqn:Pred1}
		\begin{align}
		\hat{\myVec{s}}_{t|t-1} &= f(\hat{\myVec{s}}_{t-1}), \\ 
		\hat{\myVec{x}}_{t|t-1} &= h(\hat{\myVec{s}}_{t|t-1}), 
		\end{align}
	\end{subequations}
	and the state estimate still takes the form 
	\begin{equation}
	\label{eqn:KalmanEq}
	\hat{\Label}_{t} = \hat{\Label}_{t|t-1} +\myMat{K}_t\cdot(\Input_t - \hat{\Input}_{t|t-1}).
	\end{equation}
	Recall that in the conventional Kalman filter, the Kalman gain $\myMat{K}_t$ is computed by tracking the second order moments of the predictions in \eqref{eqn:Pred1}; when the setting is non-linear, the \ac{ekf} computes these moments using Taylor series approximations of $f(\cdot)$ and $h(\cdot)$. While this allows dealing with the non-linearity of  $f(\cdot)$ and $h(\cdot)$ (though in a sub-optimal manner), one still has to cope with the unknown distribution of the noise signals. In fact, if the noise signals are not Gaussian, a linear estimator as in \eqref{eqn:KalmanEq} (where $\myMat{K}_t$ is not a function of the inputs of the system) is likely to result in degraded performance. 
	
	\begin{figure}
		\centering
		\includegraphics[width=0.8\linewidth]{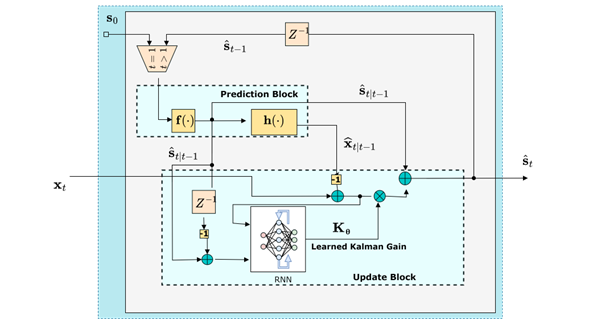}
		\caption{DNN augmented Kalman filter illustration. Here, the operation of the model-based Kalman filter is preserved in its division into prediction step (based on \eqref{eqn:Pred1}) and update step (based on \eqref{eqn:KalmanEq}), while deep learning is employed to learn from data to compute the Kalman gain, denoted $\myMat{K}_{\myVec{\theta}}$.}
		\label{fig:KalmanNet}
	\end{figure}

	Since the dependency on the noise statistics in the Kalman filter is encapsulated in the Kalman gain $\myMat{K}_t$, its computation can be replaced with a trainable \ac{dnn}, and thus \eqref{eqn:KalmanEq} is replaced with 
	\begin{equation}
	\hat{\Label}_{t} = \hat{\Label}_{t|t-1} +\myMat{K}_{\myVec{\theta}}(\Input_t,\hat{\Label}_{t-1})\cdot  (\Input_t - \hat{\Input}_{t|t-1}),
	\label{eqn:KalmanNEt}
	\end{equation}
	where $\myMat{K}_{\myVec{\theta}}(\cdot)$ is a \ac{dnn} with parameters $\myVec{\theta}$. Since $\myMat{K}_t$ is updated recursively, its learned computation is carried out with an \ac{rnn}. One can consider different architectures for implementing  $\myMat{K}_{\myVec{\theta}}(\cdot)$: the work \cite{revach2021kalmannet} proposed two architectures, one that is based on a single \ac{rnn} with preceding and subsequent \ac{fc} layers, and one that is based on three \acp{rnn}. An alternative architecture was proposed in \cite{choi2022split}, where two \acp{rnn} were used to track the covariances of $\hat{\myVec{s}}_{t|t-1}$ and of $	\hat{\myVec{x}}_{t|t-1}$, respectively, which allow to compute the model-based Kalman gain via \eqref{eqn:Kalman3}. By letting $f(\cdot;\myVec{\theta})$ be the latent state estimate computed using \eqref{eqn:KalmanNEt} with parameters $\myVec{\theta}$,  the overall system is trained end-to-end via
	\begin{equation}
	\myVec{\theta}^*=\mathop{\arg \min}_{\myVec{\theta} } \frac{1}{\Ntraining T}\sum_{i=1}^{\Ntraining}\sum_{t=1}^{T}\|
	f(\myVec{x}_{t}^i;\myVec{\theta}) - \myVec{s}_{t}^i\|^2_2.
	\label{eqn:KalmanNEtObj}
	\end{equation}
	The resulting system is illustrated in Fig.~\ref{fig:KalmanNet}.
\end{example}

\begin{figure}
	\centering
	\includegraphics[width=0.6\columnwidth]{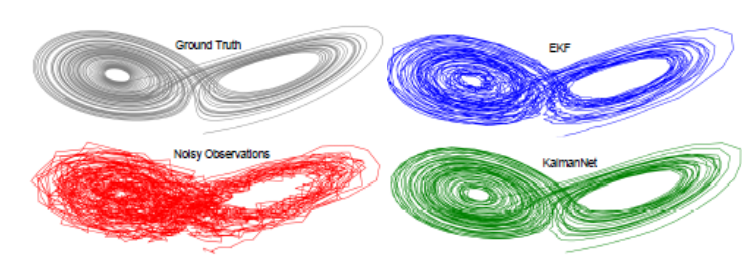}
	\caption{Tracking a single trajectory using the \ac{dnn}-aided KalmanNet compared with the model-based EKF (reproduced from \cite{revach2021kalmannet}).}
	\label{fig:KNet} 
\end{figure}

The \ac{dnn}-aided Kalman filter of Example~\ref{exm:KNet}, referred to as {\em KalmanNet} \cite{revach2021kalmannet}, learns to track dynamic systems from data while preserving the operation of the model-based Kalman filter and its variants. The ability to successfully track complex trajectories is illustrated in Fig.~\ref{fig:KNet}, which considers  tracking of the three-dimensional Lorenz attractor chaotic system, where the model is trained using trajectories of length $T=200$ and evaluated on $3000$ samples. The gains of \ac{dnn}-aided inference are not only in performance, but also in accuracy and run-time. This is showcased in Table~\ref{tbl:decimation}, where the \ac{dnn}-aided KalmanNet is compared with the direct application of an \ac{rnn} trained with the same data, as well as model-based variants of the Kalman filter deisnged for non-linear settings -- the \ac{ekf}; unscented Kalman filter (UKF); and particle filter (PF). We observe in Table~\ref{tbl:decimation} that \ac{dnn}-aided inference leverages data to simultaneously improve \ac{mse} along with inference speed, as well as generalization compared with conventional \acp{dnn}.

\begin{table}
	\begin{center}
		\begin{tabular}{|c|c|c|c|c|c|c|}
			\hline
			&  EKF&  UKF &   PF & KalmanNet & \ac{rnn}\\
			\hline
			MSE {\rm{[dB]}} &  -6.432 & -5.683 & -5.337 & {\bf -11.284} & 17.355\\
			\hline
			\textrm{Run-time}{\rm{[sec]}} &  5.440
			& 6.072 & 62.946
			& {\bf 4.699} & 2.291\\
			\hline
		\end{tabular}
		\caption{\ac{mse} performance and run-time of the \ac{dnn}-aided KalmanNet, end-to-end \ac{rnn}, and the model-based EKF, UKF, and PF.}
		\label{tbl:decimation}
	\end{center} 
\end{table}

Additional gains of the \ac{dnn}-aided inference based design of Example~\ref{exm:KNet} include   the ability to train in an unsupervised manner, as well as the extraction of uncertainty in the estimate. The former follows from the fact that Kalman filter internally predicts the next observation as $\hat{\Input}_{t|t-1}$. Hence, one can train the model to produce accurate predictions of the future observations, exploiting its awareness of the state evolution and observation functions. This is achieved by replacing \eqref{eqn:KalmanNEtObj} with \cite{revach2021unsupervised} 
	\begin{equation}
\myVec{\theta}^*=\mathop{\arg \min}_{\myVec{\theta} } \frac{1}{\Ntraining T}\sum_{i=1}^{\Ntraining}\sum_{t=2}^{T}\|
\hat{\Input}_{t|t-1}(\myVec{x}_{t-1}^i;\myVec{\theta}) - \Input_{t}^i\|^2_2, 
\label{eqn:KalmanNEtObjUsup}
\end{equation}
where in \eqref{eqn:KalmanNEtObjUsup} $\hat{\Input}_{t|t-1}(\myVec{x}_{t-1}^i;\myVec{\theta})$ is the input predicted at time step $t$ based on the inputs up to time $t-1$ using KalmanNet with parameters $\myVec{\theta}$.

\begin{figure}
	\centering
	\includegraphics[width=0.7\columnwidth]{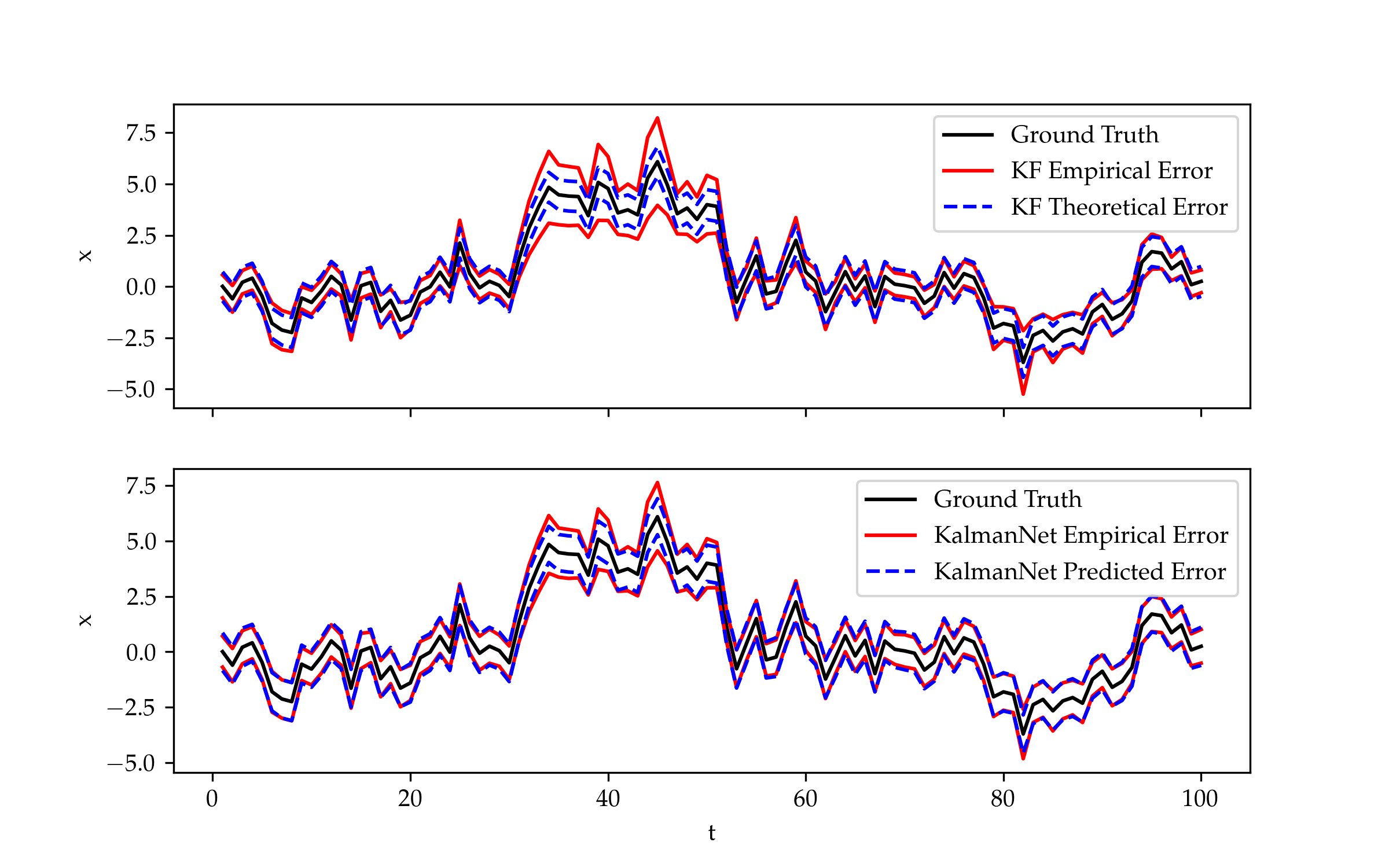}
	\caption{Tracking a single trajectory along with the error regions predicted by the model-based Kalman filter (upper plot) compared with KalmanNet (lower plot) when tracking based on   a mismatched linear state-space model (reproduced from \cite{klein2022uncertainty}).} 
	\label{fig:confidence_mismatch} 
\end{figure}

The second gain implies that the \ac{dnn} augmented KalmanNet can  recover not only the state $\Label_t$, but also provide an estimate of the error covariance matrix. This follows from the fact that the Kalman gain matrix can be in some conditions mapped into the error covariance matrix ($\myMat{\Sigma}_{t|t}$ in \eqref{eqn:SigmaKal}). In particular, 
when the state-space model has linear observations $\myMat{H}$ with full column rank, i.e.,  $\tilde{\myMat{H}}=\big(\myMat{H}^T\myMat{H}\big)^{-1}$ exists, then, filtering with Kalman gain $\myMat{K}_t$  results in estimation with error covariance  \cite[Thm. 1]{klein2022uncertainty} 
\begin{align}
\myMat{\Sigma}_{t|t}=\left(\myMat{I}-\myMat{K}_t\cdot\myMat{H}\right)&\cdot\tilde{\myMat{H}}
\myMat{H}^T\left(\myMat{I}-\myMat{H}\cdot\myMat{K}_{t}\right)^{-1}\myMat{H} \myMat{K}_{t} \myMat{H}\tilde{\myMat{H}}.  
\label{eqn:KN_covariance}
\end{align}
Consequently, one can use \eqref{eqn:KN_covariance} to map the learned Kalman gain $\myMat{K}_{\myVec{\theta}}(\Input_t,\hat{\Label}_{t-1})$ into an estimate of the error covariance. This yields a reliable characterization of the error which is robust to mismatches in the state-space model, as illustrated in Fig.~\ref{fig:confidence_mismatch}, where KalmanNet produces a better estimate of the uncertainty in its estimate compared with a Kalman filter with both models operating with a mismatched state evolution model.

Example~\ref{exm:KNet} is shown to overcome non-linearities and mismatches in the state-space model, outperforming the classical Kalman filter while retaining its data efficiency and interpretability. This is achieved by augmenting the operation of the Kalman filter with a dedicated \ac{dnn} model for computing the specific part of the algorithm flow which encapsulates its domain knowledge sensitivity, while converting the algorithm into a  discriminative model that is trainable end-to-end. A similar rationale can be applied to augmenting subspace methods for \ac{doa} estimation, detailed in Examples~\ref{exm:Subspace}-\ref{exm:RootMUSIC}, overcoming their associated limitations to, e.g., coherent and narrowband sources.

\begin{example}
	\label{exm:SubspaceNet}
	Consider the \ac{doa} estimation setting detailed in Example~\ref{exm:DoA}. 
	As discussed in Example~\ref{exm:Subspace}, when the sources are narrowband and non-coherent,  the covariance of the observations can be decomposed into orthogonal signal and noise subspaces, from which the \acp{doa} can be extracted via, e.g., RootMUSIC (Example~\ref{exm:RootMUSIC}). 
	
	When the above assumptions do not necessarily hold, yet one has access to a data set comprised of sequences of observations and their corresponding \ac{doa} vectors, i.e., $\Data = \{\Input_1^i,\ldots,\Input_{T_i}^i, \Label^i\}_{i=1}^{\Ntraining}$,  deep learning can be utilized to learn to compute a {\em surrogate covariance} which obeys the desired subspace decomposition. In this case, the empirical covariance in \eqref{eqn:MUSICCov} is replaced with a  mapping $\myMat{C}_{\myVec{\theta}}(\Input_1,\ldots,\Input_{T})$, where $\myMat{C}_{\myVec{\theta}}(\cdot)$ is a \ac{dnn} with parameters $\myVec{\theta}$. Since the \ac{dnn} is applied to a time sequence whose duration $T$ may vary between sequences, a possible implementation uses \acp{rnn}~\cite{merkofer2022deep}. Alternatively, one can set  $\myMat{C}_{\myVec{\theta}}(\cdot)$ to be a convolutional autoencoder applied to the empirical autocorrelation of $\Input_t$, as proposed in \cite{Shmuel2023deep}.

	While there is no ground truth surrogate covariance (i.e., a covariance matrix that can be decomposed into orthogonal signal and noise subspaces), the \ac{dnn} can be trained as part of the RootMUSIC algorithm, i.e., to optimize the ability of the subspace method to decompose the learned covariance into orthogonal signal and noise subspaces. 
	By letting $f(\cdot;\myVec{\theta})$ be the \acp{doa} estimated by applying RootMUSIC to the covariance matrix computed using $\myMat{C}_{\myVec{\theta}}(\cdot)$, the resulting architecture is trained as a discriminative model via
	\begin{equation}
	\myVec{\theta}^*=\mathop{\arg \min}_{\myVec{\theta} } \frac{1}{\Ntraining }\sum_{i=1}^{\Ntraining}\|
f(\Input_1^i,\ldots,\Input_{T_i}^i;\myVec{\theta}) - \myVec{s}_{t}^i\|^2_2. 
\label{eqn:SubspaceNEtObj}
	\end{equation}	
\end{example}

In Example~\ref{exm:SubspaceNet}, the subspace method used to map the learned surrogate covariance into an estimate of the \acp{doa} is RootMUSIC. The selection of this specific subspace method is not arbitrary, but is specifically based on the fact that RootMUSIC translates the estimated covariance into \acp{doa} by seeking the roots of a polynomial, which is a {\em differentiable} mapping. This differentiability is key when converting an algorithm into a discriminative model, as deep learning training methods that are used to tune this model utilize gradient methods which inherently rely on differentiability of the mapping. For this reason, RootMUSIC is selected during training, as opposed to \ac{music}, whose inference rule is based on the non-differentiable procedure of finding peaks in the spatial spectrum. 

\begin{figure}
		\centering
	\includegraphics[width=0.8\linewidth]{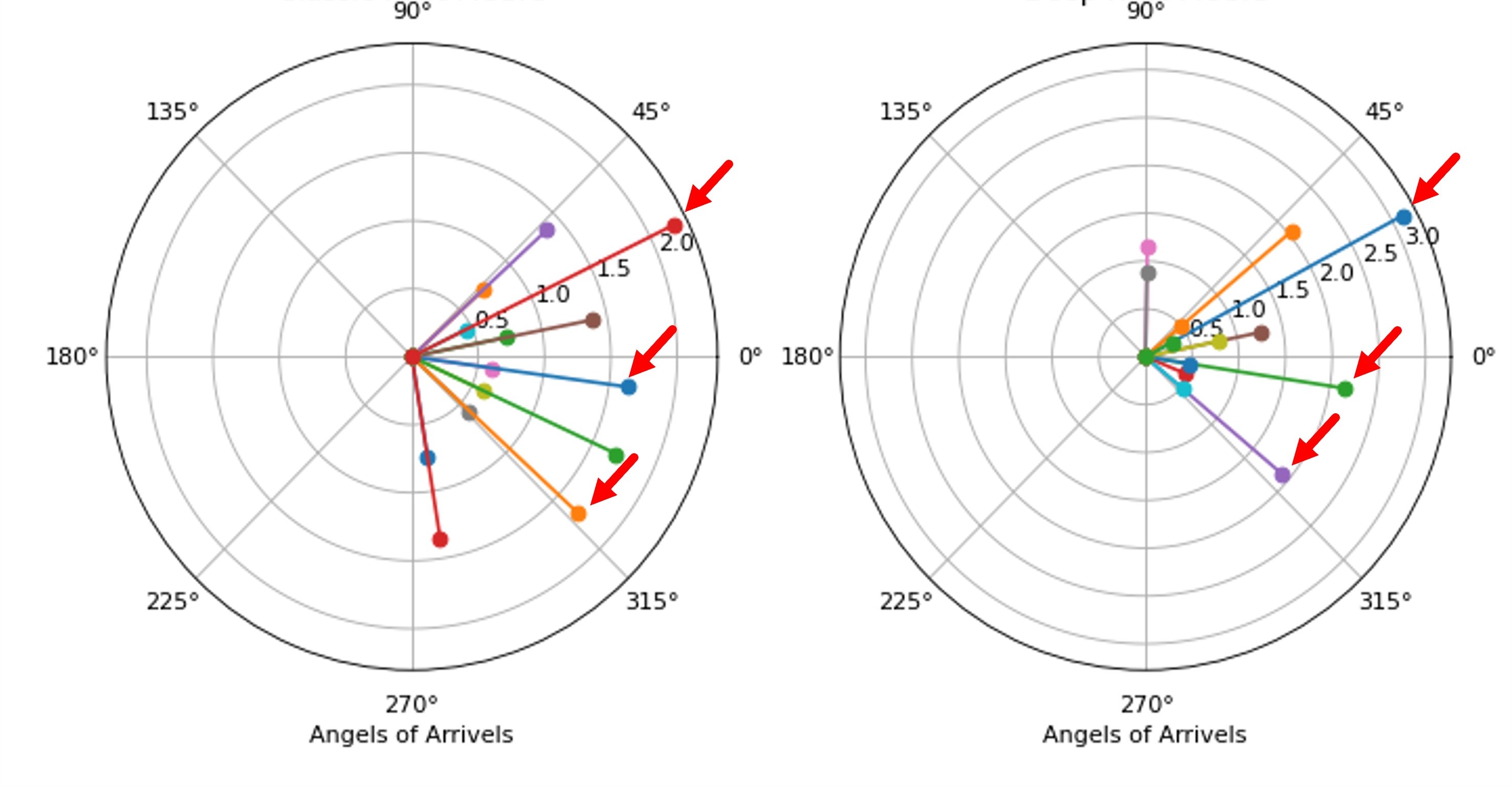}
	\caption{RootMuSIC spectrum obtained when applying classic RootMUSIC (left) and its \ac{dnn}-aided implementation (right) for recovering coherent sources at angles $12.5^\circ,89^\circ$. It is clearly observed that the \ac{dnn}-aided version pushes roots that do not correspond to true angles to be more distant from the unit circle (marked with red arrows), thus reducing the chance that they will be mistaken for true \acp{doa}. }
	\label{fig:DRMusic1}
\end{figure}

\begin{figure}
	\centering
	\includegraphics[width=0.8\linewidth]{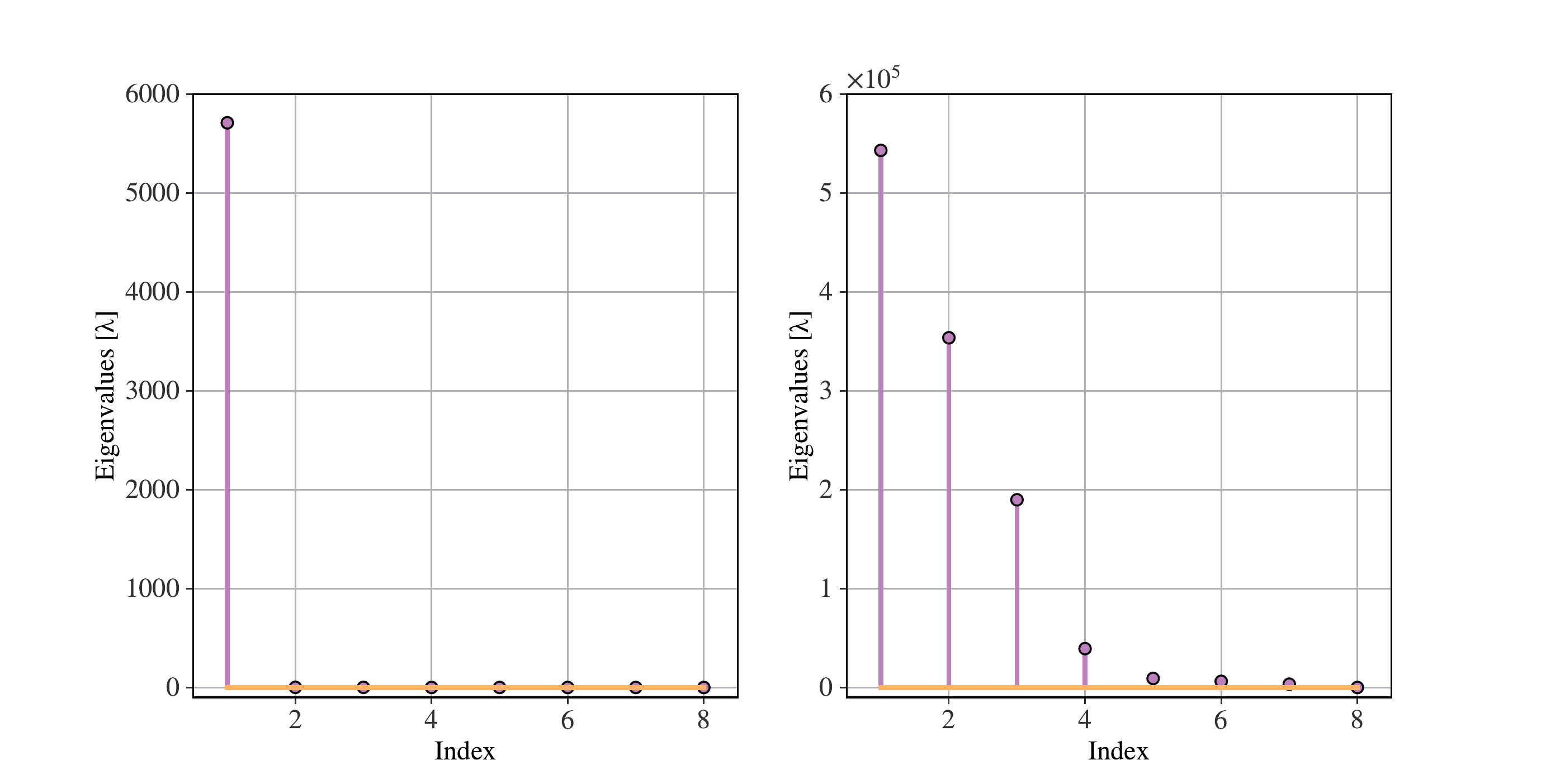}
	\caption{Eigenvalues of the covariance matrix obtained via conventional covariance estimation (left) and using an augmented \ac{dnn} (right) for measurements corresponding to three coherent sources. It is observed that the \ac{dnn}-aided implementation indeed yields a covariance matrix from which one can clearly identify the three eigenmodes corresponding to the signal subspace, as opposed to the conventional computation. }
	\label{fig:Eigenvalues_pdf}
\end{figure}

\begin{figure}
	\centering 
	\begin{subfigure}{0.45\textwidth}
		\centering
		{\includegraphics[width=\columnwidth]{MUSIC_Normalized.pdf}} 
		\caption{ \ac{music} spectrum. 
		}
		\label{fig:MUSIC_F} 	
	\end{subfigure}
	$\quad$
	\begin{subfigure}{0.45\textwidth}
		\centering
		{\includegraphics[width=\columnwidth]{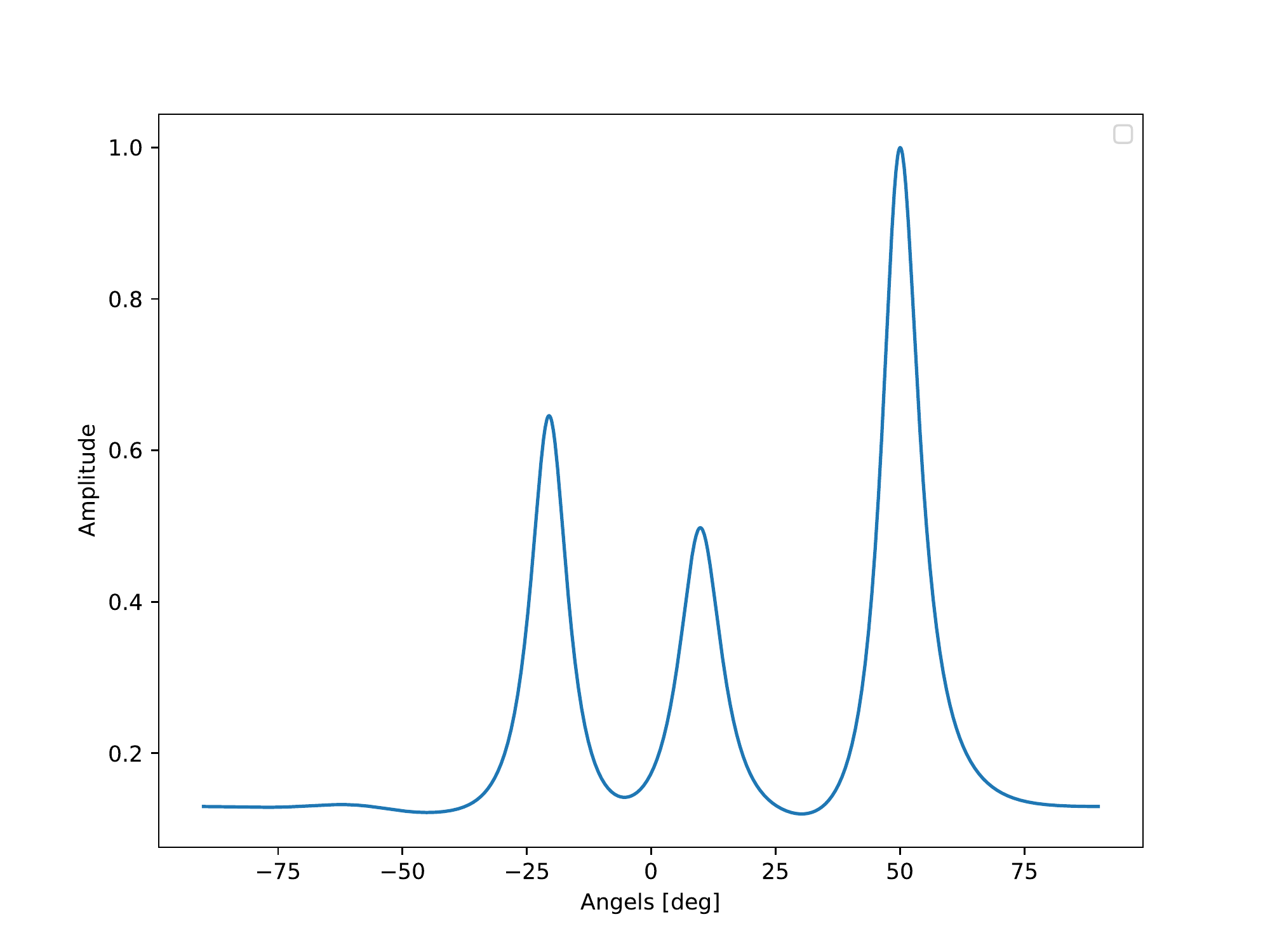}} 
		\caption{ \ac{dnn}-aided \ac{music} spectrum.
		}
		\label{fig:MUSIC_Faug} 
	\end{subfigure}
	\caption{The spectrum obtained by MUSIC when applied for recovering coherent sources located at angles $-22^\circ, 12^\circ, 50^\circ$.}
	\label{fig:MUSIC}
\end{figure}

Example~\ref{exm:SubspaceNet} thus trains its \ac{dnn} to produce surrogate covariances that are decomposable into orthogonal subspaces using RootMUSIC.  Specifically, the architecture employed to compute $\myMat{C}_{\myVec{\theta}}(\Input_1,\ldots,\Input_{T})$ is comprised of a convolutional autoencoder with three conlvolutional layers and three deconvolution layers with anti-rectifier activations, whose input features are the empirical autocorrelation of $\Input_1,\ldots,\Input_{T}$. 
The trained \ac{dnn} yields a surrogate covariance matrix with RootMUSIC spectrum from which one can clearly identify the \acp{doa}, even when dealing with  coherent sources, as visualized in Fig.~\ref{fig:DRMusic1}. Once trained, the resulting surrogate covariance mapping is decomposable into noise and signal subspaces, as demonstrated in Fig.~\ref{fig:Eigenvalues_pdf}, and consequently can be combined with other subspace methods, such as MUSIC, as we show in Fig.~\ref{fig:MUSIC}. This reveals a key gain of \ac{dnn}-aided inference, and of model-based deep learning in general, in its ability to  preserve the interpretable nature of model-based algorithms and its associated gains while using deep learning to enable operation in domains where these classic algorithms often fail.

%
%
%
%
%
%

\subsection{External DNN Augmentation}
\label{sec:DNNAug}
\begin{figure}
	\centering
	\includegraphics[width=\columnwidth]{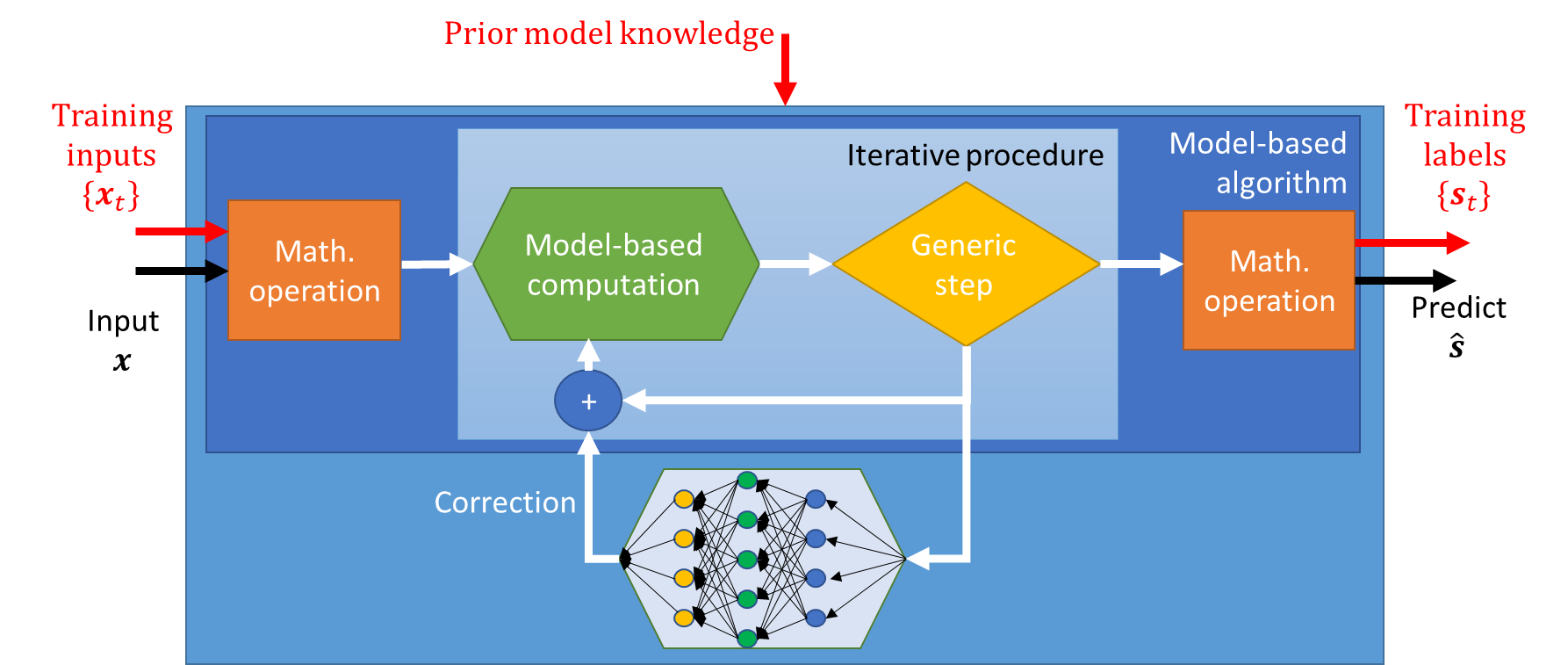}
	\caption{Neural augmentation illustration.}
	\label{fig:NeuralAug1}
\end{figure}

The \ac{dnn}-aided inference strategies detailed so far utilize model-based algorithms to carry out inference, while replacing explicit domain-specific computations with dedicated \acp{dnn}. An alternative approach  utilizes the complete model-based algorithm for inference, i.e., without embedding deep learning into its components, while using an external \ac{dnn} for correcting some of its intermediate computations.  An illustration of this approach is depicted in Fig.~\ref{fig:NeuralAug1}.

The main advantage in utilizing an external \ac{dnn} for correcting internal computations 
stems from its ability to notably improve the robustness of model-based methods to inaccurate knowledge of the underlying objective parameters. Since the model-based algorithm is individually implemented, one must posses the complete domain knowledge it requires, and thus the external correction \ac{dnn} allows the resulting  system to overcome inaccuracies in this domain knowledge by learning to correct them from data. Furthermore, the learned correction term can improve the performance of model-based algorithms in scenarios where they are sub-optimal.

The design of external \ac{dnn} augmented inference systems is comprised of the following steps:
\begin{enumerate}
	\item Choose a suitable iterative model-based method for the problem of interest, and  identify the information exchanged between the iterations, along with  the intermediate computations used to produce this information.
	\item The information exchanged between the iterations is updated with a correction term learned by a \ac{dnn}. The \ac{dnn}  is designed to combine the same quantities used by the model-based algorithm, only in a learned fashion.
	\item The overall hybrid model-based/data-driven system is trained in an end-to-end fashion, where one considers not only the algorithm outputs in the loss function, but also the intermediate outputs of the internal iterations. 
\end{enumerate}

We next demonstrate how these steps are carried out in order to augment smoothing in dynamic systems, as proposed in \cite{satorras2019combining}.

\begin{example}
	\label{exm:AugKNet}		
	Consider again a linear Gaussian state-space model, i.e., \eqref{eqn:ssmodela} where $f(\myVec{s}_t) \equiv \myMat{F}\myVec{s}_t$, $h(\myVec{s}_t) \equiv \myMat{H}\myVec{s}_t$, and the noise signals $\myVec{v}_t$ and $\myVec{w}_t$ are Gaussian with covariance matrices $\myMat{V}$ and $\myMat{W}$, respectively. Now, we are  interested in recovering a sequence of $\Blklen$ state variables $\{\myS_t\}_{t=1}^\Blklen$  from the entire observed sequence $\{\Input_t\}_{t=1}^{\Blklen}$. While the loss is still the \ac{mse}, the setting is different from the filtering problem where one wishes to use solely $\{\Input_\tau\}_{\tau\leq t}$ when recovering $\Label_t$. 
	We focus on scenarios where the state-space model, which is available to the inference system, is an inaccurate approximation of the true underlying dynamics. 
	
	By writing  $\myVec{s} =[\myS_1,\ldots,\myS_{\Blklen}]^T$, one can estimate by gradient descent optimization on the joint log likelihood function, i.e., by iterating over
	\begin{equation}
	\label{eqn:GradKal}
	\myVec{s}^{(q+1)} = \myVec{s}^{(q)} + \gamma \nabla_{\myVec{s} }\log p\left(\Input, \myVec{s}^{(q)} \right)
	\end{equation}
	where $\gamma>0$ is a step-size. 
	The state-space model \eqref{eqn:ssmodela} implies that the joint \ac{pdf} satisfies
	\begin{align}
	p\left(\Input, \myVec{s}\right) &= p\left(\Input| \myVec{s}\right)p\left(\myVec{s}\right) = \prod_t p(\Input_t | \Label_t) p(\Label_t | \Label_{t-1}).
	\end{align}
	Consequently, it holds that 
	\begin{align}
	&\frac{\partial}{\partial \Label_t} \log p\left(\Input, \myVec{s}\right) \notag \\
	&= \frac{\partial}{\partial \Label_t} \sum_\tau \log p(\Input_\tau | \Label_\tau) +\sum_\tau \log p(\Label_\tau | \Label_{\tau-1}) \notag \\
	&= \frac{\partial}{\partial \Label_t} \log p(\Input_t | \Label_t) + \frac{\partial}{\partial \Label_t}  \log p(\Label_t | \Label_{t-1})  + \frac{\partial}{\partial \Label_t} \log  p(\Label_{t+1} | \Label_{t}) \notag \\
	&= \frac{\partial}{\partial \Label_t} (\Input_t - \myMat{H}\Label_t)^T\myMat{W}^{-1} (\Input_t - \myMat{H}\Label_t) 
	+ \frac{\partial}{\partial \Label_t} (\Label_t - \myMat{F}\Label_{t-1})^T\myMat{V}^{-1}(\Label_t - \myMat{F}\Label_{t-1}) \notag \\
	&\qquad
	+ \frac{\partial}{\partial \Label_t} (\Label_{t+1} - \myMat{F}\Label_{t})^T\myMat{V}^{-1}(\Label_{t+1} - \myMat{F}\Label_{t}) \notag \\
	&= \myMat{H}^T\myMat{W}^{-1}\left( \Input_{t}- \myMat{H} \myS_{t} \right) + -\myMat{V}^{-1}\left( \myS_t - \myMat{F} \myS_{t-1}  \right) + \myMat{F}^T\myMat{V}^{-1}\left( \myS_{t+1} - \myMat{F} \myS_{t}  \right).
	\end{align}
	Consequently, the $t$th entry of the log likelihood gradient in \eqref{eqn:GradKal}, abbreviated henceforth as $\nabla^{(q)}_t$, can be obtained as 
	$\nabla^{(q)}_t = \mu^{(q)}_{\myVec{S}_{t-1}\rightarrow \myVec{S}_t} +  \mu^{(q)}_{\myVec{S}_{t+1}\rightarrow \myVec{S}_t} + \mu^{(q)}_{\myVec{X}_{t}\rightarrow \myVec{S}_t}$, where the summands, referred to as messages, are given by
	\begin{subequations}
		\label{eqn:KalSmoothQuant1}
		\begin{align}
		\mu^{(q)}_{\myVec{S}_{t-1}\rightarrow \myVec{S}_t} & =  -\myMat{V}^{-1}\left( \myS_t^{(q)} - \myMat{F} \myS_{t-1}^{(q)}  \right),  \\
		\mu^{(q)}_{\myVec{S}_{t+1}\rightarrow \myVec{S}_t} & =  \myMat{F}^T\myMat{V}^{-1}\left( \myS_{t+1}^{(q)} - \myMat{F} \myS_{t}^{(q)}  \right),  \\
		\mu^{(q)}_{\myVec{X}_{t}\rightarrow \myVec{S}_t} & =  \myMat{H}^T\myMat{W}^{-1}\left( \Input_{t}- \myMat{H} \myS_{t}^{(q)}  \right).
		\end{align}
	\end{subequations}
	The iterative procedure in \eqref{eqn:GradKal}, is repeated until convergence, and the resulting $\myVec{s}^{(q)}$ is used as the estimate. 
	
	The gradient descent formulation in \eqref{eqn:GradKal} is evaluated by the messages \eqref{eqn:KalSmoothQuant1}, which in turn rely on accurate knowledge of the state-space model \eqref{eqn:ssmodela}. To facilitate operation with inaccurate model knowledge due to, e.g., \eqref{eqn:ssmodela} being a linear approximation of a non-linear setup, one can introduce neural augmentation to learn to correct inaccurate computations of the log-likelihood gradients. This is achieved by using an external \ac{dnn} to map the messages in \eqref{eqn:KalSmoothQuant1} into a correction term, denoted $\myVec{\epsilon}^{(q+1)}$. 
	
	The learned mapping of the messages into a correction term operates in the form of a \ac{gnn}. This is implemented by maintaining an internal node variable for each variable in \eqref{eqn:KalSmoothQuant1}, denoted $h_{\myS_t}^{(q)}$ for each $\myS_{t}^{(q)}$ and  $h_{\Input_t}$ for each $\Input_t$, as well as internal message variables $m_{\myVec{V}_n\rightarrow \myVec{S}_t}^{(q)}$ for each message computed by the model-based algorithm in \eqref{eqn:KalSmoothQuant1}, i.e., $\myVec{V}_n \in \{\myVec{S}_{t+1},\myVec{S}_{t-1}, \myVec{X}_{t}\}$. The node variables $h_{\myS_t}^{(q)}$ are updated along with the model-based smoothing algorithm iterations as estimates of their corresponding variables, while the variables $h_{\Input_t}$ are obtained once from $\Input$ via a neural network.  The \ac{gnn} then maps the messages produced by the model-based Kalman smoother into its internal messages via a neural network $f_{e}(\cdot)$ which operates on the corresponding node variables, i.e., 
	\begin{equation}
	m_{\myVec{V}_n\rightarrow \myVec{S}_t}^{(q)} = f_{e}\left(h_{\myVec{v}_n}^{(q)}, h_{\myS_t}^{(q)}, \mu^{(q)}_{\myVec{V}_{n}\rightarrow \myVec{S}_t}  \right)
	\label{eqn:MapEnc1}
	\end{equation}
	where $h_{\Input_n}^{(q)} \equiv h_{\Input_n}$ for each $q$. 
	These messages are then combined and forwarded into a \ac{gru}, which produces the refined estimate of the node variables $\{h_{\myS_t}^{(q+1)}\}$ based on their corresponding messages \eqref{eqn:MapEnc1}. Finally, each updated node variable $h_{\myS_t}^{(q+1)}$ is mapped into its corresponding error term $\myVec{\epsilon}_t^{(q+1)}$ via a fourth neural network, denoted $f_d(\cdot)$. 
	
	The correction terms $\{\myVec{\epsilon}_t^{(q+1)}\}$ aggregated into the vector  $\myVec{\epsilon}^{(q+1)}$ are used  to update the log-likelihood gradients, resulting in the update equation \eqref{eqn:GradKal} replaced with
	\begin{equation}
	\label{eqn:GradKal2}
	\myVec{s}^{(q+1)} = \myVec{s}^{(q)} + \gamma \left( \nabla_{\myVec{s} }\log \PdfNew{\myVec{X}, \myVec{S}}\left(\Input, \myVec{s}^{(q)} \right) + \myVec{\epsilon}^{(q+1)} \right).
	\end{equation}
	The overall architecture is illustrated in Fig.~\ref{fig:NeuralKalman1}. 
	
	\begin{figure}
		\centering
		\includegraphics[width=\columnwidth]{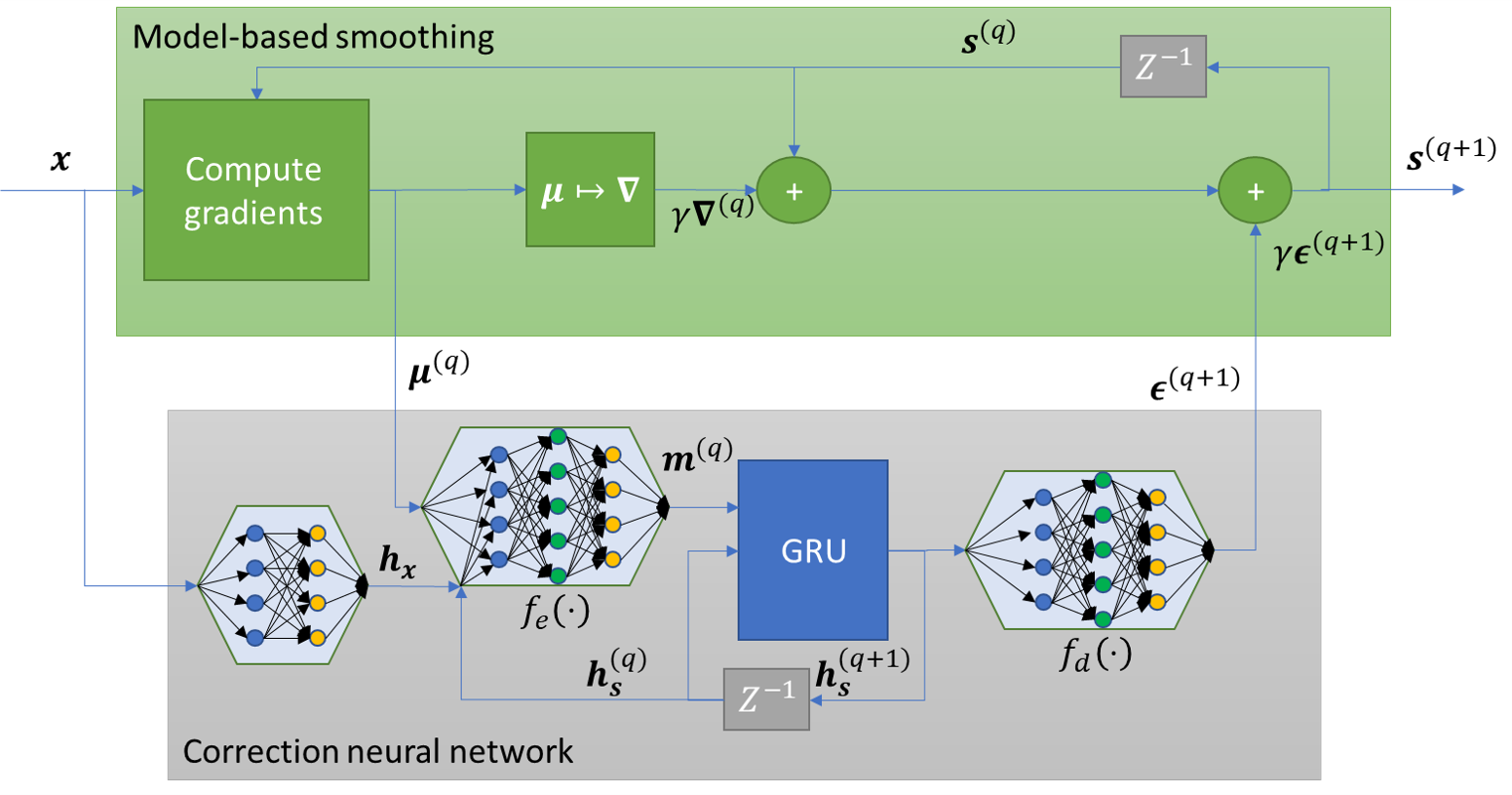}
		\caption{Neural augmented smoother illustration.}
		\label{fig:NeuralKalman1}
	\end{figure}

	Let $\myVec{\theta}$ be the  parameters of the \ac{gnn} in  Fig.~\ref{fig:NeuralKalman1}. The hybrid system is trained  end-to-end  to minimize the empirical weighted $\ell_2$ norm loss over its intermediate layers, where the contribution of each iteration to the overall loss increases as the iterative procedure progresses. In particular,  letting $\{(\myVec{s}^i, \Input^i) = ([\Label_1^i,\ldots,\Label_T^i]^T,[\Input_1^i,\ldots,\Input_T^i]^T)\}_{i=1}^{\Ntraining}$ be the training set, the loss function used to train the neural-augmented  smoother is given by
	\begin{equation}
	\label{eqn:LossKalman}
	\myVec{\theta}^*=\mathop{\arg \min}_{\myVec{\theta} }  \frac{1}{\Ntraining}\sum_{i=1}^{\Ntraining}\sum_{q=1}^{\Niter} \frac{q}{\Niter}\|\myVec{s}^i - \hat{\myVec{s}}_q(\Input^i; \myVec{\theta}) \|^2
	\end{equation}
	where $\hat{\myVec{s}}_q(\Input; \myVec{\theta})$ is the estimate produced by the $q$th iteration, i.e., via \eqref{eqn:GradKal2},  with parameters $\myVec{\theta}$ and input $\Input$.  
	
\end{example}
\newpage

\section{Summary}

\begin{tcolorbox}[width=\textwidth,colback={yellow}] 
	\begin{itemize}
		\item Model-based deep learning encompasses a family of methodology that can generally be viewed as a middle ground between highly-specific model-based methods and highly-parameterized deep learning.
		\item A repeated rationale in model-based deep learning focuses on the conversion of model-based algorithms into trainable architectures, and thus differentiability of the algorithm with respect to its internal features is key. 
		\item A straight-forward methodology is {\em learned optimization}, which uses deep learning tools to tune objective and hyperparameters from data.
		\item The popular {\em deep unfolding} methodology encompasses several different methods with varying levels of parameterization, which are all based on data-driven tuning of an iterative optimizer with a fixed number of iterations. 
		\item The strategy coined {\em \ac{dnn}-aided inference}  is not restricted to iterative optimizers, turning an algorithm into a trainable architecture by augmenting specific internal computations of the algorithm with \acp{dnn}, such that the overall flow is preserved while being amenable to end-to-end learning. An alternative approach is {\em external \ac{dnn} augmentation}, which instead of replacing an internal computation, carries it out in parallel using both a model-based module and a \ac{dnn}. 
		\item A form of model-based deep learning which uses generative learning is {\em \ac{dnn}-aided priors}, where the underlying statistical characterization is implicitly learned in a manner that can be incorporated into a dedicated optimizer.
	\end{itemize}
	
\end{tcolorbox}

\chapter{Conclusions}
\label{ch:Conclusions}
	In this monograph, we identified model-based optimization and data-driven deep learning as distinct edges of a spectrum varying in parameterization and specificity, and used this representation to provide a mapping of methodologies for combining classical model-based and data-driven inference via model-based deep learning. To conclude this overview, we pinpoint two fundamental questions -- {\em why} should one design inference rules via model-based deep learning, and {\em how} to approach such design tasks. To answer the first question, we summarize the key advantages of model-based deep learning in Section~\ref{sec:Conc_Discussion}. To clarify how should one approach a model-based deep learning design in light of the proposed categorization, we present  guidelines for selecting a design methodology for a given application, intended to facilitate the derivation of future hybrid data-driven/model-based systems. 

\section{Advantages of Model-Based Deep Learning}
\label{sec:Conc_Discussion}
The combination of traditional handcrafted algorithms with data-driven tools via model-based deep learning brings forth several key advantages. These can be divided based on their benchmark, whether model-based methods or conventional deep learning.

\paragraph{Advantages over Classical Model-Based Methods}
Compared to purely model-based schemes, the integration of deep learning effectively trades specificity for abstractness in a controllable fashion. Consequently, model-based deep learning facilitates {\em inference in complex environments}, where accurately capturing the underlying model in a closed-form mathematical expression may be infeasible. For instance, incorporating \ac{dnn}-based implicit regularization was shown to enable super-resolution without having the impose a prior (Section~\ref{sec:DNNpriors}), while integrating subspace methods with \acp{dnn} enabled their operation in setups involving coherent signals and few snapshots (Section~\ref{sec:DNNinference}). 

 The model-agnostic nature of deep learning also allows hybrid model-based/data-driven inference to achieve {\em improved resiliency to model uncertainty} compared to inferring solely based on domain knowledge. 
For example, augmenting the model-based Kalman with an \ac{rnn} can improve its performance when the state-space model does not fully reflect the true dynamics (Section~\ref{sec:DNNinference}).    

Finally, the fact that hybrid systems learn to carry out part of their inference based on data allows to infer with {\em reduced delay and less computational complexity} compared to the corresponding fully model-based methods. This reduction in inference speed is prominent when learning to optimize iterative optimizers via learned optimization (Section~\ref{sec:LearnedOpt}), and more substantially when using deep unfolding techniques (Section~\ref{sec:Unfolding}). Such improved latency can also be obtained when augmenting a model-based algorithm with a \ac{dnn} by replacing a specific time-costly computation with a trainable model of relatively low inference speed.

\paragraph{Advantages over Conventional Deep Learning}
Compared to utilizing conventional \ac{dnn} architectures for inference, model-based deep learning can be viewed as using the operation of a task-specific algorithm as an inductive bias. Proper selection of an inductive bias is known in the \ac{ml} literature to {\em facilitate learning}. This is translated into a few key advantages. First, the interleaving of deep learning into model-based algorithms yields  trainable architectures that are less prone to overfitting~\cite{shalev2014understanding}, and the lesser parameterization supports training with smaller data sets compared with highly-parameterized model-agnostic \acp{dnn}.
This gain also translates into improved {\em generalization}, as the trained architecture learns to carry out a suitable algorithm, rather than just map inputs to outputs based on the training data.
 Furthermore, when a parameterized architecture can be shown to specialize a model-based algorithm suitable for the task at hand, e.g., when using deep unfolding with learned objective parameters (Section~\ref{sec:Unfolding}), then this configuration can be used as a principled initialization, again facilitating the training procedure.

An additional core gain of model-based deep learning over conventional deep learning lies in the {\em interpretability} of the resulting inference rule. This interpretability indicates that one can assign a concrete meaning to internal features of the inference rule, and understand the operation and individual tasks of its internal building blocks. Interpretability improves trustworthiness as one can understand how decisions are made (as shown for, e.g., augmented subspace methods in Section~\ref{sec:DNNinference}), and is translated into operational gains. These include  facilitating training by properly penalizing internal features (as suggested in Section~\ref{sec:Unfolding}); robustness, particularly to noise, as this is often inherently supported by model-based methods~\cite{lavi2023learn}; provide uncertainty and confidence measures~\cite{klein2022uncertainty}; and enable quick adaptation to variations in the underlying statistical model~\cite{raviv2022online}.  Moreover, the fact that model-based deep learning follows the operation of a model-based algorithms, which is often associated with theoretical performance guarantees, facilitate the characterization of similar guarantees for some forms of model-based deep learning, as was shown in \cite{shultzman2023generalization,pu2022optimization,scarlett2022theoretical}.

\section{Choosing a Model-Based Deep Learning Strategy}
\label{sec:Conc_Choosing}
The aforementioned gains of model-based deep learning are shared at some level by all the different approaches presented in Chapter~\ref{ch:MBDL}. However, each strategy is focused on exploiting a different advantage of hybrid model-based/data-driven inference, particularly in the context of signal processing oriented applications. Consequently, to complement the mapping of model-based deep learning strategies and facilitate the implementation of future application-specific hybrid systems, we next enlist the main considerations one should take into account when seeking to combine model-based methods with data-driven tools for a given problem. 	 

\paragraph{Step 1: Domain knowledge and data  characterization:} First, one must ensure the availability of the two key ingredients in model-based deep learning, i.e., domain knowledge and data. The former corresponds to what is known \textit{a priori} about the problem at hand, in terms of statistical models and established assumptions, as well as what is unknown, or is based on some approximation that is likely to be inaccurate. The latter addresses the amount of labeled and unlabeled samples one posses in advance for the considered problem, as well as whether or not they reflect the scenario in which the system is requested to infer in practice.  

\paragraph{Step 2: Identifying a model-based method:} Based on the available domain knowledge, the next step is to identify a suitable model-based algorithm for the problem. This choice should rely on the portion of the domain knowledge which is {\em available}, and not on what is {\em unknown}, as the latter can be compensated for by integration of deep learning tools. This stage must also consider the requirements of the inference system in terms of performance, complexity, and real-time operation, as these are encapsulated in the selection of the algorithm. 
The identification of a model-based algorithm, combined with the availability of domain knowledge and data, should also indicate whether model-based deep learning mechanisms are required for the application of interest. 

\paragraph{Step 3: Implementation challenges:} Having identified a suitable model-based algorithm, the selection of the approach to combine it with deep learning should be based on the understanding of its main implementation challenges. Some representative  issues  and their relationship with the recommended model-based deep learning approaches include:
\begin{enumerate}
	\item Missing domain knowledge - model-based deep learning can implement the model-based inference algorithm when  parts of the underlying model are unknown, or alternatively, too complex to be captured analytically, by harnessing the model-agnostic nature of deep learning. In this case, the selection of the implementation approach depends on the format of the identified model-based algorithm:  Iterative optimizers are systematically converted into trainable architectures via deep unfolding (Section~\ref{sec:Unfolding}). Similarly, when the missing domain knowledge is represented as a complex search domain, \ac{dnn}-aided priors (Section~\ref{sec:DNNpriors}) can typically facilitate optimization with implicitly learned regularizers.  Finally, when the algorithm is divided into modules with a specific block identified as encompassing the missing domain knowledge, one can naturally augment its operation via \ac{dnn}-aided inference (Section~\ref{sec:DNNinference}).
	\item Inaccurate domain knowledge - model-based algorithms are typically sensitive to inaccurate knowledge of the underlying model and its parameters. In such cases, where one has access to a complete description of the underlying model up to some uncertainty, model-based deep learning can robustify the model-based algorithm and learn to achieve improved accuracy. A candidate approach to robustify model-based processing is by adding a learned correction term via external \ac{dnn} augmentation (Section~\ref{sec:DNNAug}). Alternatively, when the model-based algorithm takes an iterative form, improved resiliency can be obtained by unfolding the algorithm into a \ac{dnn} while learning either the hyperparameters or also the objective parameters (Section~\ref{sec:Unfolding}).
	\item Inference speed - model-based deep learning can learn to implement iterative inference algorithms, which typically require a large amount of iterations to converge, with reduced inference speed. This is achieved by by learning the hyperparameters of the optimizer, either on a iteration-dependent manner with fixed iterations via deep unfolding (Section~\ref{sec:Unfolding}) or by fully preserving the algorithm and tuning its hyperparameters via learned optimization (Section~\ref{sec:LearnedOpt}).
\end{enumerate}

The aforementioned implementation challenges constitute only a partial list of the considerations one should account for when selecting a model-based deep learning design approach. Additional considerations include computational capabilities during both training as well as inference; the  need to handle variations in the statistical model, which in turn translate to a possible requirement to periodically re-train the system; and the quantity and the type of available data. Nonetheless, the above division  provides systematic guidelines which one can utilize and possibly extend when seeking to implement an inference system relying on both data and domain knowledge.
Finally, we note that some of the detailed model-based deep learning strategies may be combined, and thus one can select more than a single design approach. For instance, one can interleave \ac{dnn}-aided inference via implicitly learned regularization and/or priors, with deep unfolding of the iterative optimization algorithm.

\begin{acknowledgements}
	The authors are grateful to Elad Sofer, Dor Haim Shmuel, Xiaoyong Ni, Lital Dabush, and Arbel Yaniv for their help with the numerical examples provided in this monograph. 
	We would also like to thank Stephen Boyd, Tirza Routtenberg, and Vishal Monga, 
	 for the helpful discussions leading to the formulation of the ideas summarized in this monograph. 
\end{acknowledgements}

\appendix
\chapter{Notations and Abbreviations}\label{App:Notations}

\section*{Mathematical Notation}

The following mathematical notations are used:


\begin{tabular}{l l} 	   
	$\mySet{R}$ & The set of real numbers\\ 
	$\mySet{R}^+$ & The set of non-negative real numbers\\ 
$\E\{\cdot\}$ & The stochastic expectation operator\\
$\Pr(\cdot)$ & The probability measure\\
$\log(\cdot)$ & The natural logarithm operator\\
$\exp(\cdot)$ & The natural exponent operator\\
$\det(\cdot)$ & The determinant operator\\
$\rm{sign}(\cdot)$ & The sign operator\\
$\|\cdot\|$ & The $\ell_2$ norm operator\\
$\|\cdot\|_p$ & The $\ell_p$ norm operator\\
$(\cdot)^T$ & The transpose operator\\
$(\cdot)^H$ & The Hermitian transpose operator\\
$\odot$ & Element-wise (Hadamard) product\\
$\myMat{I}$ & The identity matrix\\
$\mySet{N}(\myVec{\mu},\myMat{\Sigma})$ & The Gaussian distribution with mean $\myVec{\mu}$ and covariance $\myMat{\Sigma}$\\
	
\end{tabular} 

We use boldface lower-case and upper-case letters for vectors and matrices, respectively, i.e., $\myVec{x}$ is a vector, while $\myVec{X}$ is a matrix. Calligraphic letters, e.g., $\mySet{X}$, are used for sets. 

\section*{Abbreviations}
The following acronyms and abbreviations are used in this monograph:

\bigskip

\begin{tabular}{l l} 	   
	\acsu{admm} & \acl{admm}\\ 
	\acsu{ai} & \acl{ai}\\ 	
	\acsu{cnn} & \acl{cnn}\\ 
	\acsu{dnn} & \acl{dnn}\\ 
	\acsu{doa} & \acl{doa}\\ 
	\acsu{ekf} & \acl{ekf}\\ 
	\acsu{evd} & \acl{evd}\\ 
	\acsu{gnn} & \acl{gnn}\\ 
	\acsu{gru} & \acl{gru}\\ 
	\acsu{ista} & \acl{ista}\\  
	LISTA & learned ISTA \\
	\acsu{map} & \acl{map}\\
	\acsu{mse} & \acl{mse}\\
	\acsu{ml} & \acl{ml} \\
	\acsu{pdf} & \acl{pdf}\\ 
	\acsu{relu} & \acl{relu}\\ 
	\acsu{rnn} & \acl{rnn}\\ 
	\acsu{sgd} & \acl{sgd} 
	
\end{tabular} 
 
\backmatter  

\printbibliography

\end{document}